# Regular Partitions and Their Use in Structural Pattern Recognition

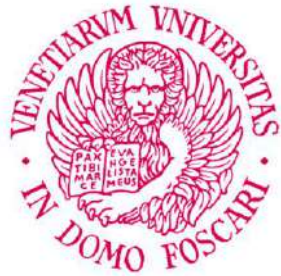

**Marco Fiorucci**

Supervisor: Prof. Marcello Pelillo

This dissertation is submitted for the degree of
*Doctor of Philosophy in Computer Science*

March 2019

*To my wife and to my parents*

# Acknowledgements

Gratus animus virtus est non solum maxima sed etiam mater virtutum omnium.

Marcus Tullius Cicero

First of all, I wish to express my sincerest recognition to my supervisor, prof. Marcello Pelillo, for being a caring and friendly guide. He always did his best to take time out of his busy schedule to talk with me: his valuable discussions helped me improve my scientific background and, in particular, taught me the importance of taking up new challenges out of my comfort zone.

My gratitude also goes to prof. Francisco Escolano from the Department of Computer Science and Artificial Intelligence of the Univesity of Alicante for granting me a research visiting position. In particular, I would like to thank him for his scientific advice, his knowledge on spectral graph theory and his friendly support all over my visiting periods in Spain. The days we spent discussing together at the university campus were very productive and pleasant.

Special thanks to Dr. Hannu Reittu for the great time I spent with him in VTT, Espoo. I will never forget our inspiring scientific walks in Finnish nature. I am so grateful to him for all the efforts he made to give me the opportunity to introduce my research work to the Department of Mathematics and Systems Analysis of Aalto University. During my stay he was so caring and became for me more than a coach at work: he made me feel at home and helped me discover and love Finnish culture.

Thanks to the reviewers of this thesis, prof. Pasquale Foggia and prof. Francesc Serratosa. I thank them for the time spent in carefully reading this work and for their useful comments and suggestions.

Furthermore, I am grateful to a number of people I had the pleasure to collaborate with during the past few years, in particular Ismail Elezi, Francesco Pelosin, Mauro Tempesta, Alessandro Torcinovich, Sebastiano Vascon, Eyasu Zemene and to all the people who joined our student branch. A special thank to the European Centre of Living Technology who has played an important role for me since my Master studies.

More personally, I want to express my gratitude to my family. Since my childhood my parents, Mario and Elmina, have always trusted me and stimulated to learn. I am also grateful to my brothers Paolo and Giacomo for all the experiences we have shared together. Thanks to all the people who have believed in me, my friends and relatives, who are still close despite most of them live hundreds of kilometers away. Among them, a special mention goes to my best friend Ennio, my uncle Vincenzo and my mother-in-law Martine. Finally, a lovely thank goes to Corinne, my wife, the person who has played the most import role all over this journey, encouraging me to follow my dreams since the very beginning and supporting me in any single step of this great adventure.

# Abstract

Recent years are characterized by an unprecedented quantity of available network data which are produced at an astonishing rate by an heterogeneous variety of interconnected sensors and devices. This high-throughput generation calls for the development of new effective methods to store, retrieve, understand and process massive network data. In this thesis, we tackle this challenge by introducing a framework to summarize large graphs based on *Szemerédi's Regularity Remma* (RL), which roughly states that any sufficiently large graph can almost entirely be partitioned into a bounded number of random-like bipartite graphs, called *regular pairs*. The partition resulting from the RL gives rise to a summary, called *reduced graph*, which inherits many of the essential structural properties of the original graph. Thus, this lemma provides us with a principled way to summarize a large graph revealing its main structural patterns, while filtering out noise, which is common in any real-world network.

We first extend an heuristic version of the RL to improve its efficiency and its robustness. We use the proposed algorithm to address graph-based clustering and image segmentation tasks. An extensive series of experiments demonstrated the effectiveness and the scalability of our approach. Along this path, we show how the notion of regular partition can provide fresh insights into old pattern recognition and machine learning problems. In addition, we analyze the practical implication of the RL in the preservation of metric information contained in large graphs. To do so, we use graph resistance-based measures to assess the quality of the obtained summaries, and to study the robustness of the proposed heuristic to natural sparsification of input proximity graphs.

In the second part of the thesis, we introduce a new heuristic algorithm which is characterized by an improvement of the summary quality both in terms of reconstruction error and of noise filtering. We use the proposed heuristic to address the graph search problem defined under a similarity measure. In our approach, all the graphs contained in a database are compressed off-line, while the query graph is compressed on-line. Thus, graph search can be performed on the summaries, and this allows to speed up the search process and to reduce storage space. To the best of our knowledge, we are the first to devise a graph search algorithm which efficiently deals with databases containing a high number of large graphs, and, moreover, it is principled robust against noise, which is always presented in real-world data.

Finally, we study the linkage among the regularity lemma, the stochastic block model and the minimum description length. This study provide us a principled way to develop a graph decomposition algorithm based on stochastic block model which is fitted using likelihood maximization. The RL is used as a prototype of the structural information which should be preserved, defining a new model space for graph-data. The minimum description length is exploited to obtain a stopping criterion which establishes when the optimal regular decomposition is found.

# Table of contents

# List of figures

# List of tables

# Preface

Chapter 1 is devoted to the introduction of the regularity lemma and the idea of using it for summarizing large graphs.

Chapter 2 describes an heuristic algorithm for finding regular partitions and its application to structural pattern recognition. This lead to a first publication in *Pattern Recognition Letters* [81]. Moreover, in this chapter is presented an analysis of the practical implication of the regularity lemma in the preservation of metric information contained in large graphs, which is reported in [39].

Chapter 3 introduces a new heuristic to summarize large graphs which is characterized by an improvement of the summary quality both in terms of reconstruction error and of noise filtering. Here, the proposed heuristic is used to address the graph search problem in order to speed up the search process and to reduce storage space. These results are under review on Pattern Recognition journal [38].

Chapter 4 is devoted to the study of the linkage among the regularity lemma, the stochastic block model and the minimum description length. It has been conducted in the last part of my Ph.D. course during a visit to the VTT Technical Research Centre of Finland. This study has been published in [85].

Finally, the Appendix of this thesis presents a system for beverage product recognition through the analysis of cooler shelf images, which has been published in [37]. This work is not linked to the main topic of the thesis and therefore it is not included as a chapter of this work.

# Introduction

We are surrounded by systems which exhibit a complex collective behavior that cannot be inferred only from the knowledge of its components. The twenty-first century, as Stephen Hawking stated, is the "century of complexity". Consider for example our brain which is composed of billions of neurons that, interacting in a coherent way, allow us to think, walk and feel; or a power grid which is made up of a huge number of interconnected devices designed to carry energy in our house and, at the same time, to ensure robustness against component failures. Hence, understanding and forecasting the behavior of these systems, called *complex systems*, is of exceptional relevance both for practical and theoretical reasons.

Beside being the century of complexity, the twenty-first century is also characterized by an unprecedented quantity of available data which are produced at an astonishing rate by an heterogeneous variety of interrelated sources. Such increasing amount of network data can play a key role in understanding the behavior of complex systems but, on the other hand, poses formidable computational problems which call for the development of new effective methods for extracting useful structural information from these massive network data. One compelling approach that has attracted increasing interest in recent years is *graph summarization*: building a concise representation of an input graph by revealing its main structural patterns. Applications range from clustering [26], to classification [105], to community detection [23], to outlier detection [96], to pattern set mining [107], to graph anonymization [49], just to name a few. The reader can refer to [67] for a survey on the applications of graph summarization.

Unfortunately, the graph summarization problem is not well-defined, i.e. any given graph can be summarized in drastically different ways, with the evaluation of the summary quality that is application dependent. For example, in the context of clustering, the reconstruction error is minimized; while in the context of retrieving, the query accuracy and time are optimized. However, Liu et al. [67] highlighted that graph summarization has the following main challenges: (a) speeding up graph analysis by performing them on the summary; (b) revealing interesting information: "the cut off between interesting and uninteresting can be difficult to determine in a principled way"; (c) complex and noisy data: noise is often contained in real-world networks; (d) evaluation of the summary quality is application dependent and it becomes even more difficult when "multi-resolution" summaries are considered; (e) summarizing dynamic graphs, i.e. graphs that change in time.

The aim of this thesis is to introduce a *principled* graph summarization framework addressing the following question:

> *How we can separate interesting structural pattern from noise in large graphs?*

## Graph summarization using regular partitions

In this thesis, we introduce a principled framework to summarize large graphs using *Szemerédi's regularity lemma* [99], which is "one of the most powerful results of extremal graph theory" [56]. Basically, it states that any sufficiently large (dense) graph can almost entirely be partitioned into a bounded number of random-like bipartite graphs, called *regular pairs*. Komlós et al. [56, 55] introduced an important result, the so-called key lemma. It states that, under certain conditions, the partition resulting from the regularity lemma gives rise to a *reduced graph*, which inherits many of the essential structural properties of the original graph. In particular, the key lemma ensures that every small subgraph of the reduced graph is also a subgraph of the original graph. These results provide us with a principled way to obtain a good description of a large graph using a small amount of data, and can be regarded as a *manifestation of the all-pervading dichotomy between structure and randomness*. Hence, in this thesis, we posit that the regularity lemma can be used to summarize large graphs revealing its main structural patterns, while filtering out noise, which is common in any real-world networks.

The original proof of the regularity lemma [99] is not constructive, yet this has not narrowed the range of its applications in such fields as extremal graph theory, number theory and combinatorics. However, Alon et al. [4] proposed a new formulation of the lemma which emphasizes the algorithmic nature of the result. Later, other algorithms have been developed which improve the original one in several respects. In particular, we mention an algorithm developed by Frieze and Kannan (1999) [44], which is based on an intriguing relation between the regularity conditions and the singular values of matrices, and Czygrinow and Rödl's (2000) [28], who proposed a new algorithmic version of Szemerédi's lemma for hypergraphs. However, the algorithmic solutions developed so far have been focused exclusively on *exact* algorithms whose worst-case complexity, although being polynomial in the size of the underlying graph, has a hidden tower-type dependence on an accuracy parameter. In fact, Gowers proved that this tower function is necessary in order to guarantee a regular partition for *all* graphs [46]. This has typically discouraged researchers from applying regular partitions to practical problems, thereby confining them to the purely theoretical realm. To make the algorithm truly applicable, [97], and later [93], instead of insisting on provably regular partitions, proposed a few simple heuristics that try to construct an approximately regular partition.

In the first part of chapter 2, we describe a graph summarization heuristic, based on Alon et al.'s algorithm 1.3, which is an improvement of the previous algorithms [97, 93], while in the second part, we analyze the *ideal density regime* where the regularity lemma can find useful applications. In particular, since this lemma is suited to deal only with dense graphs, if we are out of the ideal density regime, we have to densify the input graph before summarizing it. In the last part of chapter 2, we show how the notion of regular partition can provide fresh insights into old pattern recognition and machine learning problems by using our summarization method to address graph-based clustering and image segmentation tasks.

## Separating structure from randomness

Chapter 3 is devoted to our second contribution, namely the dichotomy between structure and randomness. Here, we present a new heuristic, based on Alon et al.'s algorithm [4], which is characterized by an improvement of the summary quality both in terms of reconstruction error and of noise filtering. In particular, we first build the reduced graph of a graph $G$, and then we "blow-up" the reduced graph to obtain a graph $G'$, called *reconstructed graph*, which is close to $G$ in term of the $l_p$-reconstruction error. We study the noise robustness of our approach and we evaluate the quality of the summaries in term of the reconstruction error, by performing an extensive series of experiments on both synthetic and real data. As far as the synthetic data are concerned, we generate graphs with a cluster structure, where the clusters are perturbed with different levels of noise. In particular, each graph is generated by adding spurious edges between cluster pairs and by dropping edges inside the different clusters. The aim of this series of experiments is to assess if the framework is able to *separate structure* *from noise*. In the ideal case, the distance between $G$ and $G'$ should be only due to the filtered noise.

Moreover, in the second part of the chapter, we use our summarization algorithm to address the *graph search* problem defined under a similarity measure. The aim of graph search is to retrieve from a database the top-$k$ graphs that are most similar to a query graph. Since noise is common in any real-world dataset, the biggest challenge in graph search is developing efficient algorithms suited for dealing with large graphs containing noise in terms of missing and adding spurious edges. In our approach, all the graphs contained in a database are compressed off-line, while the query graph is compressed on-line. Thus, graph search can be performed on the summaries, and this allows us to speed up the search process and to reduce storage space. Finally, we evaluate the usefulness of our summaries in addressing the graph search problem by performing an extensive series of experiments. In particular, we study the quality of the answers in terms of the found top-$k$ similar graphs, and the scalability both in the size of the database and in the size of the query graphs.

## Regular decomposition of large graphs

Chapter 4 is devoted to the study of the linkage among the regularity lemma, the stochastic block model and the minimum description length. This study provide us with a principled way to develop a graph decomposition algorithm based on stochastic block model, which is fitted using likelihood maximization. The stochastic block model is an important paradigm in network research [3], and it usually revolves around the concept of *communities*, which are well connected sub-graphs with only few links between each pair of them. We aim to extend stochastic block model-style concepts to other type of networks, that do not fit well to such a community structure. The regularity lemma is used as a prototype of the structural information which should be preserved, defining a new model space for graph-data. In particular, we propose an heuristic postulating that in the case of graphs and similar objects, a good *a priori* class of models should be inferred from the regularity lemma, which points to stochastic block models. The minimum description length is here exploited to obtain a stopping criterion that establishes when the optimal regular decomposition is found.

# Chapter 1

# Szemerédi's Regularity Lemma

> Too much knowledge could be a bad thing. I was lead to the Szemerédi theorem by proving a result, about squares, that Euler had already proven, and I relied on an "obvious" fact, about arithmetical progressions, that was unproved at the time. But that lead me to try and prove that formerly unproved statement about arithmetical progressions and that ultimately lead to the Szemerédi Theorem.
>
> Endré Szemerédi

In 1941, the Hungarian mathematician P. Turán provided an answer to the following innocent-looking question. What is the maximal number of edges in a graph with $n$ vertices not containing a complete subgraph of order $k$, for a given $k$? This graph is now known as a Turán graph and contains no more than $n^2(k-2)/2(k-1)$ edges. Later, in another classical paper, T. S. Motzkin and E. G. Straus (1965) provided a novel proof of Turán's theorem using a continuous characterization of the clique number of a graph. Thanks to contributions of P. Erdös, B. Bollobás, M. Simonovits, E. Szemerédi and others, Turán's study developed soon into one of the richest branches of 20th-century graph theory, known as *extremal graph theory*, which has intriguing connections with Ramsey theory, random graph theory, algebraic constructions, etc. Roughly, extremal graph theory studies how the intrinsic structure of graphs ensures certain types of properties (e.g., cliques, coloring and spanning subgraphs) under appropriate conditions (e.g., edge density and minimum degree) [16].

Among the many achievements of extremal graph theory, Szemerédi's regularity lemma is certainly one of the best known [32]. Basically, it states that every graph can be partitioned into a small number of random-like bipartite graphs, called *regular pairs*, and a few leftover edges. Szemerédi's result was introduced in the mid-seventies as an auxiliary tool for proving the celebrated Erdös-Turán conjecture on arithmetic progressions in dense sets of integers [98]. Over the past two decades, this result has been refined, extended and interpreted in several ways and has now become an indispensable tool in discrete mathematics and theoretical computer science [56, 55, 100, 68]. Interestingly, an intriguing connection has also been established between the (effective) testability of graph properties (namely, properties that are testable with a constant number of queries on a graph) and regular partitions [5]. These results provide essentially a way to obtain a good description of a large graph using a small amount of data, and can be regarded as a manifestation of the all-pervading dichotomy between structure and randomness.

Indeed, the notion of separating structure from randomness in large (and possibly dynamic) data sets is prevalent in nearly all domains of applied science and technology, as evidenced by the importance and ubiquity of clustering methods in data analysis. However, unlike standard clustering approaches, regular partitions minimize discrepancies both within and between clusters in the sense that the members of a cluster behave roughly similarly toward members of each (other or own) cluster [14, 6]. This is a new paradigm for structural decomposition, which distinguishes it radically from all prior works in data analysis. This property allows for exchangeability among members of distinct parts within the partition, which can be important in a variety of real-world scenarios.

In next section we provide the basic concepts and notations used in the rest of the thesis as well as the formal definition of graph summary.

## 1.1   Preliminary definitions

Let $G = (V, E)$ be an undirected graph without self-loop, where $V$ is the set of vertices and $E$ is the set of edges. The *edge density* of a pair of two disjoint vertex sets $C_i, C_j \subseteq V$ is defined as:

$$d(C_i, C_j) = \frac{e(C_i, C_j)}{|C_i| \, |C_j|} \tag{1.1}$$

where $e(C_i, C_j)$ denotes the number of edges of $G$ with an endpoint in $C_i$ and an endpoint in $C_j$.

Given a positive constant $\varepsilon > 0$, we say that the pair $(C_i, C_j)$ of disjoint vertex sets $C_i, C_j \subseteq V$ is $\varepsilon$-*regular* if for every $X \subseteq C_i$ and $Y \subseteq C_j$ satisfying $|X| > \varepsilon |C_i|$ and $|Y| > \varepsilon |C_j|$

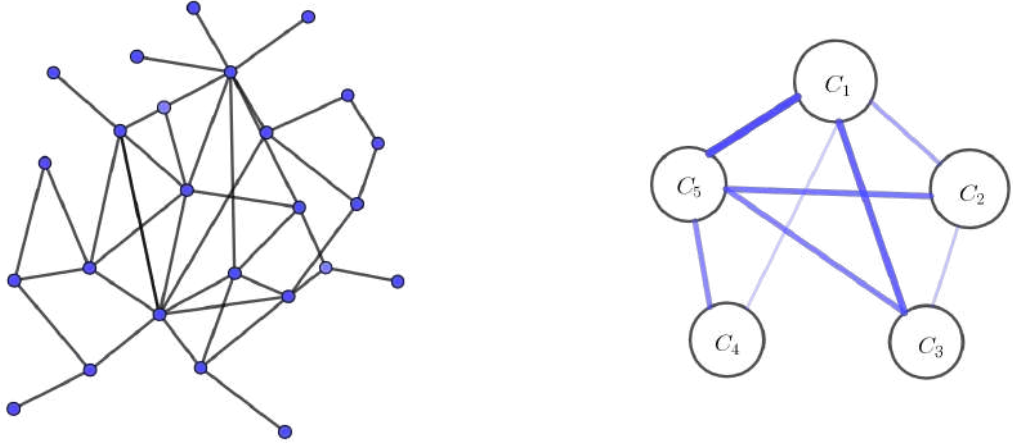

Fig. 1.1 Example of the reduced graph (summary) construction. Left: the original graph. Right: the *reduced graph* which contains eight $\varepsilon$-regular classes pairs. The density of each pair is expressed by the thickness of the edge that connects the classes of that pair. If a pair is $\varepsilon$-irregular the corresponding classes are not connected by an edge.

we have

$$\left| d(X,Y) - d(C_i,C_j) \right| < \varepsilon \ . \tag{1.2}$$

This means that the edges in an $\varepsilon$-regular pair are distributed fairly uniformly, where the deviation from the uniform distribution is controlled by the tolerance parameter $\varepsilon$.

A partition of $V$ into pairwise disjoint classes $C_0, C_1, ..., C_k$ is called *equitable* if all the classes $C_i$ ($1 \le i \le k$) have the same cardinality. The *exceptional* set $C_0$ (which may be empty) has only a technical purpose: it makes it possible that all other classes have exactly the same number of vertices.

**Definition 1** ($\varepsilon$-regular partition). *A partition $\mathscr{P} = C_0, C_1, \cdots, C_k$, with $C_0$ being the exceptional set is called $\varepsilon$-regular if:*

1. *it is equitable: $|C_1| = |C_2| = \cdots = |C_k|$;*

2. *$|C_0| < \varepsilon |V|$;*

3. *all but at most $\varepsilon k^2$ of the pairs $(C_i, C_j)$ are $\varepsilon$-regular ($1 \le i < j \le k$).*

The fist contribution of this thesis is the introduction of a summarization algorithm which, given an undirected graph $G = (V, E)$ without self-loop, iteratively builds a summary, called *reduced graph*, defined as follows.

**Definition 2** (Reduced graph). *Given an $\varepsilon$-regular partition $P = \{C_1, C_2,$ $\cdots, C_k\}$ of a graph $G = (V, E)$ and $0 \le d' \le 1$, the reduced graph of $G$ is the undirected, weighted graph $R = (V_R, E_R, w)$ where $V_R = P$, $E_R \subseteq V_R^2$ and $w : E_R \to \mathbb{R}$ is defined as follows:*

$$w((C_i, C_j)) = \begin{cases} d(C_i, C_j) & \text{if } (C_i, C_j) \text{ is } \varepsilon\text{-regular and } d(C_i, C_j) \geq d', \\ 0 & \text{otherwise.} \end{cases}$$

We are now ready to state the Regularity Lemma which provides us a principled way to develop a summarization algorithm with the aim of separating structure from noise in a large graph.

## 1.2 The Regularity Lemma

In essence, Szemerédi's Regularity Lemma states that given an $\varepsilon > 0$, every sufficiently dense graph $G$ can be approximated by the union of a bounded number of quasi-random bipartite graphs, where the deviation from randomness is controlled by the tolerance parameter $\varepsilon$. In other words, we can partition the vertex set $V$ into a bounded number of classes $C_0, C_1, ..., C_k$, such that almost every pair $(C_i, C_j)$ behaves similarly to a random bipartite graph ($1 \leq i < j \leq k$).

**Theorem 1** (Szemerédi's regularity lemma (1976))**.** *For every positive real $\varepsilon$ and for every positive integer $m$, there are positive integers $N = N(\varepsilon, m)$ and $M = M(\varepsilon, m)$ with the following property: for every graph $G = (V, E)$, with $|V| \geq N$, there is an $\varepsilon$-regular partition of $G$ into $k + 1$ classes such that $m \leq k \leq M$.*

The lemma allows us to specify a lower bound $m$ on the number of classes. A large value of $m$ ensures that the partition classes $C_i$ are sufficiently small, thereby increasing the proportion of (inter-class) edges subject to the regularity condition and reducing the intra-class ones. The upper bound $M$ on the number of partitions guarantees that for large graphs the partition sets are large too. Finally, it should be noted that a singleton partition is $\varepsilon$-regular for every value of $\varepsilon$ and $m$.

The strength of the Regularity Lemma is corroborated by the so-called Key Lemma, which is an important theoretical result introduced by Komlos et al. [**?** ]. It basically states that the reduced graph does inherit many of the essential structural properties of the original graph. Before presenting its original formulation, another kind of graph needs to be defined, namely the *fold graph*. Given an integer $t$ and a graph $R$ (which may be seen as a reduced graph), let $R(t)$ denote the graph obtained by "blowing up" each vertex $j$ of $V(R)$ to a set $A_j$ of $t$ independent vertices, and joining $u \in A_x$ to $v \in A_y$ if and only if $(x, y)$ is an edge in $R$. Thus, $R(t)$ is a graph in which every edge of $R$ is replaced by a copy of the complete bipartite graph $K_{tt}$. The following lemma shows a link between the reduced graph $R$ and $R(t)$.

**Theorem 2** (Key Lemma). *Given $d > \varepsilon > 0$, a graph $R$, and a positive integer $m$, let us construct a graph $G'$, called* reconstructed graph*, by performing the following steps:*

1. *replace every vertex of $C_i$ of $R$ by a set of $m$ vertices, where $m = |C_i|$;*

2. *if the edge $(i, j)$ of $R$ has weight $w(i, j) \geq d$ replace it with $m$ edges of weight equal to $w(i, j)$.*

*Let $H$ be a subgraph of $R(t)$ with $h$ vertices and maximum degree $\Delta > 0$, and let $\delta = d - \varepsilon$ and $\varepsilon_0 = \delta^\Delta / (2 + \Delta)$. If $\varepsilon \leq \varepsilon_0$ and $t - 1 \leq \varepsilon_0 m$, then $H$ is embeddable into $G'$ (i.e. $G'$ contains a subgraph isomorphic to $H$). In fact, we have:*

$$\left\| H \to G' \right\| > (\varepsilon_0 m)^h \tag{1.3}$$

*where $\| H \to G' \|$ denotes the number of labeled copies of $H$ in $G'$.*

The Key Lemma states that the reconstructed graph $G'$ has the same structural properties of the reduced graph $R$, since if for $t = 1$, $R(t) = R$ and if the constraint on the edge density $d$ is satisfied, the Key Lemma ensures that every small subgraph of $R$ is also a subgraph of $G'$. Hence, since the Regularity Lemma implies that any graph $G$ can be approximated by its reduced graph $R$ [**?** ], which is a graph composed by the union of random-like bipartite graphs, we posit that $G'$ is a good approximation of the original graph $G$. Thus, we can use the Regularity Lemma to build a summary $R$ of $G$, and then we can infer structural properties of $G$ by studying the same properties on $R$.

Given an $r \times r$ symmetric matrix $(p_{ij})$ with $0 \leq p_{ij} \leq 1$, and positive integers $n_1, n_2, ..., n_r$, a generalized random graph $R_n$ for $n = n_1 + n_2 + ... + n_r$ is obtained by partitioning $n$ vertices into classes $C_i$ of size $n_i$ and joining the vertices $x \in V_i$, $y \in V_j$ with probability $p_{ij}$, independently for all pairs $\{x, y\}$. Now, as pointed out by Komlós and Simonovits (1996), the regularity lemma asserts basically that every graph can be approximated by generalized random graphs. Note that, for the lemma to be useful, the graph has to to be dense. Indeed, for sparse graphs it becomes trivial as all densities of pairs tend to zero [32].

However, we mention that after the publication of Szemerédi's original lemma several variations, extensions and interpretations have been proposed in the literature. In particular, we have got weaker regularity notions [44, 68] and stronger ones [7, 100, 68], and we have also got versions for sparse graphs and matrices [45, 95] and hypergraphs [28, 42]. Interestingly, [100] provided an interpretation of the lemma in terms of information theory, while [68] offered three different analytic interpretations.

## 1.3   The first algorithmic version

The original proof of the Regularity Lemma is not constructive, but during the last decades different constructive versions have been proposed. In this thesis, we focus on the Alon et al. [4] work. In particular, they proposed a new formulation of the Regularity Lemma which emphasizes the algorithmic nature of the result.

**Theorem 3.** *(Alon et al., 1994) For every $\varepsilon > 0$ and every positive integer $t$ there is an integer $Q = Q(\varepsilon, t)$ such that every graph with $n > Q$ vertices has an $\varepsilon$-regular partition into $k + 1$ classes, where $t \leq k \leq Q$. For every fixed $\varepsilon > 0$ and $t \geq 1$ such partition can be found in $O(M(n))$ sequential time, where $M(n) = O(n^{2.376})$ is the time for multiplying two $n \times n$ matrices with 0,1 entries over the integers. It can also be found in time $O(\log n)$ on an EREW PRAM with a polynomial number of parallel processors.*

A sketch of the proof is then presented. Let $H$ be a bipartite graph with equal color classes $|A| = |B| = n$. Let us define the average degree $\bar{d}$ of $H$ as:

$$\bar{d}(A, B) = \frac{1}{2n} \sum_{i \in A \cup B} deg(i)$$

where $deg(i)$ is the degree of vertex $i$.

For two distinct vertices $y_1, y_2 \in B$ the *neighbourhood deviation* of $y_1$ and $y_2$ is defined as:

$$\sigma(y_1, y_2) = |N(y_1) \cap N(y_2)| - \frac{\bar{d}^2}{n}$$

where $N(x)$ is the set of neighbours of vertex $x$. For a subset $Y \subset B$ the *deviation* of $Y$ is defined as:

$$\sigma(Y) = \frac{\sum_{y_1, y_2 \in Y} \sigma(y_1, y_2)}{|Y|^2}$$

Let $0 < \varepsilon < 1/16$, it can be proved that, if there exists $Y \subset B$, $|Y| > \varepsilon n$ such that $\sigma(Y) \geq \frac{\varepsilon^3}{2} n$, then at least one of the following cases occurs:

1. $\bar{d} < \varepsilon^3 n$ ($H$ is $\varepsilon$-regular);

2. there exists in $B$ a set of more than $\frac{1}{8} \varepsilon^4 n$ vertices whose degrees deviate from $\bar{d}$ by at least $\varepsilon^4 n$ ($H$ is $\varepsilon$-irregular);

3. there are subsets $A' \subset A$, $B' \subset B$, $|A'| \geq \frac{\varepsilon^4}{n} n$, $|B'| \geq \frac{\varepsilon^4}{n} n$ such that $|\bar{d}(A', B') - \bar{d}(A, B)| \geq \varepsilon^4$ ($H$ is $\varepsilon$-irregular).

Note that one can easily check if condition 1 holds in time $O(n^2)$. Similarly, it is trivial to check if condition 2 holds in $O(n^2)$ time, and in case it holds to exhibit the required subset of $B$ establishing this fact. If the first two conditions are not verified, the third condition must be checked. To this end, we have to find the subsets $A', B'$, called *certificates*, that witness the irregularity of the bipartite graph $H$. To address this task, we first select a subset of $B$ whose vertex degrees "deviate" the most from the average degree $\bar{d}$ of $H$. More formally: for each $y_0 \in B$ with $|deg(y_0) - \bar{d}| < \varepsilon^4 n$ we find the vertex set $B_{y_0} = \{y \in B \,|\, \sigma(y_0, y) \geq 2\varepsilon^4 n\}$. The proof provided by Alon et al. guarantees the existence of at least one such $y_0$ for which $|B_{y_0}| \geq \frac{\varepsilon^4}{4} n$. Thus, the subsets $B' = B_{y_0}$ and $A' = N(y_0)$ are the required certificates. These two subsets represent the collection of vertices that contribute more to the irregularity of the pair $(A, B)$. The sets $\bar{A}' = A \setminus A'$, $\bar{B}' = B \setminus B'$ are called *complements*. Since the computation of the quantities $\sigma(y, y')$, for $y, y' \in B$, can be done by squaring the adjacency matrix of $H$, the overall complexity of this algorithms is $O(M(n)) = O(n^{2.376})$.

In order to understand the final partitioning algorithm we need the following two lemmas.

**Lemma 1** (Alon et al., 1994). *Let $H$ be a bipartite graph with equal classes $|A| = |B| = n$. Let $2n^{-\frac{1}{4}} < \varepsilon < \frac{1}{16}$. There is an $O(n^{2.376})$ algorithm which verifies that $H$ is $\varepsilon$-regular or finds two subsets $A' \subseteq A$ and $B' \subseteq B$ such that $|A'| \geq \frac{\varepsilon^4}{4} n$, $|B'| \geq \frac{\varepsilon^4}{4} n$, and $|d(A', B') - d(A, B)| \geq \varepsilon^4$.*

It is quite easy to check that the regularity condition can be rephrased in terms of the average degree of $H$. Indeed, it can be seen that if $d < \varepsilon^3 n$, then $H$ is $\varepsilon$-regular, and this can be tested in $O(n^2)$ time. Next, it is necessary to count the number of vertices in $B$ whose degrees deviate from $d$ by at least $\varepsilon^4 n$. Again, this operation takes $O(n^2)$ time. If the number of deviating vertices is more than $\frac{\varepsilon^4 n}{8}$, then the degrees of at least half of them deviate in the same direction and if we let $B'$ be such a set of vertices and $A' = A$ we are done. Otherwise, it can be shown that there must exist $Y \subseteq B$ such that $|Y| \geq \varepsilon n$ and $\sigma(Y) \geq \frac{\varepsilon^3 n}{2}$. Hence, our previous discussion shows that the required subsets $A'$ and $B'$ can be found in $O(n^{2.376})$ time.

Given an equitable partition $P$ of a graph $G = (V, E)$ into classes $C_0, C_1 \ldots C_k$, [99] defines a measure called *index of partition*:

$$ind(P) = \frac{1}{k^2} \sum_{s=1}^{k} \sum_{t=s+1}^{k} d(C_s, C_t)^2 \ . \tag{1.4}$$

Since $0 \leq d(C_s, C_t) \leq 1$, $1 \leq s, t \leq k$, it can be seen that $ind(P) \leq 1/2$.

The following lemma is the core of Szemerédi's original proof.

**Lemma 2** (Szemerédi, 1976). *Fix $k$ and $\gamma$ and let $G = (V, E)$ be a graph with $n$ vertices. Let $P$ be an equitable partition of $V$ into classes $C_0, C_1, \ldots, C_k$. Assume $|C_1| > 4^{2k}$ and $4^k > 600\gamma^{-5}$. Given proofs that more than $\gamma k^2$ pairs $(C_r, C_s)$ are not $\gamma$-regular, then one can find in $O(n)$* *time a partition P' (which is a refinement of P) into $1 + k4^k$ classes, with the exceptional class of cardinality at most $|C_0| + \frac{n}{4^k}$ and such that*

$$ind(P') \geq ind(P) + \frac{\gamma^5}{20}. \tag{1.5}$$

The idea formalized in the previous lemma is that, if a partition violates the regularity condition, then it can be refined by a new partition and, in this case, the index of partition measure can be improved. On the other hand, the new partition adds only few elements to the current exceptional set so that, in the end, its cardinality will respect the definition of equitable partition.

We are now in a position to sketch the complete partitioning algorithm. The procedure is divided into two main steps: in the first step all the constants needed during the next computation are set; in the second one, the partition is iteratively created. An iteration is called *refinement step*, because, at each iteration, the current partition is closer to a regular one.

Given any $\varepsilon > 0$ and a positive integer $t$, we define the constants $N = N(\varepsilon, t)$ and $T = T(\varepsilon, t)$ as follows; let $b$ be the least positive integer such that

$$4^b > 600(\frac{\varepsilon^4}{16})^{-5}, b \geq t. \tag{1.6}$$

Let $f$ be the integer valued function defined inductively as

$$f(0) = b, f(i+1) = f(i)4^{f(i)}. \tag{1.7}$$

Put $T = f(\lceil 10(\frac{\varepsilon^4}{16})^{-5} \rceil)$ and $N = max\{T4^{2T}, \frac{32T}{\varepsilon^5}\}$.

Finally, we can now present Alon et al.'s algorithm, which provides a way to find an $\varepsilon$-regular partition. The procedure is divided into two main steps: in the first step all the constants needed during the next computation are set; in the second one, the partition is iteratively created. An iteration is called *refinement step*, because, at each iteration, the current partition is closer to a regular one.

**Alon et al.'s Algorithm**

1. Create the initial partition: arbitrarily divide the vertices of $G$ into an equitable partition $\mathscr{P}_1$ with classes $C_0, C_1, \cdots, C_b$ where $|C_i| = \lfloor \frac{n}{b} \rfloor$.

2. Check Regularity: for every pair $(C_r, C_s)$ of $\mathscr{P}_i$, verify if it is $\varepsilon$-regular or find two certificates $C'_r \subset C_r, C'_s \subset C_s, |C'_r| \geq \frac{\varepsilon^4}{16}|C_1|, |C'_s| \geq \frac{\varepsilon^4}{16}|C_1|$ such that $|\bar{d}(C'_r, C'_s) - \bar{d}(C_r, C_s)| \geq \varepsilon^4$.

3. Count regular pairs: if there are at most $\varepsilon \binom{k_i}{2}$ pairs that are not $\varepsilon$-regular, then stop. $\mathscr{P}_i$ is an $\varepsilon$-regular partition.

4. Refine: apply the refinement algorithm and obtain a partition $\mathscr{P}'$ with $1 + k_i 4^{k_i}$ classes.

5. Go to step 2.

The algorithm described above for finding a regular partition was the first one proposed in the literature. Even if the above mentioned algorithm has polynomial worst case complexity in the size of $G$, there is a hidden tower-type dependence on an accuracy parameter. Unfortunately, Gowers [46] proved that this tower function is necessary in order to guarantee a regular partition for *all* graphs. This implies that, in order to have a faithful approximation, the original graph size should be astronomically big. This has typically discouraged researchers from applying regular partitions to practical problems, thereby confining them to the purely theoretical realm. To make the algorithm truly applicable, [97], and later [93], instead of insisting on provably regular partitions, proposed a few simple heuristics that try to construct an approximately regular partition.

In the next chapter, we will describe an improved version of the previous heuristic algorithms, and we will present a study of the density regime where the regularity lemma can find useful applications. Moreover, we will show how the notion of regular partition can provide fresh insights into old pattern recognition and machine learning problems by using our summarization method to address graph-based clustering and image segmentation tasks.

# Chapter 2

# The Regularity Lemma and Its Use in Pattern Recognition

In mathematics the primary subject-matter is not the individual mathematical objects but rather the structures in which they are arranged.

Michael D. Resnik

A crucial role in the development of machine learning and pattern recognition is played by the tractability of large graphs, which is intrinsically limited by their size. In order to overcome this limit, the input graph can be summarized into a reduced version exploiting the regularity lemma, which provides us with a principled way to obtain a good description of a large graph using a small amount of data.

In the first part of this chapter, we present the main limitations that prevent the practical applications of the (exact) Alon et al.'s algorithm 1.3. In particular, even if this algorithm has polynomial worst case complexity in the size of the input graph, there is a hidden tower-type dependence on an accuracy parameter. To make the algorithm truly applicable, we then introduce a few heuristics for finding an approximately regular partition, which will be used to construct a summary of the input graph. In the second part of this chapter, we analyze the *ideal density regime* of the input graph where the regularity lemma can find useful applications. In particular, since this lemma is suited to deal only with dense graphs, if we are out of the ideal density regime, we have to densify the input graph before summarizing it. Finally, we show how the notion of regular partition can provide fresh insights into old pattern recognition and machine learning problems by using our summarization method to address graph-based clustering and image segmentation tasks.

## 2.1   An heuristic to summarize large graphs

For the sake of clarity, we report the constructive version of the regularity lemma proposed by Alon et al.'s [4], which has been described in the previous chapter.

**Alon et al.'s Algorithm**

1. Create the initial partition: arbitrarily divide the vertices of $G$ into an equitable partition $\mathscr{P}_1$ with classes $C_0, C_1, \cdots, C_b$ where $|C_i| = \lfloor \frac{n}{b} \rfloor$.

2. Check Regularity: for every pair $(C_r, C_s)$ of $\mathscr{P}_i$, verify if it is $\varepsilon$-regular or find two certificates $C'_r \subset C_r$, $C'_s \subset C_s$, $|C'_r| \geq \frac{\varepsilon^4}{16}|C_1|$, $|C'_s| \geq \frac{\varepsilon^4}{16}|C_1|$ such that $|\bar{d}(C'_r, C'_s) - \bar{d}(C_r, C_s)| \geq \varepsilon^4$.

3. Count regular pairs: if there are at most $\varepsilon \binom{k_i}{2}$ pairs that are not $\varepsilon$-regular, then stop. $\mathscr{P}_i$ is an $\varepsilon$-regular partition.

4. Refine: apply the refinement algorithm and obtain a partition $\mathscr{P}'$ with $1 + k_i 4^{k_i}$ classes.

5. Go to step 2.

The main limitations which prevent the application of the above algorithm to practical problems concern Step 2 and Step 4.

Indeed , in Step 2, the algorithm checks the regularity of all classes pairs, and outputs the number of irregular pairs (*#irr_pairs*). In particular, to check the regularity of a pair of classes $(C_r, C_s)$, the following three conditions are used:

1. $\bar{d} < \varepsilon^3 n$ ($H$ is $\varepsilon$-regular);

2. there exists in $C_s$ a set of more than $\frac{1}{8}\varepsilon^4 n$ vertices whose degrees deviate from $\bar{d}$ by at least $\varepsilon^4 n$ ($H$ is $\varepsilon$-irregular);

3. there are subsets $C'_r \subset C_r$, $C'_s \subset C_s$, $|C'_r| \geq \frac{\varepsilon^4}{n}n$, $|C'_s| \geq \frac{\varepsilon^4}{n}n$ such that $|\bar{d}(C'_r, C'_s) - \bar{d}(C_r, C_s)| \geq \varepsilon^4$ ($H$ is $\varepsilon$-irregular).

where $H = (C_r, C_s)$ is the bipartite graph with equal color classes $C_r, C_s$ such that $|C_r| = |C_s| = n$; and $\bar{d}$ is the average degree of $H$. Given a pair $(C_r, C_s)$, condition 1 verifies if it is $\varepsilon$-regular, otherwise conditions 2 and 3 are used to obtain the *certificates* $C'_r$ and $C'_s$ that witness the irregularity. The main obstacle concerning the implementation of condition 3 is the necessity to scan over *almost all possible subsets* of $C_s$. To make the implementation of condition 3 feasible, given a class $C_s$, we select in a *greedy way* a set $Y' \subseteq C_s$ with the highest deviation $\sigma(Y')$ (the deviation is defined in 1.3). To do so, the nodes of $C_s$ are sorted by bipartite degree, and $Y'$ is built by adding $\frac{\varepsilon^4}{4}n$ nodes with the highest degree. At each iteration of the greedy algorithm, the node with a degree that deviates more from the average degree is added to the candidate certificates. This last operations is repeated until the subset $C_s'$ that satisfies condition 3 is found. This almost guarantees to put in a candidate certificates the nodes that have a *connectivity pattern* that deviates from the one characterizing the majority of the nodes which belong to $C_s$.

As far Step 4 is concerned, here an irregular partition $\mathscr{P}_\varepsilon^i$ is *refined* by a new partition such $\mathscr{P}_\varepsilon^{i+1}$, such that the index of partition measure (*sze_idx* defined in 1.4) is increased. This step poses the main obstacle towards a practical version of Alon et al.'s algorithm involving the creation of an exponentially large number of subclasses at each iteration. Indeed, as we have said, Step 2 finds all possible irregular pairs in the graph. As a consequence, each class may be involved with up to $(k-1)$ irregular pairs, $k$ being the number of classes in the current partition $\mathscr{P}_\varepsilon^i$, thereby leading to an exponential growth. To avoid the problem, for each class, one can limit the number of irregular pairs containing it to at most one, possibly chosen randomly among all irregular pairs. This simple modification allows one to divide the classes into a constant, rather than exponential, number of subclasses $l$ (typically $2 \le l \le 7$). Despite the crude approximation this seems to work well in practice.

The devised heuristic algorithm takes as input two main parameters, the tolerant parameter $\varepsilon$ and the minimum compression rate $c\_min$, that acts as a stopping criterion in the refinement process. The pseudocode of the proposed algorithm is reported in Algorithm 1. The overall complexity of our summarization algorithms is $O(M(n)) = O(n^{2.376})$, which is dominated by the verification of condition 3.

## 2.2 Analysis of the ideal density regime

In this section, we analyze the *ideal density regime* where the regularity lemma can find useful applications. This ideal density regime is defined as the range of densities of the input graph $G$ such that our heuristic algorithm outputs an expanded graph $G'$ preserving the main topological properties of $G$. If we are out of this ideal range, we have to densify the graph before applying the regularity lemma. In particular, we combine the use of the regularity lemma and the key lemma in the following way:

- Start with a graph $G = (V, E)$ and apply the algorithm 1, finding a regular partition $\mathscr{P}$;

- Construct the reduced graph $R$ of $G$, w.r.t. the partition $\mathscr{P}$;

- Build a reconstructed graph $G'$ using the definition of the fold graph (see **??**).

---

**Algorithm 1** The Summarization Algorithm

**Input**:
- $\varepsilon$ is the tolerant parameter 1.2;
- $G = (V, E)$ is an undirected simple graph (the input graph);
- $c\_min$ is the minimum compression rate, expressed as $k/|V|$

**Output**:
- $\mathscr{P}$ is a regular partition of $G$, where $|\mathscr{P}| = k$.

---

1: **procedure** APPROXALON($\varepsilon$, c_min, $G = (V, E)$)
2:      partitions = empty list
3:      $\mathscr{P}_\varepsilon^1$ = Create initial random partition from $G$
4:      **while** True **do**
5:          #irr_pairs = CHECKPAIRSREGULARITY($\mathscr{P}_\varepsilon^i$)
6:          **if** #irr_pairs $> \varepsilon \binom{k}{2}$ or COMPRESSRATE($\mathscr{P}_\varepsilon^i$) $<$ c_min **then**
7:              break
8:          **else**
9:              $\mathscr{P}_\varepsilon^{i+1}$ = REFINEMENT($\mathscr{P}_\varepsilon^i$)
10:              **if** $\mathscr{P}_\varepsilon^{i+1}$ is $\varepsilon$-regular **then**
11:                  partitions.add($\mathscr{P}_\varepsilon^{i+1}$)
12:              **else**
13:                  break
14:      Select best partition $\mathscr{P}^*$ with maximum *sze_idx* from list partitions

---

Among the many topological measures we test the *effective resistance* (or equivalently the scaled commute time), one of the most important metrics between the vertices in the graph, which has been very recently questioned. In [106] argued that this measure is meaningless for large graphs. However, recent experimental results show that the graph can be pre-processed (densified) to provide some informative estimation of this metric [35, 34].

The effective resistance is a metric between the vertices in $G$, whose stability is theoretically constrained by the size of $G$. In particular, von Luxburg et al. [106] derived the following bound for any connected, undirected graph that is not bipartite:

$$\left| \frac{1}{vol(G)} C_{ij} - \left( \frac{1}{d_i} + \frac{1}{d_j} \right) \right| \leq \frac{1}{\lambda_2} \frac{2}{d_{min}} \tag{2.1}$$

where $C_{ij}$ is the commute time between vertices $i$ and $j$, $vol(G)$ is the volume of the graph, $\lambda_2$ is the so called *spectral gap* and $d_{min}$ is the minimum degree in $G$. Since $C_{ij} = vol(G)R_{ij}$, where $R_{ij}$ is the effective resistance between $i$ and $j$, this bound leads to $R_{ij} \approx \frac{1}{d_i} + \frac{1}{d_j}$. This means that, in large graphs, effective resistances do only depend on local properties, i.e. degrees.

However, Escolano et al. [35] showed that densifying $G$ significantly decreases the spectral gap, which in turn enlarges the von Luxburg bound. As a result, effective resistances do not depend only on local properties and become meaningful for large graphs provided that these graphs have been properly densified. As defined in [48] and revisited in [34], *graph densification* aims to significantly increase the number of edges in $G$ while preserving its properties as much as possible.

One of the most interesting properties of large graphs is their fraction of *sparse cuts*, that are cuts where the number of pairs of vertices involved in edges is a small fraction of the overall number of pairs associated with any subset $S \subset V$, i.e. sparse cuts stretch the graphs, thus leading to small conductance values, which in turn reduce the spectral gap. This is exactly what is accomplished by the state-of-the-art strategies for graph densification, including anchor graphs [66].

In light of these observations, our experiments aim to answer two questions:

- *Phase transition:* What is the expected behavior of our heuristic algorithm when the input graph is locally sparse?

- *Commute times preservation:* Given a densified graph $G$, to what extent does our algorithm preserve its metrics in the expanded graph $G'$?

To address them, we perform experiments both on synthetic and real-world datasets. Experiments on synthetic datasets allow us to control the degree of both intra-cluster and inter-cluster sparsity. On the other hand, the use of real-world datasets, such as NIST, leads to understand the so called *global density scenario*. Reaching this scenario in realistic data sets may require a proper densification, but once it is provided, the regularity lemma becomes a powerful structural summarization method.

### 2.2.1   Experimental results

Since we are exploring the practical effect of combining regularity and key lemmas to preserve metrics in large graphs, our performance measure relies on the so called relative deviation between the measured effective resistance and the von Luxburg et al. local prediction [106]:

$$RelDev(i,j) = \frac{\left| R_{ij} - \left( \frac{1}{d_i} + \frac{1}{d_j} \right) \right|}{R_{ij}}. \tag{2.2}$$

The larger $RelDev(i,j)$ the better the performance. For a graph, we retain the average $RelDev(i,j)$, although the maximum and minimum deviations can be used as well.

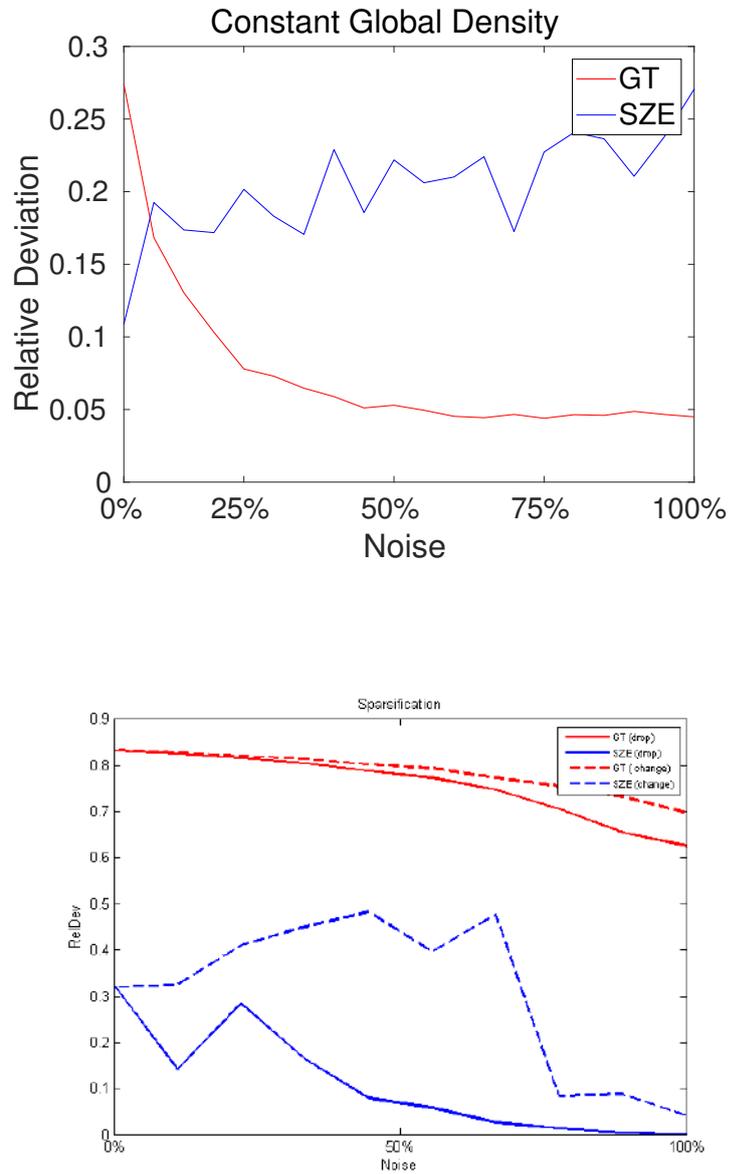

Fig. 2.1 Top: experiments 1. Bottom: experiment 2 ($n = 200$, $k = 10$ classes).

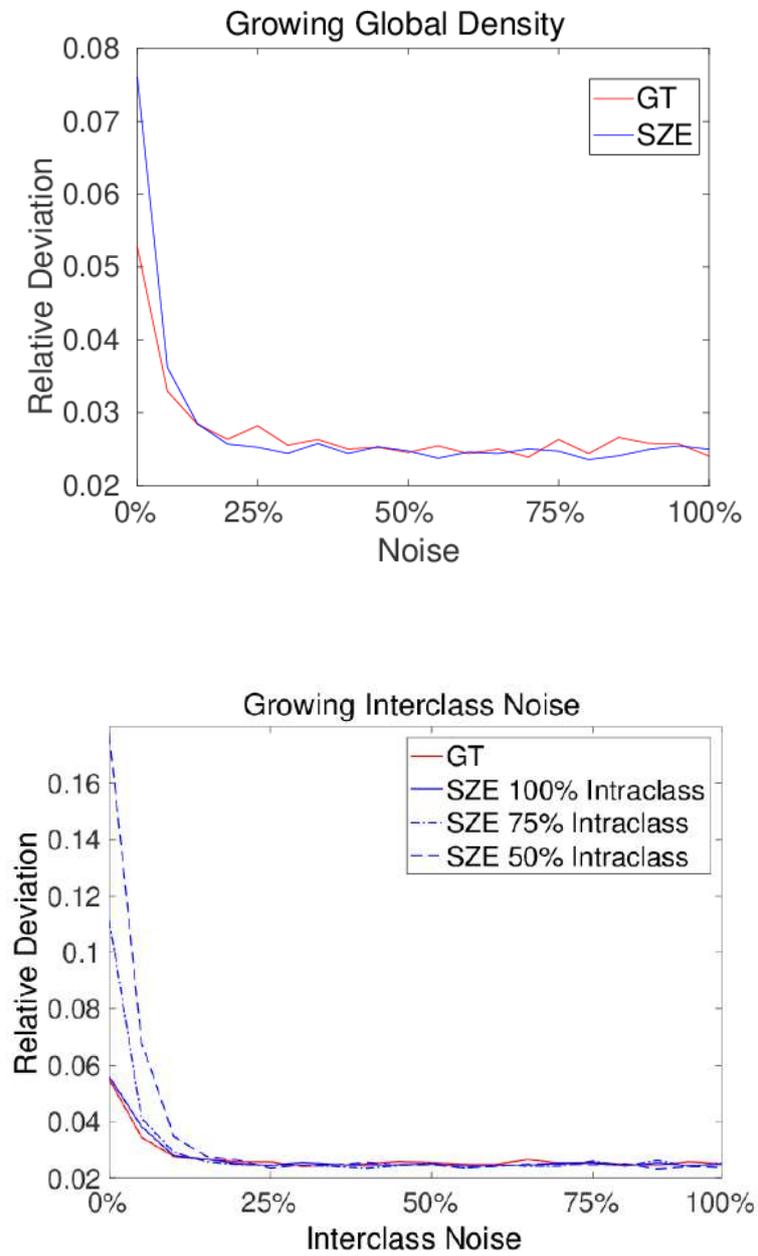

Fig. 2.2 Experiment 3 ($n = 200$, $k = 10$ classes).

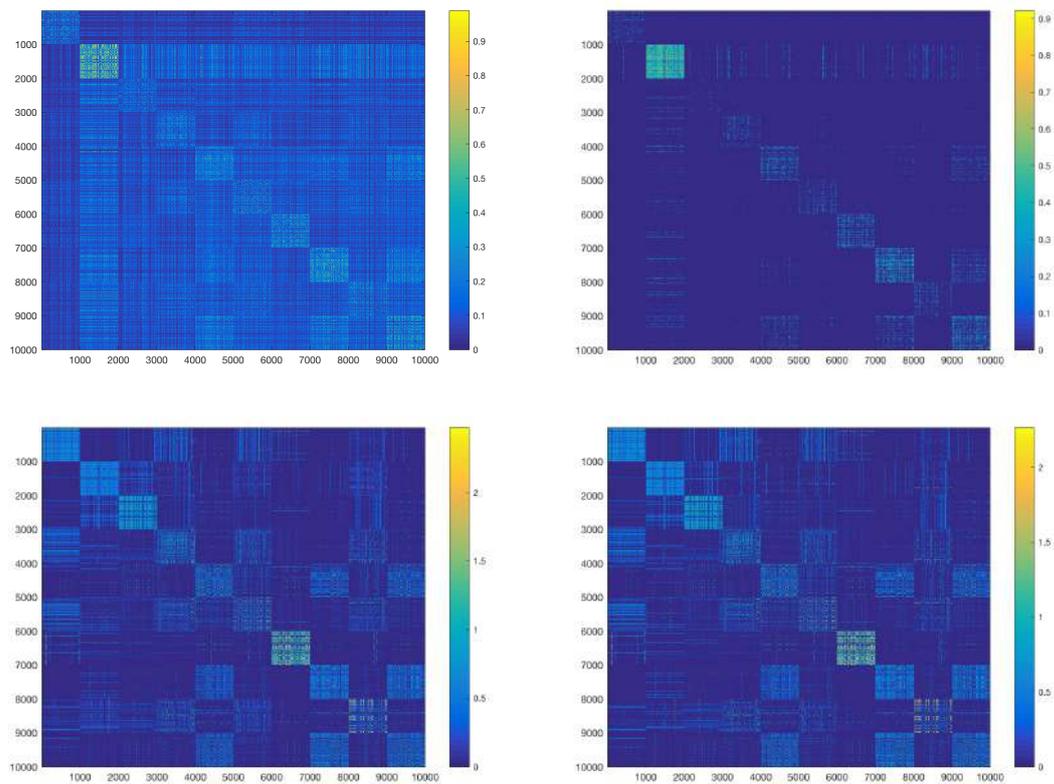

Fig. 2.3 Reconstruction from R. From left to right: Original similarity matrix $W$ with $\sigma = 0.0248$, its reconstruction after compressing-decompressing, sparse matrix obtained by densifying $W$ and its reconstruction.

**Synthetic experiments**    For these experiments we designed a ground truth (GT) consisting of $k$ cliques linked by $O(n)$ edges. Inter-cluster links in the GT were only allowed between class $k$ and $k+1$, for $k = 1, \cdots, k-1$. Then, each experiment consisted of modifying the GT by either removing intra-cluster edges (sparsification) and/or adding inter-cluster edges and then looking at the reconstructed GT after the application of our heuristic partition algorithm followed by the expansion of the obtained reduced graph (key lemma). We refer to this two stage approach as SZE.

**Experiment 1**    (*Constant global density*).   We first proceeded to incrementally sparsify the cliques while adding the same amount of inter-cluster edges that are removed. This procedure assures the constancy of the global density. Since in these conditions the relative deviation provided by the expanded graph is quite stable, we can state the our heuristic algorithm produces partitions that preserve many of the structural properties of the input graph. However, the performances of the uncompressed-decompressed GT decay along this process Fig.2.1(top).

**Experiment 2**    (*Only sparsification*).   Sparsifying the cliques without introducing inter-cluster edges typically leads to an inconsistent partition, since it is difficult to find regular pairs. So SZE *RelDev* is outperformed by that of the GT without compression. This is an argument in favor of using graph densification with approximate cut-preservation as a preconditioner of the regularity lemma. However, this is only required in cases where the amount of inter-cluster noise is negligible. In Fig. 2.1 (bottom), we show two cases: deleting inter-cluster edges (solid plots) *vs* replacing these edges by a constant weight $w = 0.2$ (dotted plots). Inter-cluster completion (dotted-plots) increases the global density and this contributes to significantly increase the performances of our heuristic algorithm, although it is always outperformed by the uncompressed corrupted GT.

**Experiment 3**    (*Selective increase of the global density*).   In this experiment, we increase the global density of the GT as follows. For Fig. 2.2 (top), each noise level $x$ means the fraction of intra-cluster edges removed, while the same fraction of inter-cluster edges is increased. Herein, the density of $x$ is $D(x) = (1-x)\#_{In} + x\#_{Out}$, where $\#_{In}$ is the maximum number of intra-cluster links and $\#_{Out}$ is the maximum number of inter-cluster links. Since $\#_{Out} \gg \#_{In}$, we have that $D(x)$ increases with $x$. However, only moderate increases of $D(x)$ lead to a better estimation of commute times with SZE, since adding many inter-cluster links destroys the cluster structure.

However, in Fig. 2.2 (bottom), we show the impact of increasing the fraction $x'$ of inter-cluster noise (add edges) while the intra-cluster fraction is fixed. We overlay three results for SZE: after retaining 50%, 75% and 100% of $\#_{In}$. We obtain that SZE contributes better to the estimation of commute times for small fractions on $\#_{In}$, which is consistent with Experiment 2. Hence, the optimal configuration for SZE is given by low inter-cluster noise and moderate sparsified clusters.

As a conclusion of the synthetic ex periments, we can state that our heuristic algorithm is robust against a high amount of intra-clustering sparsification provided that a certain number of inter-cluster edges exists. This answers the first question (*phase transition*). It also partially ensures the preservation of commute times provided that the density is high enough or it is kept constant during a sparsification process, which answers to the second question (*commute times preservation*).

**Experiments with real-world dataset**    (*NIST*). When analyzing real datasets, NIST (herein we use $10,000$ samples with $d = 86$) provides a nice amount of intra-cluster sparsity and inter-cluster noise (both due to ambiguities). We compare our two stage approach (SZE) either applied to the original graph (for a given $\sigma$) or to an anchor graph obtained with a nested MDL strategy relying on our EBEM clustering method [9]. In Fig. 2.3, we show a NIST similarity matrix $W$ (with $O(10^7)$ edges) obtained using the negative exponentiation method. Even with $\sigma = 0.0248$ we obtain a dense matrix due to inter-cluster noise. Let $R(W)$ be the reduced graph of $W$. After expanding this graph we obtain a locally dense matrix, which suggests that our algorithm plays the role of a cut densifier. We also show the behaviour of compression-decompression for densified matrices in Fig. 2.3. The third graph in this figure corresponds to $D(W)$, namely the selective densification of $W$ (with $O(2 \times 10^6)$ edges). From $R(D(W))$ the key lemma leads to a reconstruction with a similar density but with more structured inter-cluster noise. Finally, it is worth noting that the compression rate in both cases is close to 75%.

## 2.2.2   Concluding remarks

In this section, we have explored the interplay between regular partitions and graph densification. Our synthetic experiments show that the proposed heuristic version of Alon et al.'s algorithm is quite robust against intra-cluster sparsification provided that the graph is globally dense. This behavior has a good impact in similarity matrices obtained from negative exponentiation, since this implementation of the regularity lemma plays the role of a selective densifier. Regarding the effect of summarization-reconstruction in non-densified matrices, the reconstruction preserves the structure of the input matrix. This result suggests that graph densification acts as a pre-conditioner to obtain reliable regular partitions.

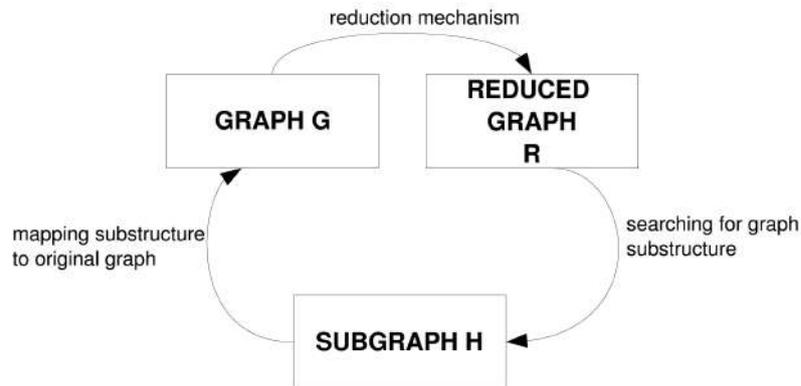

Fig. 2.4 Reduction strategy to find significant substructures in a graph. From [97].

## 2.3   Using the regularity lemma for pairwise clustering

Sperotto and Pelillo reported arguably the first practical application of the regularity lemma and related algorithms [97]. The original motivation was to study how to take advantage of the information provided by Szemerédi's partitions in a pairwise clustering context. Here, the regularity lemma is used as a preclustering strategy, in an attempt to work on a more compact, yet informative, structure. In fact, this structure is well known in extremal graph theory and is commonly referred to as the *reduced graph*. Some important auxiliary results, such as the so-called key lemma 2 or the Blow Up lemma [57], reveal that this graph does inherit many of the essential structural properties of the original graph.

As described in [56], a common and helpful combined use of the reduced graph and the key lemma is as follows (see Figure 2.4):

- Start with a graph $G = (V, E)$ and apply the regularity lemma, finding a regular partition $P$;

- Construct the reduced graph $R$ of $G$, w.r.t. the partition $P$;

- Analyze the properties of $R$, in particular its subgraphs;

- As it is assured by Theorem 2, every small subgraph of $R$ is also a subgraph of $G$.

In summary, a direct consequence of the key lemma is that it is possible to search for significant substructures in a reduced graph $R$ in order to find common subgraphs of $R$ and the original graph.

Now, returning to the clustering problem, the approach developed in [97] consists in a two-phase procedure. In the first phase, the input graph is decomposed into small pieces using Szemerédi's partitioning process and the corresponding (weighted) reduced graph is constructed, the weights of which reflect edge-densities between class pairs of the original partition. Next, a standard graph-based clustering procedure is run on the reduced graph and the solution found is mapped back into original graph to obtain the final groups. Experiments conducted on standard benchmark datasets confirmed the effectiveness of the proposed approach both in terms of quality and speed.

Note that this approach differs from other attempts aimed at reducing the complexity of pairwise grouping processes, such as [10, 40, 78], as the algorithm performs no sampling of the original data but works instead on a derived structure which does retain the important features of the original one.

The ideas put forward in [97] were recently developed and expanded by [93] and [76], who confirmed the results obtained in the original paper. [27] have recently applied these ideas to improve the efficiency of edge detection algorithms. They compared the accuracy and the efficiency obtained using the regularity-based approach with that obtained with a combination of a factorization-based compression algorithm and quantum walks. They achieved a huge speed up, from an average of 2 hours for an image of $125 \times 83$ pixels (10375 vertices) to 2 minutes with factorization-based compression and of 38 seconds with regularity compression.

### 2.3.1 An example application: Image segmentation

To give a taste of how the two-phase strategy outlined in the previous section works, here we present some fresh experimental results on the problem of segmenting gray-level images. Each image is abstracted in terms of an edge-weighted graph where vertices represent pixels and edge-weights reflect the similarity between pixels. As customary, the similarity between pixels, say $i$ and $j$, is measured as a function of the distance between the corresponding brightness values, namely, $w(i,j) = exp(-((I(i) - I(j))^2 / \sigma^2)$, where $I(i)$ is the normalized intensity value of pixel $i$ and $\sigma$ is a scale parameter.

We took twelve images from Berkeley's BSDS500 dataset [8] and, after resizing them to $81 \times 121$ pixels, we segmented them using two well-known clustering algorithms, namely Dominant Sets (DS) [77] and Spectral Clustering (SC) [74]. The results obtained were then compared with those produced by the two-phase strategy, which consists of first compressing the original graph using regular partitions and then using the clustering algorithm (either DS of CS) on the reduced graph [97].

Two well-known measures were used to assess the quality of the corresponding segmentations, namely the Probabilistic Rand Index (PRI) [102] and the Variance of Information (VI) [69]. The PRI is defined as:

$$PRI(S, \{G_k\}) = \frac{1}{T} \sum_{i<j} [c_{ij} p_{ij} + (1 - c_{ij})(1 - p_{ij})] \tag{2.3}$$

where $S$ is the segmentation of the test image, $\{G_k\}$ is a set of ground-truth segmentations, $c_{ij}$ is the event that pixels $i$ and $j$ have the same label, $p_{ij}$ its probability, and $T$ is the total number of pixel pairs. The PRI takes values in $[0, 1]$, where $PRI = 1$ means that the test image segmentation matches the ground truths perfectly.

The VI measure is defined as:

$$VI(S, S') = H(S) + H(S') - 2I(S, S') \tag{2.4}$$

where $H$ and $I$ represent the entropy and mutual information between the test image segmentation $S$ and a ground truth segmentation $S'$, respectively. The VI is a metric which measures the distance between two segmentations. It takes values in $[0, log_2 n]$, where $n$ is the total number of pixels, and $VI = 0$ means a perfect match.

The results are shown in Figures 2.5 and 2.6, while Figure 2.7 shows the actual segmentations obtained for a few representative images. Note that the results of the two-stage procedure are comparable with those obtained by applying the corresponding clustering algorithms directly on the original images, and sometimes are even better. Considering that in all cases, the Szemerédi algorithm obtained a compression rate well above 99% (see Table 2.1 for details), this is in our opinion impressive. Note that these results are consistent with those reported in [97, 93, 76, 27].

We also investigated the behavior of the index of partition $ind(P)$ defined in (1.4), during the evolution of the Szemerédi compression algorithm. Remember that this measure is known to increase at each iteration of Alon et al.'s (exact) algorithm described in Section 1.3, and it is precisely this monotonic behavior which guarantees the correctness of the algorithm (and, in fact, of Szemerédi's lemma itself). With our heuristic modifications, however, there is no such guarantee, and hence it is reassuring to see that in all cases the index does in fact increase at each step, as shown in Table 2.2, thereby suggesting that the simple heuristics described in [97, 93] appear not to alter the essential features of the original algorithm. A similar behavior was also reported in [93].

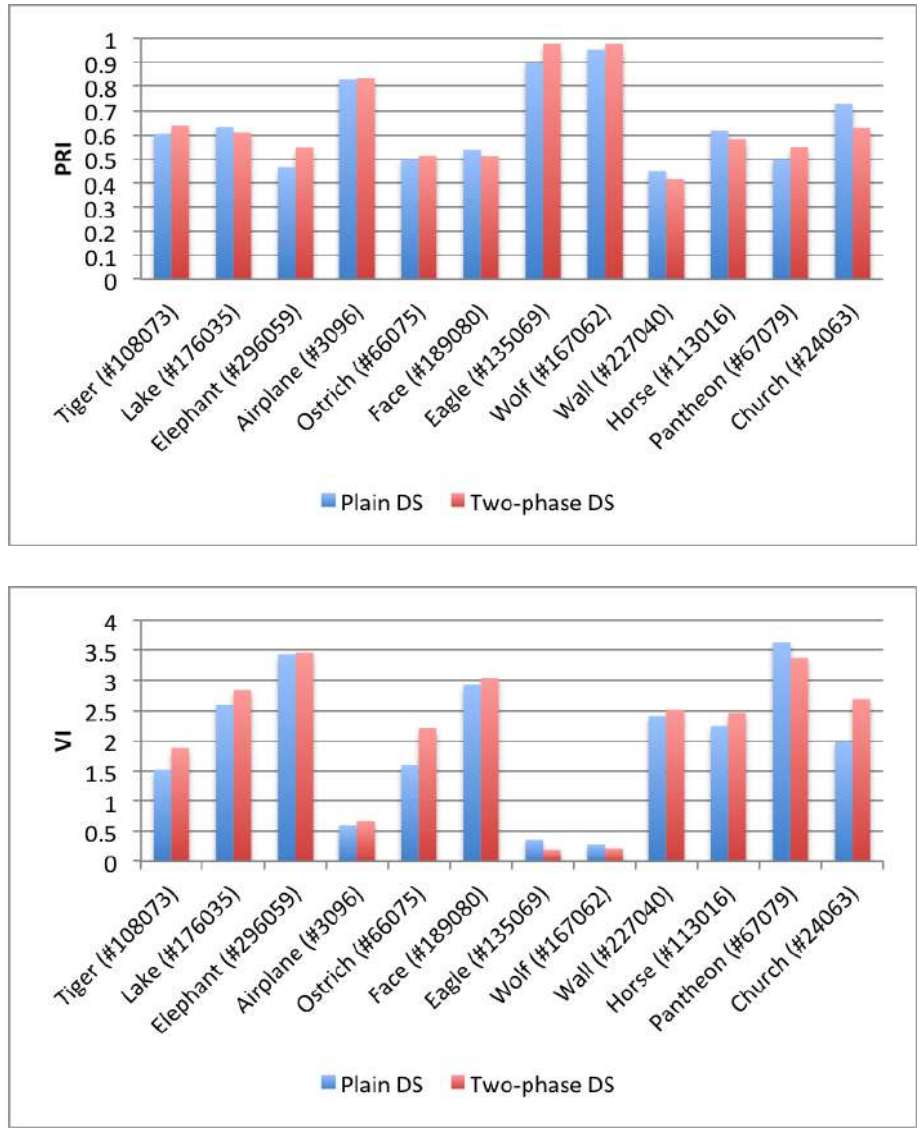

Fig. 2.5 Image segmentation results on twelve images taken from the Berkeley dataset using plain Dominant Sets (Plain DS) and the two-phase strategy described in Section 2.3 (Two-phase DS). Top: Probabilistic Rand index (PRI). Bottom: Variance of Information (VI). (See text for explanation). The numbers in parentheses on the x-axes represent the image identifiers within the dataset.

## 2.3.2 Related works

Besides the use of the regularity lemma described above, in the past few years there have been other algorithms explicitly inspired by the notion of a regular partition which we briefly describe below.

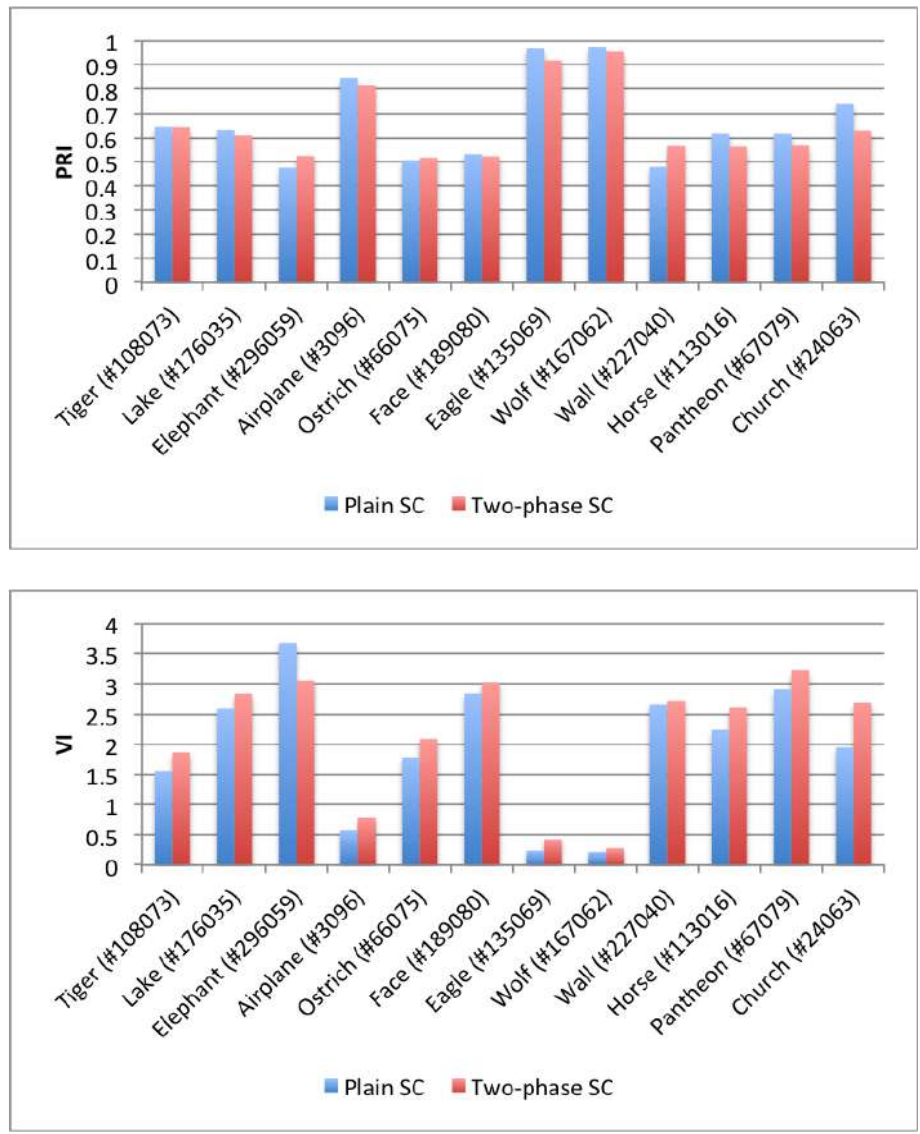

Fig. 2.6 Image segmentation results on twelve images taken from the Berkeley dataset using plain Spectral Clustering (Plain SC) and the two-phase strategy described in Section 2.3 (Two-phase SC). Top: Probabilistic Rand index (PRI). Bottom: Variance of Information (VI). (See text for explanation). The numbers in parentheses on the x-axes represent the image identifiers within the dataset.

[72] introduced a method inspired by Szemerédi's regularity lemma to predict missing connections in cerebral cortex networks. To do so, they proposed a probabilistic approach where every vertex is assigned to one of $k$ groups based on its outgoing and incoming edges, and the probabilistic description of connections between and inside vertex groups are determined by the cluster affiliations of the vertices involved. In particular, they introduced

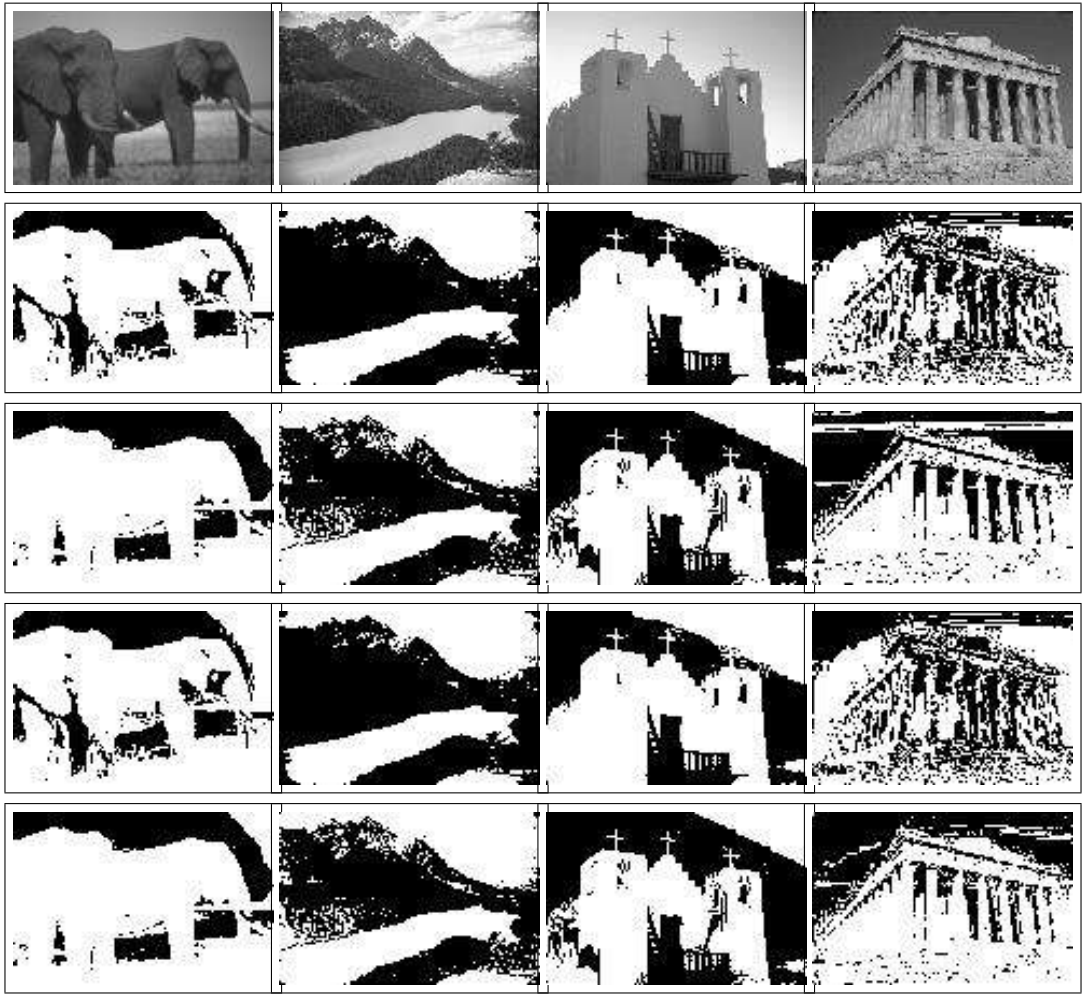

Fig. 2.7 Comparing the segmentation results of plain clustering (DS/CS) and the two-phase approach. First row: original images. Second row: results of plain Dominant Set (DS) clustering. Third row: results of the two-phase Szemerédi+DS strategy. Fourth row: results of plain Spectral Clustering (SC). Fifth row: results of the two-phase Szemerédi+SC strategy.

a parametrized stochastic graph model, called the preference model, which is able to take into account the amount of uncertainty present in the data in terms of uncharted connections. Their method was tested on a network containing 45 vertices and 463 directed edges among them. The comparison of their experimental results with the original data showed that their algorithm is able to reconstruct the original visual cortical network with an higher accuracy compared to state-of-the-art methods.

These good results motivated [79] to develop a method inspired by the notion of regular partition to analyse an experimental peer-to-peer system. This network is modeled as directed weighted graph, where an edge direction indicates a client-server relation and a weight is the proportion of all chunks obtained from such link (edge) during the whole experiment. Their aim was to understand the peer's behavior. In particular, they want to group peers with similar behavior with respect to downloading and uploading characteristics in the same cluster. Their approach exploits max likelihood estimation to extract a partition where all cluster pairs are as much as possible random bipartite subgraphs. Their method was tested on a small network of 48 vertices of a p2p experimental network. The results showed that their algorithm detected some hidden statistical properties of the network. They pointed out that for larger systems, sharper results could be expected, although for larger networks an algorithmic version of Szemerédi's Regularity Lemma could be more plausible solution.

| Image | Original size | Size of reduced graph | Compression rate |
|---|---|---|---|
| Tiger | 9801 | 16 | 99.84% |
| Lake | 9801 | 4 | 99.96% |
| Elephant | 9801 | 64 | 99.35% |
| Airplane | 9801 | 16 | 99.84% |
| Ostrich | 9801 | 16 | 99.84% |
| Face | 9801 | 32 | 99.67% |
| Eagle | 9801 | 64 | 99.35% |
| Wolf | 9801 | 16 | 99.84% |
| Wall | 9801 | 8 | 99.92% |
| Horse | 9801 | 8 | 99.92% |
| Pantheon | 9801 | 8 | 99.92% |
| Church | 9801 | 4 | 99.96% |

Table 2.1 Sizes of the reduced graphs after running the Szemerédi compression algorithm, and corresponding compression rates, for all images used.

More recently, Szemerédi's lemma inspired [84], who developed a variant of stochastic block models [80] for clustering multivariate discrete time series. To this end, they introduced a counterpart of Szemerédi regular partition, called regular decomposition, which is a partition of vertices into $k$ sets is such a way that structure between sets and inside sets are random-like. In particular, the number of clusters $k$ increases at each iteration of their algorithm as long as large clusters are created. The stopping criterion is provided by means of Rissanen's minimum description length (MDL) principle. This choice is driven by the regularity lemma: the algorithm searches for large regular structure, corresponding to a local MDL optimum with the smallest value of $k$. The application of their method to real-life electric smart meter customer has given structures which are more informative than of the structures which are obtained by means of a traditional clustering method as $k$-means.

| Image | $ind(P_1)$ | $ind(P_2)$ | $ind(P_3)$ | $ind(P_4)$ |
|---|---|---|---|---|
| Tiger | 0.142 | 0.217 | 0.272 | 0.317 |
| Lake | 0.004 | 0.085 | 0.129 | 0.173 |
| Elephant | 0.071 | 0.154 | 0.204 | 0.248 |
| Airplane | 0.205 | 0.306 | 0.363 | 0.408 |
| Ostrich | 0.154 | 0.231 | 0.279 | 0.319 |
| Face | 0.006 | 0.061 | 0.102 | 0.135 |
| Eagle | 0.213 | 0.318 | 0.376 | 0.417 |
| Wolf | 0.014 | 0.103 | 0.181 | 0.215 |
| Wall | 0.201 | 0.304 | 0.362 | 0.399 |
| Horse | 0.078 | 0.169 | 0.214 | 0.252 |
| Pantheon | 0.049 | 0.101 | 0.151 | 0.179 |
| Church | 0.009 | 0.081 | 0.126 | 0.167 |

Table 2.2 Behavior of the index of partition (1.4) in the first four steps of the Szemerédi compression stage, for all images used.

Finally, we mention the recent work of [17] who have introduced a local algorithm for correlation clustering to deal with huge datasets. In particular, they took inspiration from the PTAS for dense MaxCut of [43] and used low-rank approximations to the adjacency matrix of the graph. The algorithm searches a weakly regular partition for the graph in sub-linear time to get a good approximate clustering. They pointed out that their algorithm could be naturally adapted to distributed and streaming systems to improve their latency or memory usage. Thus, it can be used to detect communities in large-scale evolving graphs.

### 2.3.3    Concluding remarks

In this section, we discussed the relevance of the regularity lemma in the context of structural pattern recognition. We focused, in particular, on graph-based clustering and image segmentation, and we showed how the notion of a regular partition and associated algorithms can provide fresh insights into old pattern recognition and machine learning problems. Preliminary results on some real-world data seem to suggest that, although Szemerédi's lemma is a result concerning very large graphs, 'regular-like' structures can appear already in suprisingly small-scale graphs. The strength of regular partitions, though, is expected to reveal itself in larger and larger graphs and, if confirmed, this would pave the way for a principled approach to big data analysis. Presently, virtually all standard methods for dealing with big data are based on classical clustering techniques, such as $k$-means or variations thereof. Regular-like partitions could offer a different, more principled perspective to the problem by providing more informative structures than traditional clustering methods.

# Chapter 3

# Separating Structure from Noise in Large Graphs

> There is no property absolutely essential
> to any one thing. The same property
> which figures as the essence of a thing
> on one occasion becomes a very
> inessential feature upon another.
>
> William James

*How can we separate structural information from noise in large graphs?* To address this fundamental question, we present a new heuristic algorithm which is characterized by an improvement of the summary quality both in terms of reconstruction error and of noise filtering. In this chapter, we use our new heuristic to first build a summary of a graph $G$, and then we "blow-up" the summary to obtain a graph $G'$, called reconstructed graph, which is close to $G$ in terms of the $l_p$-reconstruction error. We study the noise robustness of our approach in terms of the reconstruction error by performing an extensive series of experiments on both synthetic and real-world data. As far as the synthetic data are concerned, we generate graphs with a cluster structure, where the clusters are perturbed with different levels of noise. As far as the real-world data are concerned, we add spurious edges in accord with different noise probabilities. The aim of this series of experiments is to assess if the framework is able to separate structure from noise. In the ideal case, the distance between $G$ and $G'$ should be only due to the filtered noise.

Moreover, in the second part of the chapter, we use our summarization algorithm to address the *graph search* problem defined under a similarity measure. The aim of graph search is to retrieve from a database the top-$k$ graphs that are most similar to a query graph.

Since noise is common in any real-world dataset, the biggest challenge in graph search is developing efficient algorithms suited for dealing with large graphs containing noise in terms of missing and adding spurious edges. In our approach, all the graphs contained in a database are compressed off-line, while the query graph is compressed on-line. Thus, graph search can be performed on the summaries, and this allows us to speed up the search process and to reduce storage space. Finally, we evaluate the usefulness of our summaries in addressing the graph search problem by performing an extensive series of experiments. In particular, we study the quality of the answers in terms of the found top-$k$ similar graphs and the scalability both in the size of the database and in the size of the query graphs.

## 3.1   Related works

The first contribution presented in this chapter is the introduction of a principled framework for summarizing large graphs with the aim of preserving their main structural patterns. Previous related works presented methods which mainly built summaries by grouping the vertices into subsets, such that the vertices within the same subset share some topological properties. The works in [94, 73] introduced methods for partitioning the vertices into non-overlapping clusters, so that vertices within the same cluster are more connected than vertices belonging to different clusters. A graph summary can be constructed by considering each cluster as a *supernode*, and by connecting each pair of supernodes with a *superedge* of weight equals to the sum of the cross-cluster edges. However, since graph summarization and clustering have different goals, this approach is suited only if the input graph has a strong community structure. In [60], the summary is generated by greedily grouping vertices, such that the normalized reconstruction error between the adjacency matrix of the input graph and the adjacency matrix of the *reconstructed graph* is minimized. Since in their work they exploited heuristic algorithms, they can not give any guarantees on the quality of the summary. The work in [88] proposed a method of building a summary with quality guaranty by minimizing the $l_p$-*reconstruction error* between the adjacency matrix of the input graph and the adjacency matrix of the reconstructed graph. Since both approaches aim to minimize a distance measure between the input and the reconstructed graph, they are not the best choice for summarizing noise graphs. By contrast, our goal is to develop a graph summarization algorithm which is robust against noise. For a more detailed picture on how the field has evolved previously, we refer the interested reader to the survey of Liu et al. [67].

The second contribution presented in this chapter consists in addressing the graph search problem using the proposed summarization framework. Locating the occurrences of a query graph in a large database is a problem which has been approached in two main different ways, based on subgraph isomorphism and approximate graph matching respectively. Ullman [101] posed one of the first milestones in subgraph isomorphism. He proposed an algorithm which decreases the computational complexity of the matching process by reducing the search space with backtracking. Recently, Carletti et al. [21] introduced an algorithm for graph and subgraph isomorphism which scales better than Ullmann's one. In particular, Carletti et al's algorithm, which may be considered as the state-of-the-art in exact subgraph matching, can process graphs of size up to ten thousand nodes. However, since subgraph isomorphism is a NP-complete problem, the algorithms based on exact matching are prohibitively expensive for querying against a database which contains large graphs. Moreover, due to the noise contained in real-world data, it is common to mismatch two graphs which have the same structure but different levels of noise. Indeed, these contributions are focused on exact matching and, even if they proposed efficient solutions, they are not noise robust. By contrast, our goal is to develop an efficient graph search algorithm which is robust against noise. Hence, approaches based on approximate graph matching are more suitable for addressing the graph search problem. Indeed, in this category lies the most effective graph similarity search algorithms. Most of the time, the searching phase is conducted under the *graph edit distance* (*GED*) constraint [64, 110, 111]. The graph edit distance $GED(g_1, g_2)$ is defined as the minimum number of edit operations (adding, deletion and substitution) that modify $g_1$ step-by-step to $g_2$ (or vice versa). In [110] and in [109] the authors underline the robustness of *GED* against noise due to its error-tolerant capability. Unfortunately, the *GED* computation is NP-hard, and most existing solutions adopt a *filtering-verification* technique. In particular, first, they use a pruning strategy to filter out false positive matches, and then verify the remaining candidates by computing *GED*. In this context, the work of Liang and Zhao [64] represents the state-of-the-art. They provide a partition-based *GED* lower bound to improve the filter capability, and a multi-layered indexing approach to filter out false positives in an efficient way. Their algorithm can deal with databases with a high number of graphs, but cannot handle large graphs due to the complexity of *GED* computation. Instead, our algorithm is designed to scale both in the size of the databases and in the size of the graphs.

## 3.2   The summarization algorithm

In chapter 2, we have pointed out that the main limitations which prevent the application of Alon et al.'s algorithm to practical problems concern Step 2 and Step 4. To make the algorithm truly applicable, we introduced a greedy algorithm that allow us to overcome the limitations posed by Step 2. For the sake of clarity, we report briefly the heuristic for finding the certificates that witness the irregularity of a pair of classes. Given a class $C_s$, we select in a greedy way a set $Y' \subseteq C_s$ with the highest deviation $\sigma(Y')$ (the deviation is defined in 1.3). To do so, the nodes of $C_s$ are sorted by bipartite degree, and $Y'$ is built by adding $\frac{\varepsilon^4}{4}n$ nodes with the highest degree. At each iteration of the greedy algorithm, the node with a degree that deviates more from the average degree is added to the candidate certificates. This last operation is repeated until the subset $C'_s$ that satisfies condition 3 is found. This almost guarantees to put in a candidate certificate the nodes that have a *connectivity pattern* that deviates from the one characterizing the majority of the nodes which belong to $C_s$.

As far Step 4 is concerned, we provided a simple heuristic to deal with the tower-type dependence on the accuracy parameter $\varepsilon$. In particular, for each class, one can limit the number of irregular pairs containing it to at most one, possibly chosen randomly among all irregular pairs.

In this chapter, we introduce a new refinement algorithm (REFINEMENT in Algorithm 2) with the aim of refining irregular pairs into more regular new pairs. In particular, the refinement heuristic starts by randomly selecting a class, then iteratively processes all the others.

- If $C_i$ is $\varepsilon$-regular with all the others, the procedure sorts the nodes of $C_i$ by their *internal degree*, i.e. the degree calculated with respect to the nodes of the same class, obtaining the following sorted sequence of nodes $v_1, v_2, v_3, v_4, v_5, v_6, \cdots, v_{|C_i|}$. The next step splits (UNZIP) this sequence into two sets $C_i^1 = \{v_1, v_3, v_5, \cdots, n_{|C_i|-1}\}$ and $C_i^2 = \{v_2, v_4, v_6, \cdots, v_{|C_i|}\}$. The latter sets are part of the refined partition $\mathscr{P}_\varepsilon^{i+1}$.

- If $C_i$ forms an irregular pair with other classes, the heuristic selects the candidate $C_j$ that shares the most similar internal structure with $C_i$ by maximizing $S = d(C_i, C_j) + (1 - |d(C_i, C_i) - d(C_j, C_j)|)$, where $d(C_i, C_i) = e(C_i, C_i)/|C_i|^2$ is the *internal density*.

  After selecting the best matching class $C_j$, we are ready to split the pair $(C_i, C_j)$ in 4 new classes $C_i^1, C_i^2, C_j^1, C_j^2$ based on the internal densities of the certificates $C'_i$ and $C'_j$.

  - In particular, a SPARSIFICATION procedure is applied when the internal density of a certificate is below a given threshold. This procedure randomly splits the certificate into two new classes. In order to match the equi-cardinality property, the new classes are filled up to $|C_i|/2$ by adding the remaining nodes from the corresponding complement. We choose the nodes that share the minimum number of connections with the new classes.

  - On the other hand, if the internal density of a certificate is above a given threshold, then a DENSIFICATION procedure is applied. In particular, the heuristics sorts the nodes of the certificate by their internal degree and UNZIP the set into two new classes. Also in this case, we fill the new sets up to $|C_i|/2$ by adding the remaining nodes from the corresponding complement by choosing the nodes which share the major number of connections with the new classes.

The pseudocode of the summarization algorithm is reported in Algorithm 2, while the procedure REFINEMENT is reported in Algorithm 3. The overall complexity of our summarization algorithms is $O(M(n)) = O(n^{2.376})$, which is dominated by the verification of Condition 3.

---

**Algorithm 2** The Summarization Algorithm

**Input**:
- $\varepsilon$ is the tolerant parameter 1.2;
- $G = (V, E)$ is an undirected simple graph (the input graph);
- $c\_min$ is the minimum compression rate, expressed as $k/|V|$

**Output**:
- $\mathscr{P}$ is a regular partition of $G$, where $|\mathscr{P}| = k$.

---

1: **procedure** APPROXALON($\varepsilon$, c_min, $G = (V, E)$)
2:     partitions = empty list
3:     $\mathscr{P}_\varepsilon^1$ = Create initial random partition from $G$
4:     **while** True **do**
5:         #irr_pairs = CHECKPAIRSREGULARITY($\mathscr{P}_\varepsilon^i$) (see 2.1)
6:         **if** #irr_pairs $> \varepsilon\binom{k}{2}$ or COMPRESSRATE($\mathscr{P}_\varepsilon^i$) < c_min **then**
7:             break
8:         **else**
9:             $\mathscr{P}_\varepsilon^{i+1}$ = REFINEMENT($\mathscr{P}_\varepsilon^i$)
10:            **if** $\mathscr{P}_\varepsilon^{i+1}$ is $\varepsilon$-regular **then**
11:                partitions.add($\mathscr{P}_\varepsilon^{i+1}$)
12:            **else**
13:                break
14:    Select best partition $\mathscr{P}^*$ with maximum $sze\_idx$ from list partitions

---

## 3.3   Graph search using summaries

In this section, we discuss how to use our summarization framework to efficiently address the graph search problem defined under a similarity measure. The aim of graph search is to retrieve from a database the top-$k$ graphs that are most similar to a query graph.

**Problem definition** We consider a graph database $\mathscr{D}$ containing a high number of simple undirected graphs $g_j \in \mathscr{D}$, $j = 1 \cdot |\mathscr{D}|$, and, for the sake of generality, we allow the edges to be weighted.

---

**Algorithm 3** Refinement step performed at the $i$-th iteration of the summarization algorithm 1. Statements 5,10 and 12 may add a node to $C_0$. $\mathscr{P}_\varepsilon^i$ is the partition at iteration $i$ of the summarization algorithm

---

1: **procedure** REFINEMENT($\mathscr{P}_\varepsilon^i$)
2:     **for** each class $C_i$ in $\mathscr{P}_\varepsilon^i$ **do**
3:         **if** $C_i$ is $\varepsilon$-regular with all the other classes **then**
4:             $C_i = $ SORT_BY_INDEGREE($C_i$)
5:             $C_i^1, C_i^2 = $UNZIP($C_i$)
6:         **else**
7:             Select $C_j$ with most similar internal structure
8:             Get certificates $(A', B')$ and complements $(\bar{A}', \bar{B}')$ of $C_i, C_j$
9:             **if** $d(A', A') < 0.5$ **then**
10:                $C_i^1, C_i^2 = $ SPARSIFICATION($A', \bar{A}' \cup \bar{B}'$)
11:             **else**
12:                $C_i^1, C_i^2 = $ DENSIFICATION($B', \bar{B}' \cup \bar{B}'$)
13:             Perform step 9,10,11,12 for $B'$
14:     **if** $|C_0| > \varepsilon n$ and $|C_0| > |\mathscr{P}_\varepsilon^{i+1}|$ **then**
15:         Uniformly distribute nodes of $C_0$ between all the classes
16:     **else**
17:         **return** ($\mathscr{P}_\varepsilon^{i+1}, irregular$)
18:     **return** ($\mathscr{P}_\varepsilon^{i+1}, regular$)

---

**Problem 1** (Graph search). *Given a graph database $\mathscr{D} = \{g_1, g_2, \cdot, g_{|D|}\}$, a query graph $q$, and a positive integer $k$, the graph similarity search problem is to find the top-$k$ graphs in $\mathscr{D}$ that are most similar to $q$ according to a similarity measure.*

As far as the similarity measure is concerned, the most used one is the *graph edit distance* (*GED*) due to its generality, broad applicability and noise robustness [64, 109]. However, since the *GED* computation is NP-hard, it is not suited to deal with large graphs. To overcome this limitation, we use the *spectral distance* [52], which is computed by comparing the eigenvalues of the two graphs being matched. The choice of this measure is motivated by the work of Van Dam and Haemers [103], who show that graphs with similar spectral properties generally share similar structural patterns. In this thesis, we introduce a slightly modified version of the spectral distance with the aim of increasing its range of applicability to pair of graphs that violates the assumption of the Theorem 1 in [103], which states a precise order between the eigenvalues of the two graphs being matched.

**Definition 3** (Spectral distance). *Given two simple undirected weighted graphs $G_1 = (V_1, W_1)$ with $|V_1| = n_1$, and $G_2 = (V_2, W_2)$ with $V_2 = n_2$. Let us denote the corresponding spectras as $0 = \lambda_1^{(1)} \leq \lambda_2^{(1)}, \leq \cdot \leq \lambda_{n_1}^{(1)}$ and $0 = \lambda_1^{(2)} \leq \lambda_2^{(2)}, \leq \cdot \leq \lambda_{n_2}^{(2)}$. We may assume without loss of generality that $n_2 > n_1$. The spectral distance is then defined as*

$$SD(G_1, G_2, l) = \frac{1}{k} \left( \sum_{i=1}^{l} |\lambda_i^{(2)} - \lambda_i^{(1)}| + \sum_{i=l+1}^{n1} |\lambda_{i+n_2-k}^{(2)} - \lambda_i^{(1)}| \right) \tag{3.1}$$

**Using the summaries** In our approach, all the graphs contained in a database are summarized off-line, while the query graph is summarized on-line by means of our summarization framework. Thus, graph search can be performed on graph summaries, and this allows us to speed up the search process and to reduce storage space. In particular, for each graph $g_j$ of a database $\mathscr{D}$, we store two different informations: the summary $r_j$ of $g_j$ and the eigenvalues $eig_{r_j}$ of $r_j$. We then summarized on-line the query graph $q$ obtaining its summary $r_q$. Finally, we compute the spectral distance between $r_q$ and each graphs summaries $r_j \in \mathscr{D}$. The desired top-$k$ graphs will be obtained by selecting, from $\mathscr{D}$, the $k$ graphs corresponding to the $k$ smallest value of the spectral distance previously computed. The pseudocode of our approach to graphs search is reported in Algorithm 4.

## 3.4 Experimental Evaluation

In this section, we evaluate our summarization algorithm both on synthetic graphs and on real-world networks to assess:

---

**Algorithm 4** Graph Search Using The Summary

---

1: **procedure** ADDGRAPHTODATABASE($g, \mathscr{D}$)
2:     $r =$ Summarize $g$
3:     $eig_r =$ Calculate the eigenvalues of the adj. matrix of $r$
4:     Store $(r, eig_r)$ in $\mathscr{D}$
5: **procedure** 2-STAGEGRAPHSEARCH($q, \mathscr{D}$)
6:     $r_q =$ Summarize $q$
7:     $eig_{r_q} =$ Calculate the eigenvalues of the adj. matrix of $r_q$
8:     $sd\_array = \emptyset$
9:     **for** $r_j$ in $\mathscr{D}$ **do**
10:         $sd =$ Spectral Distance($r_j, r_q, eig_{r_j}, eig_{r_q}$)
11:         Append $sd$ to $sd\_array$
12:     Order $sd\_array$
13:     **return** first $k$ results of $sd\_array$ and their relative graphs.

---

- the ability of the proposed algorithm to separate structure from noise;

- the usefulness of the summaries in retrieving from a database the top-$k$ graphs that are most similar to a query graph.

### 3.4.1 Experimental Settings

In our experiments we used both synthetic graphs and real-world networks. We generated synthetic graphs with a cluster structure, where the clusters are perturbed with different levels of noise. In particular, each graph is generated by adding spurious edges between cluster pairs and by dropping edges inside each cluster. Figure 3.1 provides a concrete example with a visual explanation. The pseudocode of the algorithm used to generate the synthetic datasets is reported in Algorithm 5.

As far as the real-world networks are concerned, we used two different datasets which have been taken from two famous repositories: the Stanford Large Network Dataset Collection SNAP [62] and the Konect repository of the University Koblenz-Landau Konect [59]. In particular, we used the following networks: Facebook [63], Email-Eu-core [108][61], Openflights[2], and Reactome [53]. Our algorithm is implemented in Python 3.6.3 [1] and the experiments are performed on an Intel Core i5 @ 2.60GHz HP Pavilion 15 Notebook with 8GB of RAM (DDR3 Synchronous 1600 MHz) running Arch-Linux with kernel version 4.14.4-1.

---

[1]The implementation is available from https://github.com/MarcoFiorucci/graph-summarization-using-regular-partitions

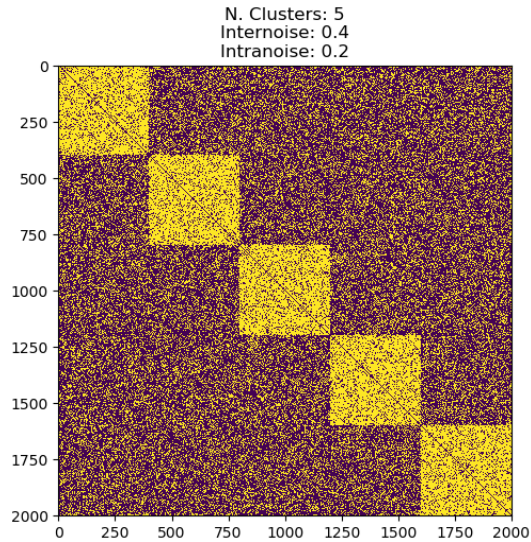

Fig. 3.1 The adjacency matrix of an undirected synthetic graph of 2000 vertices. The graph is generated by corrupting 5 cliques as described in Algorithm 5. In particular, the *intra-cluster noise probability* is 0.2 and the *inter-cluster noise probability* is 0.4.

---

**Algorithm 5** Synthetic graph generator.

**Input parameters**:

- $n$ is the size of the desired graph $G$;

- *num_c* is the number of clusters contained in $G$;

- $\eta_1$ is the probability of adding a spurious edge between a pair of clusters (inter-cluster noise probability);

- $\eta_2$ is the probability of dropping an edge inside a cluster (intra-cluster noise probability).

**Output**: $G$.

---

1: **procedure** SYNTHGRAPHGEN($n$, *num_c*, $\eta_1$, $\eta_2$)

2:     $G$ = Generate Erdős Rényi graph of size $n$ using $\eta_1$ as edge probability

3:     *clust_dim* = $n$/*num_c*

4:     **for** $i$ in *num_c* **do**

5:         Select *clust_dim* nodes from $G$ and create cluster $c_i$ with them

6:         For each edge in $c_i$ drop it with probability $\eta_2$

7:     **return** $G$

---

### 3.4.2  Graph Summarization

We performed experiments on both synthetic graphs and on real-world networks to assess the ability of the proposed algorithm to separate structure from noise. As evaluation criterion, we used the reconstruction error, which is expressed in terms of normalized $l_p$ norm computed between the similarity matrix of an input graph $G$ and the similarity matrix of the corresponding reconstructed graph $G'$.

**Definition 4** (Reconstruction error). *Given the similarity matrix of the input graph $A_G$ and the similarity matrix of the reconstructed graph $A_{G'}$, the reconstruction error is defined as follows:*

$$l_p(\mathbf{A_G}, \mathbf{A_{G'}}) = (\sum_{i=1}^{n} \sum_{j=1}^{n} (A_G(i,j) - A_{G'}(i,j))^p)^{\frac{1}{p}}$$

We decided to use the reconstruction error in order to compare our results with the ones presented by Riondato et al. [88], who evaluated the summary quality using these measures. This choice is due to the fact that their algorithm summarizes a graph by minimizing the reconstruction error. However, they pointed out that the reconstruction error has some shortcomings. In particular, given an unweighted graph $G$, it is possible to produce an uninteresting summary with only one supenode corresponding to the vertex set and $l_1$ reconstruction error at most $n^2$. On the other hand, if we obtained an useful summary, where each pair of vertices belonging to a supernode share an high number of common neighbors, then we get a low (say $o(n^2)$) $l_1$ reconstruction error: this is a desirable behavior because low values of $l_1$ correspond to high quality summaries. Unfortunately, such low values are often obtained only with summaries having an high number of supernodes. This prevents to adopt the reconstruction error as a general measure to assess the summary quality.

As far the summarization and reconstruction steps are concerned, we proceeded, in all the experiments, in the following way: we applied our summarization algorithm (see Algorithm 2) to summarize an input graph $G$. We then "blow-up" the summary in order to obtain the reconstructed graph $G'$, which preserves the main structure carried by the input graph (Figure 3.2).

**Noise Robustness Evaluation**  We study the ability of the proposed algorithm to separate structure from noise in graphs performing an extensive series of experiments on both synthetic graphs and real-world networks. As far synthetic graph experiments are concerned, we generated a graph $G$ by corrupting the clusters of $GT$ in the following way: we added spurious edges between each cluster pair with probability $\eta_1$, and we dropped edges inside each cluster with probability $\eta_2$ (see Algorithm 5). As far as the real-world networks experiments are concerned, we added spurious edges with probability *noise probability* to a original graph $GT$ obtaining an input graph $G$.

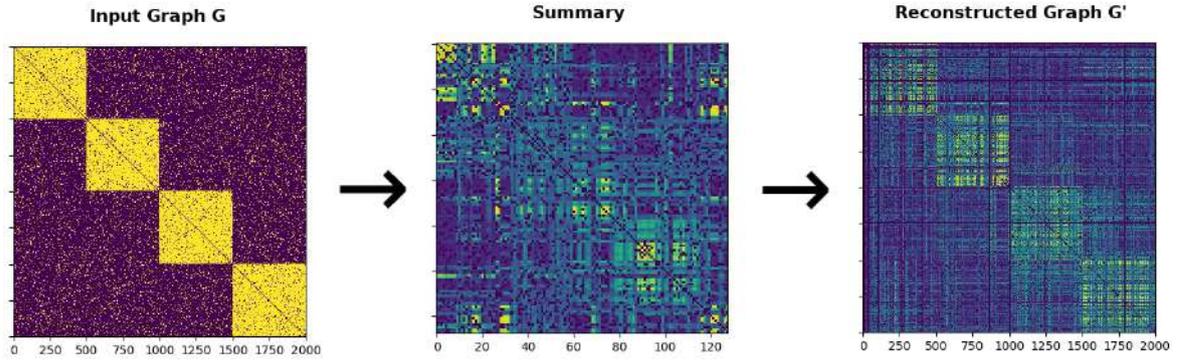

Fig. 3.2 We summarized an input graph $G$ by using the proposed algorithm 0. We then "blow up" the summary to obtain the reconstructed graph $G'$.

In the ideal case, the distance between $G$ and the corresponding reconstructed graph $G'$ should be only due to the filtered noise, while the distance between $GT$ an $G'$ should be closed to zero. Hence, we computed the reconstruction error $l_2(G', GT)$ to assess the robustness of our summarization framework against noise.

**Experiment 1.** We generated synthetic graphs of different sizes, spanning from $10^3$ up to $10^4$ nodes. We synthesized 250 graphs by considering, for each of the 10 different sizes, all the 25 combinations of the following noise probabilities:

- the probability $\eta_1$ of adding a spurious edge between a pair of clusters, called *inter-cluster noise probability*, which assumes values in $\{0.1, 0.2, 0.3, 0.4, 0.5\}$;

- the probability $\eta_2$ of dropping an edge inside each cluster, called *intra-cluster noise probability*, which assumes values in $\{0.1, 0.2, 0.3, 0.4, 0.5\}$.

Let's consider a synthetic graph $G_{n,(\eta_1,\eta_2)}$, where n is its size, and $(\eta_1, \eta_2)$ corresponds to one of the 25 pairs of noise probabilities. For each $G_{n,(\eta_1,\eta_2)}$ we obtained the reconstructed graph $G'_{n,(\eta_1,\eta_2)}$, and we then computed the reconstruction error $l_2(G'_{n,(\eta_1,\eta_2)}, GT)$. Given a size $n$, we computed the median $m_n$ of $\{l_2(G'_{n,(0.1,0.1)}, GT), l_2(G'_{n,(0.1,0.2)}, GT), \cdots, l_2(G'_{n,(0.5,0.5)}, GT)\}$. We reported in figure 3.3 the 10 medians computed using our summarization framework and the corresponding medians obtained by applying Riondato et al.'s algorithm [88]. We can see how our framework outperforms the state-of-the-art summarization algorithm in terms of robustness against noise.

**Experiment 2.** The aim of this experiment is to study separately the robustness against the inter-cluster and the intra-cluster noise. Let's consider the probability of dropping an edge inside each cluster $\eta_2$ equals to 0.2 and the graph size $n$ equals to $10^4$. To asses the inter-cluster noise robustness, we generated synthetic graphs $G_{10^4,(\eta_1,0.2)}$, where $\eta_1$ assumes values in $\{0.1, 0.15, 0.2, 0.25, 0.3, 0.35, 0.4, 0.45, 0.5\}$, and we computed the reconstruction errors $l_2(G'_{10^4,(\eta_1,0.2)}, GT)$. As far the intra-cluster noise is concerned, we chose the probability of adding a spurious edges between each pair of clusters $\eta_1 = 0.2$, and the graph size $i = 10^4$. We then generated synthetic graphs $G_{10^4,(0.2,\eta_2)}$, where $\eta_2$ assumes values in $\{0.1, 0.15, 0.2, 0.25, 0.3, 0.35, 0.4, 0.45, 0.5\}$, and we computed the reconstruction errors $l_2(G'_{10^4,(0.2,\eta_2)}, GT)$. Figure 3.4 illustrates the comparison between our results with those obtained by applying Riondato et al.'s algorithm [88]. This results are in accord to those presented in figure 3.3, and provides an experimental verification of the ability of our method to separate structure from noise in graphs.

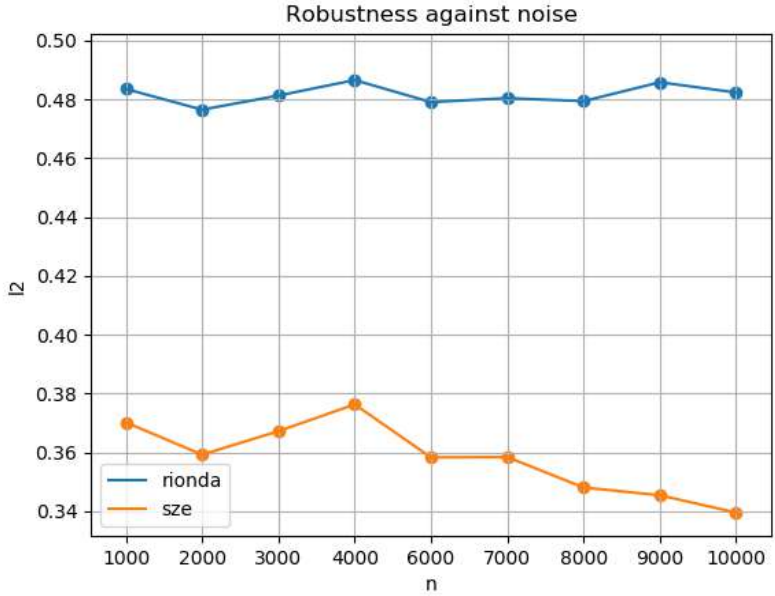

Fig. 3.3 The plot shows the medians computed, for each size $n$, from the 25 values of $l_2(G'_{n,(\eta_1,\eta_2)}, GT)$, where $(\eta_1, \eta_2)$ corresponds to one of the 25 pairs of noise probabilities. The curve "rionda" is obtained by using Riondato et al.'s algorithm [88], while the curve "sze" is obtained by applying our summarization framework.

**Experiment 3.** We added spurious edges with probability *noise probability* to an original real-world network *GT* obtaining an input graph *G*. The *noise probability* assumes values in $\{0.01, 0.02, 0.03, 0.04, 0.05, 0.06, 0.07, 0.08, 0.09, 0.1\}$. We applied this procedure on real-word-networks, which have been taken from the Stanford Large Network Dataset Collection SNAP [62] and from the Konect repository of the University Koblenz-Landau

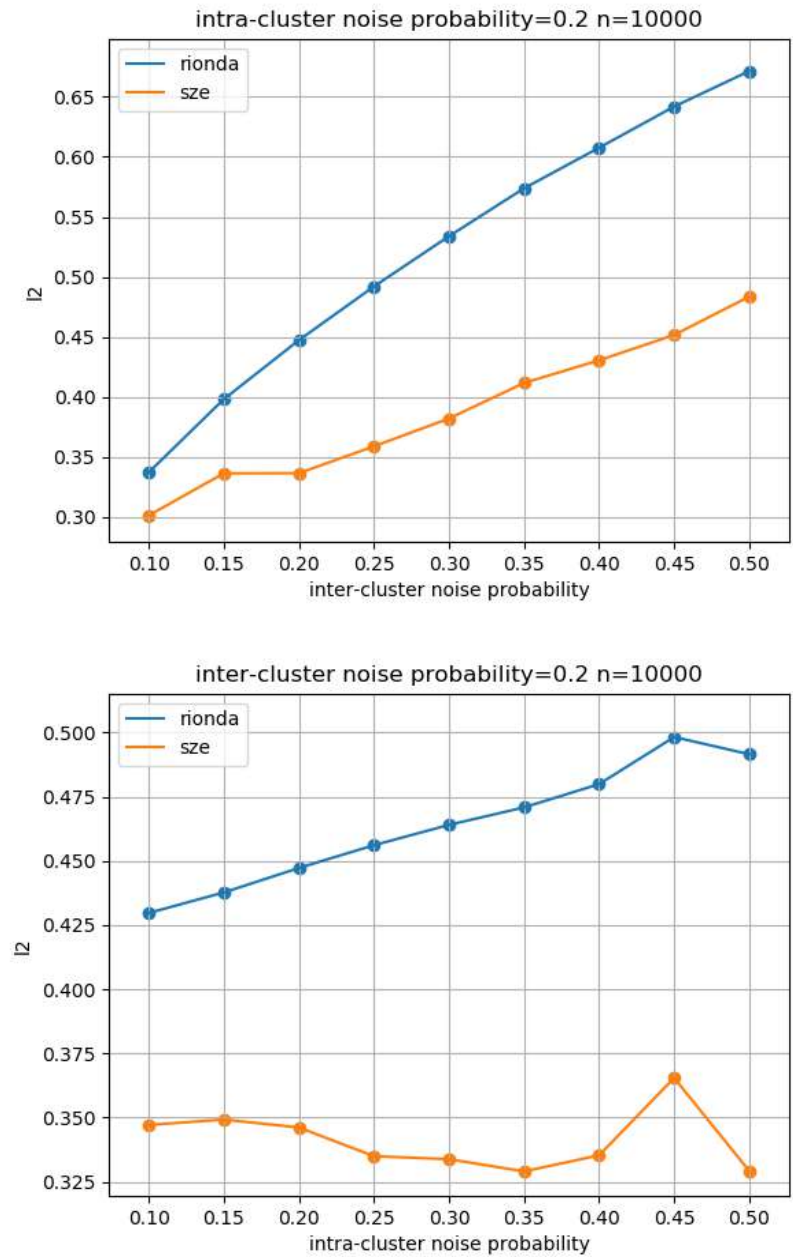

Fig. 3.4 The first plot represents $l_2(G'_{10^4,(0.2,\eta_2)}, GT)$ versus the intra-noise probability. The second plot represents $l_2(G'_{10^4,(\eta_1,0.2)}, GT)$ versus the inter-noise probability. The curve "rionda" is obtained by using Riondato et al.'s algorithm, while the curve "sze" is obtained by applying our summarization framework.

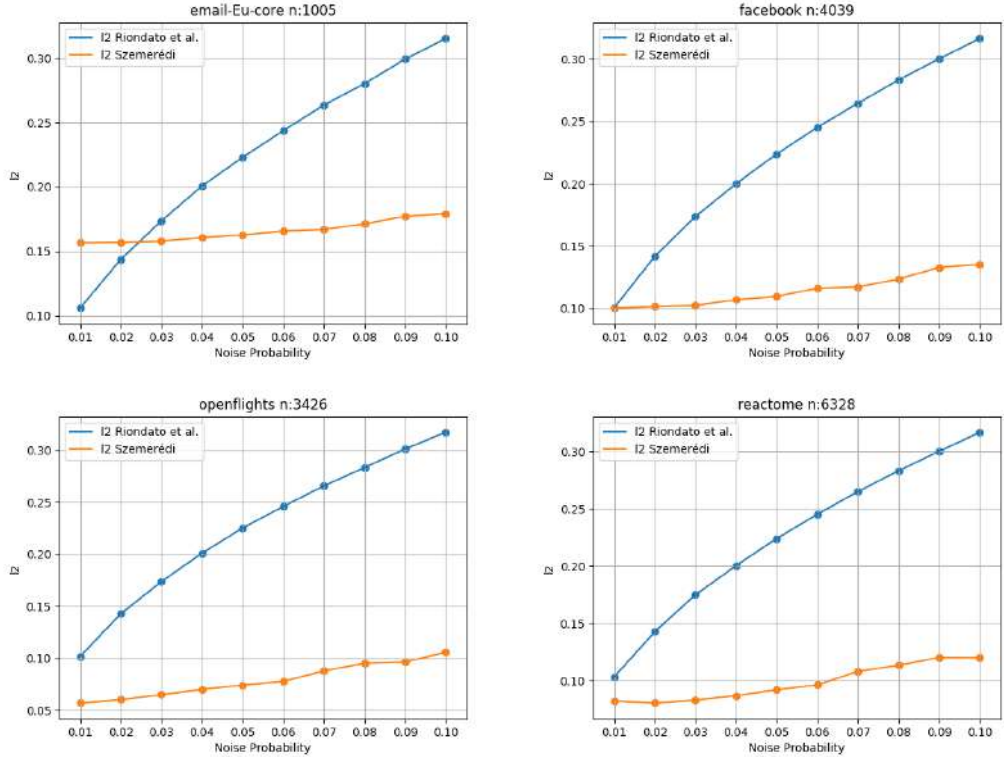

Fig. 3.5 These plots represent the median of the $l_2(G', GT)$ versus the noise probability. We run 20 experiments for each value of the noise probability. The curve "rionda" is obtained by using Riondato et al.'s algorithm, while the curve "sze" is obtained by applying our summarization framework.

Konect [59]. Since our framework is based on the Regularity Lemma, which is suited to deal only with dense graphs, we expect to obtain low quality summaries from sparse real-world networks. However, as shown in figure 3.5, our framework outperforms the state-of-the-art summarization algorithm in terms of robustness against noise providing good quality summary even on sparse real-world networks. In particular, we can see how the quality increases with the size of the input graph, which is in accord with the assumptions of the Regularity Lemma.

### 3.4.3 Graph Search

We performed extensive experiments on synthetic datasets to assess the usefulness of the summaries in retrieving, from a database, the top-$k$ graphs that are most similar to a query graph. To this end, we evaluate the quality of the answer in terms of the found top-$k$ similar graphs, and we evaluate the scalability both in the size of the database and in the size of the graphs.

**Quality Evaluation** We conducted the following experiment: we compared graph search on the summaries with the baseline approach, in which the spectral distance is computed between no preprocessing graphs. The aim of the experiment is to show that pre-summarizing the graphs in the databases increases the noise robustness of search process. We created a database $\mathscr{D}$ contained synthetic graphs, which have different structures corrupting with different levels of noise (see algorithm 5). In particular, each graph is generated by combining the following three factors: five different possible number of clusters $\{4, 8, 12, 16, 20\}$, six different possible levels of intra-cluster noise probability $\{0.05, 0.1, 0.15, 0.2, 0.25, 0.3\}$ and six different possible levels of inter-cluster noise probability $\{0.05, 0.1, 0.15, 0.2, 0.25, 0.3\}$. Given a size $n$, we generated 180 graphs considering all the possible combinations of these three parameters. As described in Algorithm 3.3, we stored in $\mathscr{D}$: the eigenvalues of the 180 synthetic graphs, their summaries and the corresponding eigenvalues. Finally, we then grouped the graphs in $\mathscr{D}$ into five groups. Each group $\omega_i$, with $i = 1, 2, 3, 4, 5$, is composed by 36 graphs that are generated by corrupting the same cluster structure with different combinations of intra-cluster and inter-cluster noise probability. Hence, all the graphs belonging to a given group $\omega_i$ are similar, since they have the same main structure.

More formally, we constructed a set $Q$ of five query graphs by randomly sampling one graph from each group $\omega_i$. Then, we first computed the spectral distance between $q_i \in Q$ and every graph in the database $\mathscr{D}$. We then calculated the $AP@k$ for each query $q_i$ by considering relevant the graphs belonging to $\omega_i$ i.e. the same group of $q_i$. Finally, we computed the $MAP@k$ score by averaging the *average precision $AP@k(q_i)$* of the five graphs in $Q$.

We repeated the same procedure using the summaries of the 180 synthetic graphs contained in $\mathscr{D}$. The aim of this experiment is to compare the quality obtained using our approach with that obtained by computing the spectral distance between original graphs. We performed the experiment by considering the following graph sizes $n = 1500, 2000, 3000, 7000$.

The $AP@k(q_i)$ and the $MAP@k$ are defined as follows.

**Definition 5** (Average precision). *Given a query $q \in Q$, a set of relevant graphs $\omega_i$ (graphs that share the same structure with q). Let us consider the output top-k graphs in a database $\mathscr{D}$ ordered by crescent spectral distance. We define the average precision at k as follows.*

$$AP@k(q) = \frac{1}{|w_i|} \cdot \sum_{j=1}^{k} Precision(j) \cdot Relevance(j) \qquad (3.2)$$

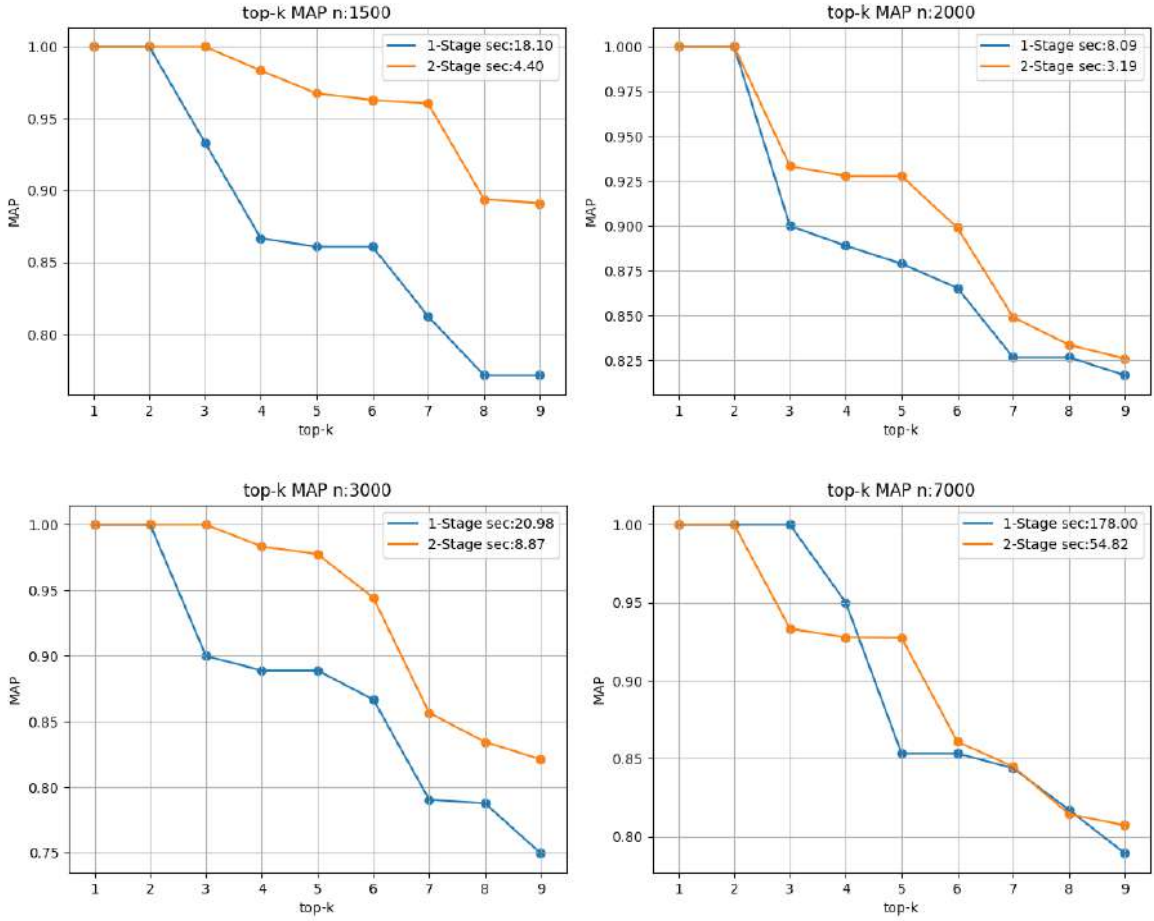

Fig. 3.6 The MAP@k of the top-*k* graphs given as output in a database of 180 graphs. Four different sizes of the graphs contained in the database have been considered: $n = 1500, 2000, 3000, 7000$.

where *Precision(j) is the relevant proportion of the found top-k graphs, while Relevance(j) is* 1 *if the considered graph is part of* $\omega_i$ *and is* 0 *otherwise. Finally,* $|\omega_i|$ *is the number of relevant graphs.*

**Definition 6** (Mean average precision). *Given a query set Q, the mean average precision is defined as follows.*

$$MAP(Q) = \frac{1}{|Q|} \cdot \sum_{q_i \in Q} AP@k(q_i) \tag{3.3}$$

In particular, the higher is the value of the $MAP \in [0, 1]$, the higher is the quality of the proposed graph search algorithm. Figure 3.6 shows that the proposed summarization based approach improved the query quality.

**Scalability**   In order to evaluate the scalability of our approach, we conducted two different experiments. In the first one, we investigated the time required to perform a single query as the dimension of the database $\mathscr{D}$ grows. In the second one, we investigated the query time in function of the size of the query graph.

In the first experiment, we fixed the size of all the graphs to be $n = 2000$. We then generated the graphs in $\mathscr{D}$ using all the possible combinations of the following factors: three different numbers of clusters $\{4, 12, 20\}$, six different levels of intra-cluster noise probability $\{0.05, 0.1, 0.15, 0.2, 0.25, 0.3\}$, and six different levels of inter-cluster noise probability $\{0.05, 0.1, 0.15, 0.2, 0.25, 0.3\}$. The combination of these three parameters allow us to generate 108 graphs. We then copied them enough times to reach a database cardinality spanning from $10^3$ up to $10^4$ graphs.

The query time is calculated as follows:

$$t = t\_s(q) + t\_eig(r_q) + t\_SD(eig_{r_q}, eig_{r_j})    j = 1, \cdot, |\mathscr{D}|. \tag{3.4}$$

where $t\_s(q)$ is the time required to summarize the query graph $q$ giving us $r_q$; $t\_eig(r_q)$ is the time required to calculate the eigenvectors of $r_q$; and $t\_SD(eig_{r_q}, eig_{r_j})$ is the time required to calculate the spectral distances between $r_q$ and each graph summary $r_j$ contained in $\mathscr{D}$. We reported, on the left part of figure 3.7, the computed $t_i$ versus the cardinality of $\mathscr{D}_i$.

In the second experiment, we generated different databases $\mathscr{D}_i$, containing 10000 graphs. All the graphs in $\mathscr{D}_i$ have the same size and have been created analogously as the previous experiment. We then constructed a query graph $q_i$ of the same size of the graphs in $\mathscr{D}_i$, and we measured the query time $t_i$ as we did for the previous experiment. We reported, on the right part of figure 3.7, the computed $t_i$ versus the size of the graph query $q_i$. Figure 3.7 provides us an experimental verification of the scalability of our approach both in the size of the databases and in the size of the query graph.

## 3.5   Concluding remarks

In this chapter, we introduced a new graph summarization heuristic, which is characterized by an improvement of the summary quality both in terms of reconstruction error and of noise filtering. We have successfully validated our framework both on synthetic and real-world graphs showing that surpass the state-of-the-art in term of noise robustness. In the second part of the paper, we presented an algorithm to address the graph similarity search problem exploiting our summaries. In particular, the proposed method is tailored for efficiently dealing with databases containing a high number of large graphs, and, moreover, it is principled robust against noise, which is always presented in real-world data. This achievement seems of particular interest since, to the best of our knowledge, we are the first to devise a graph search algorithm which satisfies all the above requirements together. As future works, we intend to extend our summarization algorithm to deal with labeled and evolving graphs.

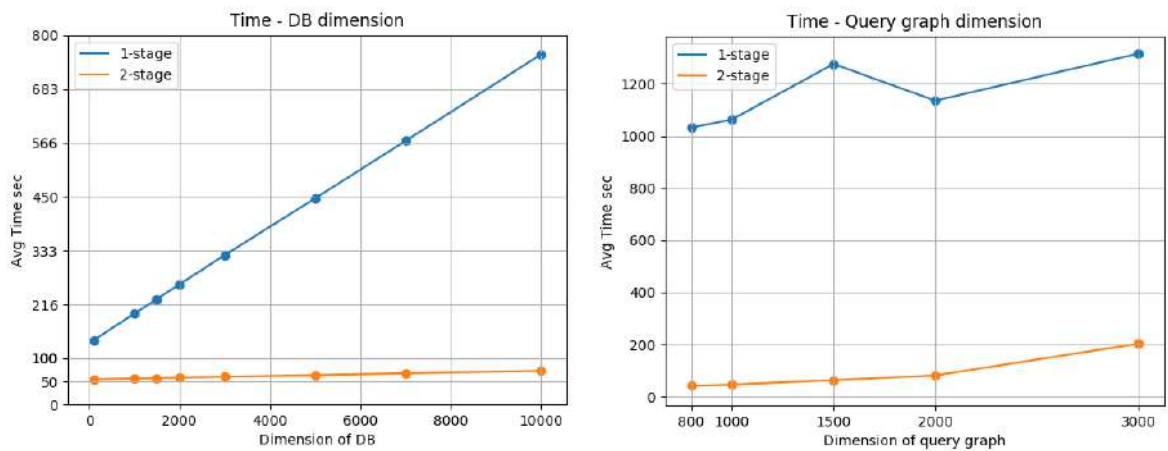

Fig. 3.7 On the left: the plot shows the time expressed in seconds to perform a query as we increase the size of the database. On the right: the time expressed in minutes for retrieving the top-*k* graphs in a database composed of 10000 graphs as we increase the dimension of the query graph.

# Chapter 4

# Graph Decomposition Using Stochastic Block Model

> That is the structure of scientific
> revolutions: normal science with a
> paradigm and a dedication to solving
> puzzles; followed by serious anomalies,
> which lead to a crisis; and finally
> resolution of the crisis by a new
> paradigm.
>
> ———————————————————
> Ian Hacking

Graphs are a useful abstraction of data sets with pairwise relations. In case of very large graph data sets, extracting structural properties and performing basic data processing tasks may be computationally infeasible using raw data stored as edge list, whereas the adjacency matrix may be too large to be stored in central memory. One obvious problem with sampling is the sparsity. The sparsity means that if we pick up two nodes at random, we usually observe no relation between them and it is impossible to create any meaningful low-dimensional model of the network. This prevents using uniform sampling as a tool for learning large and sparse graph structures.

To address this problem, we suggest to use, instead of the sparse adjacency matrix, the shortest path length matrix, whose elements are the shortest paths between pair of nodes. Hence, we map a sparse adjacency matrix to a still informative dense matrix. Of course, also in this case, it is not possible to construct the whole matrix for a very large network. However, it sufficient to get a relatively good estimate of the distance between any pair of nodes which belong to a given sample. Indeed, in recent experiments good estimates for distance matrix were reported for a billion-node graphs [82].

In this chapter, we study the linkage among the Regularity Lemma, the Stochastic Block Model (SBM) and the Minimum Description Length (MDL) with the aim of developing a graph decomposition algorithm suited to deal with sparse graphs. The SBM is an important paradigm in network research [3], which usually revolves around the concept of 'communities'. We aim to extend SBM-style concepts to other type of networks that do not fit well to such a community structure. For instance, in case of web graphs, Internet and p2p networks, we would expect a quite different structure, which is likely characterized by a power-law degree distributions and by hierarchy of sub-networks forming 'tiers'.

It would desirable to have a *decomposition* of such networks into some sub-networks different from communities, yet helping in having a big-picture of such networks. In particular, our aim is to find other, more general, type of redundancy in large sparse networks. In case of dense networks, we could search for a 'regular structure' in the sense of Szemerédi's Regularity Lemma. As we said in the previous chapters, this lemma indicates that large dense graphs have a decomposition of nodes into bounded number of groups, where most of the pairs are almost like dense random bipartite graphs. However, since the network distance matrix of a sparse network is a kind of dense weighted graph, the Regularity Lemma can be exploited to develop the desired decomposition algorithm. In particular, the Regularity Lemma is used as a prototype of the structural information which should be preserved, defining a new model space for graph-data. The best model is selected by fitting a SBM using likelihood maximization, while the MDL is exploited to obtain a stopping criterion which establishes when the optimal regular decomposition is found. This chapter is organized as follows. We first provide an overview of the Stochastic Block Model and of the Minimum Description Length. *This overview is a quote from the paper of Reittu et al. [86].* We then move to describe the decomposition algorithm and, finally, we present some experimental results on real-world networks.

## 4.1   Related works

Considerable progress towards statistical inference of sparse graphs has recently been achieved, cit. [20, 22] and references therein. Most of these methods are based on counting cycles and paths in the observed graph, possibly with some added randomness to split data and reduce noise sensitivity. Instead of cycle and path counts, here we suggest an alternative approach based on observed graph distances from a set of reference nodes to a set of target nodes. Such distances form a dense matrix. Of course, also in this case it may not be possible to have a complete matrix for very large networks. What is required is that for any given pair of nodes belonging to a sample, it is possible to have a relatively good estimate of distance between nodes. This is also a nontrivial task requiring an efficient solution, see e.g. [25]. In recent experiments good estimates for distance matrices were reported for a billion-node graphs [13]. Our sampling based approach only requires a sparse sample of the full distance matrix. When the number of reference nodes is bounded, the overall computational complexity of the proposed algorithm is linear in number of target nodes. We discuss two different sampling schemes of the reference nodes. The first is uniform sampling, which is a feasible method for graphs with light-tailed degree distributions such as those generated by stochastic block models. The second sampling scheme is nonuniform and biased towards sampling nodes with high betweenness centrality, designed to be suitable for scale-free graphs with heavy-tailed degree distributions. A crucial step is to obtain a low-rank approximation of the distance matrix based on its sample. For this we suggest to use a suitable variant of the regular decomposition (RD) method developed in [72, 79, 84, 86]. RD can be used for dense graphs and matrices and it shows good scalability and tolerance to noise and missing data. Because the observed distance matrix is dense, RD should work. The method permutes the rows of the matrix into few groups so that each column within each group of the matrix is close to a constant. We call such row groups *regular*. The regular groups form a partition of the node set. Each group of the partition induces a subgraph, and together these subgraphs form a decomposition of the graph into subgraphs and connectivity patterns between them. This decomposition is the main output of the method. The hypothesis of this paper is that the graph decomposition reveals structure of the sparse and large graphs. For instance, it should reveal small but dense subgraphs in sparse graphs as well as sets of similar nodes that form communities.

As a theoretical latent model we consider stochastic block models (SBM). SBM is an important paradigm in network research, see e.g.[3]. Usually SBM revolves around the concept of communities that are well connected subgraphs with only few links between the communities. We also look for other types of structures different from the community structure. For instance, in case of web graphs, Internet, peer-to-peer networks etc., we would expect quite different structure characterized, say, by a power-law degree distribution and hierarchy of subnetworks forming tiers that are used in routing messages in the network. The proposed distance based structuring might give valuable information about the large scale structure of many real-life networks and scale into enormous network sizes.

Our approach is stimulated by the Regularity Lemma which indicates that large dense graphs have decomposition of nodes into a bounded number of groups where most of the pairs are almost like random bipartite graphs. The structure encoded by the Regularity

Lemma ignores all intra-group relations. In our regular decomposition both of these aspects are used and both inter-group and intra-group relations matter.

As a benchmark we consider the famous planted bipartition model [30]. It is a random graph and a special case of SBM. As ground truth, there are two communities of nodes with equal number of nodes in each and with two parameters. First parameter is the link probability between nodes inside each community and the second one, the link probability between nodes in different communities. The links are drawn randomly and independently from each other. For such a model, it is known that there is a sharp transition, or 'critical point', of detectability of such a structure depending on the difference between the two parameters [30, 70]. The critical point is also located in the area of very sparse graphs, when expected degree is bounded. This example is suitable for testing our method because: having a ground truth, graph sparsity, bounded average degree and the proven sharp threshold. The preliminary results we report here, are promising. It seems that our algorithm is effective right up to the threshold in the limit of large scale. Moreover simulations indicate that such a structure can be found from very sparse and bounded size samples of the distance matrix.

Besides this benchmark, we demonstrate our method using real-life sparse networks. The first example is a Gnutella peer-to-peer file sharing network, and the second is an undirected Internet's autonomous system network. Both of them heavy-tailed degree distributions [104]. These graphs are not enormous. However, we treat them as if they were very large. Meaning that they are analyzed by using only a small fraction of the full information in the distance matrix. The computations were run in few nodes of a HPC cluster. Using this facility with 2000 cores and 40 terabytes of memory, it is possible to run experiments with much bigger data sets in the near future. *Sections 4.2 and 4.3 are quoted from Reittu et al. [86]*

## 4.2 The Stochastic Block Model

The notion of an $\varepsilon$-regular partition is purely combinatorial. The stochastic block model closest to this notion is the following.

**Definition 7.** *Let $V$ be a finite set and $\xi = \{A_1, \cdots, A_k\}$ a partition of $V$. A stochastic block model is a random graph $G = (V, E)$ with the following structure:*

- *There is a symmetric $k \times k$ matrix $D = (d_{ij})_{i,j=1}^k$ of real numbers $d_{ij} \in [0, 1]$ satisfying the* irreducibility condition *that no two rows are equal, i.e.*

$$\text{for all } i, \ j, \ i < j, \text{ there is } q_{ij} \in \{1, \cdots, k\} \text{ such that } d_{iq_{ij}} \neq d_{jq_{ij}}; \quad (4.1)$$

- *For every pair $\{v,w\}$ of distinct nodes of $V$ such that $v \in A_i$, $w \in A_j$, let $e_{vw} = e_{wv}$ be a Bernoulli random variable with parameter $d_{ij}$, assuming that all $e_{vw}$'s are independent. The edges of $G$ are*

$$E = \{\{v,w\} : v,w \in V,\ v \neq w,\ e_{vw} = 1\}.$$

Note that the case of the trivial partition $\xi = \{V\}$ yields to the classical random graph with edge probability $d_{11}$. A graph sequence $G_n = (V_n, E_n)$, presenting copies of the same stochastic block model in different sizes, can be constructed as follows.

**Construction 1.** *Let $\gamma_1, \ldots, \gamma_k$ be positive real numbers such that $\sum_{i=1}^{k} \gamma_i = 1$. Divide the interval $(0,1]$ into $k$ segments*

$$I_1 = (0, \gamma_1],\ I_2 = (\gamma_1, \gamma_1 + \gamma_2], \cdots, I_k = \left( \sum_{i=1}^{k-1} \gamma_i, 1 \right],$$

*and denote $\Gamma = \{I_1, \ldots, I_k\}$. For $n = 1, 2, \cdots$, let the vertices of $G_n$ be*

$$V_n = \{\frac{i}{n}\}, \quad i \in \{1, \cdots, n\}.$$

*For each $n$, let $\xi_n$ be the partition of $V_n$ into the blocks*

$$A_i^{(n)} = I_i \cap V_n, \quad i = 1, \cdots, k.$$

For small $n$, we may obtain several empty copies of the empty set numbered as blocks. However, from some $n_0$ on, all blocks are non-empty and $\xi_n = \{A_1^{(n)}, \ldots, A_k^{(n)}\}$ is a genuine partition of $V_n$. We can then generate stochastic block models based on $(V_n, \xi_n, D)$ according to Definition 7.

**Remark 1.** *A slightly different kind of stochastic block model can be defined by drawing first the sizes of blocks $A_i^{(n)}$ as independent Poisson($\gamma_i n$) random variables and proceeding then with the matrix $D$ as before. The additional level of randomness, regarding the block sizes, is however of no interest in the present paper.*

Next, we define the notion of a Poissonian block model in complete analogy with Definition 7.

**Definition 8.** *Let $V$ be a finite set of vertices, $n = |V|$, and let $\xi = \{A_1, \cdots, A_k\}$ be a partition of $V$. The* symmetric Poissonian block model *is a symmetric random $n \times n$ matrix $E$ with the following structure:*

- There is a symmetric $k \times k$ matrix $\Lambda = (\lambda_{ij})_{i,j=1}^{k}$ of non-negative real numbers satisfying the irreducibility condition *that no two rows are equal, i.e.,*

$$\text{for all } i, \ j, \ i < j, \text{ there is } q_{ij} \in \{1, \ldots, k\} \text{ such that } \lambda_{iq_{ij}} \neq \lambda_{jq_{ij}}; \quad (4.2)$$

- For every unordered pair $\{v, w\}$ of distinct nodes of $V$ such that $v \in A_i$, $w \in A_j$, let $e_{vw} = e_{wv}$ be a Poisson random variable with parameter $\lambda_{ij}$, assuming that all $e_{vw}$'s are independent. The matrix elements of $E$ are $e_{vw}$ for $v \neq w$, and $e_{vv} = 0$ for the diagonal elements.

Thanks to the independence assumption, the sums $\sum_{u \in A} \sum_{v \in B} e_{uv}$ are Poisson distributed for any $A, B \in \xi$.

**Remark 2.** *The rest of the technical contents of this chapter focus on the simple binary and Poissonian models of Definitions 7 and 8. However, the following extensions are straightforward:*

- bipartite graphs: *this is just a subset of simple graphs;*

- $m \times n$ matrices with independent Poissonian elements: *a matrix can be seen as consisting of edge weights of a bipartite graph, where the parts are the index sets of the rows and columns of the matrix, respectively;*

- directed graphs: *a directed graph can be presented as a bipartite graph consisting of two parts of equal size, presenting the input and output ports of each node.*

The following construction is the key to extend the minimum description length methodology for data that has the most common form of a large non-negative matrix.

**Construction 2.** *Let $C = (c_{ij})$ be a non-negative $m \times n$ matrix. Let $N$ be a (rather large) integer and denote $J = \{1, \ldots, N\}$. Let*

$$V = J_1^{(1)} \cup \cdots \cup J_m^{(1)}, \quad W = J_1^{(2)} \cup \cdots \cup J_n^{(2)},$$

$J_i^{(1)} = \{(\iota, i)\} \times J$. *Define a bipartite Poissonian block model $\mathscr{P}_N(C) = (V \cup W, \{J_{\cdot}^{(\cdot)}\}, C)$ with blocks $J_{\cdot}^{(\cdot)}$ and mean matrix $C$.*

Although we have not studied this at the technical level, it is natural to expect that, with large $N$, a partition of $\mathscr{P}_N(C)$ with minimum description length would with high probability keep the blocks $J_{\cdot}^{(\cdot)}$ unbroken. Because the regular decomposition algorithm for Poissonian block models operates only on means over blocks, Construction 2 is a strong heuristic argument that this algorithm applies as such to regular decomposition of non-negative matrices.

## 4.3   The Minimum Description Length

The Minimum Description Length (MDL) Principle was introduced by Jorma Rissanen, inspired by Kolmogorov's complexity theory, and an extensive presentation can be found in Grünwald's monography [47], and in [91]. The basic idea is the following: a set $\mathscr{D}$ of data is optimally explained by a model $\mathscr{M}$, when the combined description of the (i) model and (ii) the data as interpreted in this model is as concise as possible. By *description* we mean here a code that specifies an object uniquely. The principle is best illustrated by our actual case, simple graphs. A graph $G = (V, E)$ with $|V| = n$ can always be encoded as a binary string of length $\binom{n}{2} = n(n-1)/2$, where each binary variable corresponds to a node pair and a value 1 (resp. 0) indicates an edge (resp. absense of an edge). Thus, the MDL of $G$ is always at most $\binom{n}{2}$. However, $G$ may have a structure whose disclosure would allow a much shorter description.

Our heuristic postulate is that in the case of graphs and similar objects a good *a priori* class of models should be inferred from the regularity lemma, which points to stochastic block models.

**Definition 9.** *Denote by $\mathscr{M}_{n/k}$ the set of irreducible stochastic block models $(V, \xi, D)$ with*

- $|V| = n$,

- $|\xi| = k$, *and, denoting* $\xi = \{V_1, \ldots, V_k\}$,

- *for* $i, j \in \{1, \ldots, k\}$,

$$d_{ij} = \frac{h_{ij}}{|V_i||V_j|}, \ h_{ij} \in \mathbb{N}, \quad d_{ii} = \frac{h_{ii}}{\binom{|V_i|}{2}}, \ h_{ii} \in \mathbb{N}.$$

*The condition in the last bullet entails that each modelling space* $\mathscr{M}_{n/k}$ *is finite.*

**Remark 3.** *Without the irreducibility condition (4.1), there would not be a bijection between stochastic block models and their parameterizations.*

The models in $\mathscr{M}_{n/k}$ are parameterized by $\Theta_k = (\xi, D)$. A good model for a graph $G$ is the one that gives maximal probability for $G$ and is called *the maximum likelihood model*.

We denote the parameter of this model

$$\hat{\Theta}_k(G) := \underset{\Theta_k \in \mathcal{M}_{n/k}}{\arg\max}(P(G \mid \Theta_k)), \tag{4.3}$$

where $P(G \mid \Theta_k)$ denotes the probability that the probabilistic model specified by $\Theta_k$ produces $G$. One part of likelihood optimization is trivial: when a partition $\xi$ is selected for a given graph $G$, the optimal link probabilities are the empirical link densities:

$$d_{ij} = \frac{e(V_i, V_j)}{|V_i||V_j|}, i \neq j, \quad d_{ii} = \frac{e(V_i)}{\binom{|V_i|}{2}}. \tag{4.4}$$

Thus, the nontrivial part is to find the optimal partition for the given graph. This is the focus of the next sections.

### 4.3.1 Two-part MDL for simple graphs

Let us denote the set of all simple graphs with $n$ nodes as

$$\Omega_n = \{G : G = (V, E) \text{ is a graph, } |V| = n\}.$$

A prefix (binary) coding of a finite set $\Omega$ is an injective mapping

$$C : \Omega \to \cup_{s \geq 1} \{0, 1\}^s \tag{4.5}$$

such that no code is a prefix of another code. Recall the following theorem from information theory:

**Theorem 4.** *(Kraft's Inequality) For an m-element alphabet there exists a binary prefix coding scheme with code lengths $l_1, l_2, \cdots, l_m$ iff the code lengths satisfy: $\sum_{i=1,\cdots,m} 2^{-l_i} \leq 1$.*

An important application of Theorem 4 is the following: if letters are drawn from an alphabet with probabilities $p_1, p_2, \cdots, p_m$, then there exists a prefix coding with code lengths $\lceil -\log p_1 \rceil, \cdots, \lceil -\log p_m \rceil$, and such a coding scheme is optimal in the sense that it minimizes the expected code length (in this section, the logarithms are in base 2). In particular, any probability distribution $P$ on the graph space $\Omega_n$ indicates that there exists a prefix coding that assigns codes to elements of $G \in \Omega_n$ with lengths equal to $\lceil -\log P(\{G\}) \rceil$. The code length $l(\cdot)$ is the number of binary digits in the code of the corresponding graph. In case of a large set $\Omega$, most such codes are long and as a result the ceiling function can be omitted, a case we assume in sequel. A good model results in good compression, meaning that a graph can be described by much less bits than there are elements in the adjacency matrix. An incompressible case corresponds to the uniform distribution on $\Omega_n$ and results in code length $-\log\left(1/\mid\Omega\mid\right) = \binom{n}{2}$, equivalent to writing down all elements of the adjacency matrix. For every graph $G$ from $\Omega_n$ and model $P$ we can associate an encoding with code length distribution $-\log P(\cdot\mid\hat{\Theta}_k(G))$. However, this is not all, since in order to be able to decode we must know what particular probabilistic model $P$ is used. This means that also $\hat{\Theta}_k(G)$ must be prefix encoded, with some code-length $L(\hat{\Theta}_k(G))$. We end up with the following *description length*:

$$l(G) = \lceil -\log P(G\mid\hat{\Theta}_k(G))\rceil + L(\hat{\Theta}_k(G)). \tag{4.6}$$

Eq. (4.6) presents the so-called *two-part MDL*, [47]. In an asymptotic regime with $n \to \infty$, we get an analytic expression of the refined MDL. A simple way of estimating $L(\hat{\Theta}_k(G))$ is just to map injectively every model in $\mathscr{M}_{n/k}$ to an integer and then encode integers with $l^*(\mid\mathscr{M}_{n/k}(G)\mid)$ as an upper bound of the code-length. Here

$$l^*(m) = \max(0,\log(m)) + \max(0,\log\log(m)) + \cdots, \ m \in \mathbb{N}, \tag{4.7}$$

gives, as shown by Rissanen, the shortest length prefix coding for integers (see [47, 90]). The size of the graph must also be encoded with $l^*(n)$ bits (it is assumed that there is a way of defining an upper bound of the models with given $n$). In this point, it is necessary that the modeling space is finite. This results in

**Theorem 5.** *For any graph $G \in \Omega_n$, there exists a prefix coding with code-length*

$$l(G) = \lceil -\log P(G\mid\hat{\Theta}_k(G))\rceil + m$$
$$\leq m_k := l^*(n) + l^*\left(S2(n,k)\left(\binom{n-k+2}{2}+1\right)^{\binom{k}{2}+k}+1\right) + 1,$$

*where $S2(n,k)$ is the Stirling number of the second kind.*

*Proof.* The expression in (4.6) corresponds to a concatenation of two binary codes. The $L$-part is the length of a code for maximum likelihood parameters (in the case of a non-unique maximum, we take, say, the one with smallest number in the enumeration of all such models). The corresponding code is called the *parametric code*. The parametric code uniquely encodes the maximum-likelihood model. To create such an encoding, we just enumerate all possible models, given in Definition 9, and use the integer to fix the model. The length of a prefix code corresponding to an integer is the $l^*$-function computed for that integer, and we add 1 to handle the ceiling function. To obtain an upper bound for the parametric code length $m_k$, we find an upper bound for the number of models in the modeling space. The number of models is upper-bounded by the product of the integers, each presenting the number of partitions of an $n$-element set into $k$ non-empty sets (blocks), which equals $S2(n, k)$. We can view the blocks of a partition as the nodes of a 'reduced multi-graph' (in a multi-graph, there can be several links between a node pair, as well as self-loops). The range of multi-links is between zero and $\binom{n-k+2}{2}$: if we consider a pair of blocks (or one block internally), there can be at most $n - (k - 2)$ nodes in such a pair (in one set, slightly less), since there must be at least $k - 2$ nodes in the other blocks of the partition. Obviously, in such a subgraph of $n - (k - 2)$ nodes there can be at most $\binom{n-k+2}{2}$ links. Thus, the number of values each multi-link can take is upper-bounded by $\binom{n-k+2}{2} + 1$. Since the number of node pairs in the reduced multi-graph is $\binom{k}{2} + k$, we obtain the second multiplier in the argument of $l^*$ in the proposition. Finally, we should show that the coding of the graph is prefix. We concatenate both parts into one code that has the prescribed length and put first the prefix code of the integer that defines the parameters of the maximum likelihood model. When we start to decode from the beginning of the entire code, we first obtain a code of an integer, because we used a prefix coding for integers. At this stage we are able to define the probabilistic model that was used to create the other part of the code, corresponding to the probability distribution $P(\cdot \mid \hat{\Theta}_k(G))$. Using this information we can decode the graph $G$. It remains to show that the concatenated code itself is prefix. Assume the opposite: some prefix of such a code is prefix to some other similar code, say, the first code is a prefix to the second one. However, the parametric code was prefix, so both codes must correspond to the same model. Since the first two-part code is a prefix to the second, they both share the same parametric part, and the code for the graph of the first is a prefix of the second one. But this is impossible, since the encoding for graphs within the same model is prefix. This contradiction shows that the two-part coding is prefix. □

Finally, we call

$$\mathcal{M}_n := \bigcup_{1 \le k \le n} \mathcal{M}_{n/k} \tag{4.8}$$

the *full regular decomposition modeling space of* $\Omega_n$.

### 4.3.2    Two-part MDL for matrices

In this section, we consider input data in the form of a $n \times m$ matrix $A = (a_{ij})$ with non-negative entries. With such a matrix we associate a random bipartite multi-graph. The set of rows and the set of columns form a bipartition. Between row $i$ and column $j$ there is a random number of links that are distributed according to Poisson distribution with mean $a_{i,j}$. Such a model was introduced in [75] and it has been used in various tasks in complex network analysis. The aim of this model is to back up, heuristically, a corresponding practical algorithm for regular decomposition of matrices. Our approach is closely related to but slightly different from the Poissonian block model. Assume that $A$ is used to generate random $n \times m$ matrices $X$ with independent integer-valued elements following $Poisson(a_{ij})$ distributions. The target is to find a regular decomposition model that minimizes the expected description length of such random matrices.

We propose the following modelling spaces:

**Definition 10.** *For integers $k_1$, $k_2$ from ranges $1 \leq k_1 \leq n$ and $1 \leq k_2 \leq m$, the parameters of a model $\Theta_{k_1,k_2}$ in the modelling space $\mathscr{M}_{k_1,k_2}$ for an integer matrix $X$ are partition of rows into $k_1$ non-empty sets $V = (V_1, \cdots V_{k_1})$ and partition of columns into $k_2$ non-empty sets $U = (U_1, \cdots, U_{k_2})$ and $k_1 \times k_2$ block average matrix $P$, with elements $(P)_{\alpha,\beta} := \sum_{i \in V_\alpha, j \in U_\beta} \frac{x_{i,j}}{|V_\alpha||U_\beta|}$.*

Thanks to the addition rule of Poisson distributions, the likelihood of $X$ in a model $\Theta_{k_1,k_2} \in \mathscr{M}_{k_1,k_2}$, corresponds to probabilistic models where the elements of $X$ are independent and Poisson distributed with parameters $x_{i,j} \sim Poisson(P_{\alpha(i),\beta(j)})$, where $i \in V_{\alpha(i)}$, $j \in U_{\beta(j)}$ in the model $\Theta_{k_1,k_2}$. The corresponding likelihood is denoted as $P(X \mid \Theta_{k_1,k_2})$, the actual probability of $X$ is denoted as $P(X \mid A)$. The maximum likelihood model is found from the program that maximizes the expected log-likelihood:

$$
\begin{aligned}
\Theta^*_{k_1,k_2} &= \underset{\Theta_{k_1,k_2} \in \mathscr{M}_{k_1,k_2}}{\arg\max} \ \sum_X P(X \mid A) \log P(X \mid \Theta_{k_1,k_2}) \\
&= \underset{\Theta_{k_1,k_2} \in \mathscr{M}_{k_1,k_2}}{\arg\max} \ \left( \sum_X P(X|A) \log \frac{P(X|\Theta_{k_1,k_2})}{P(X|A)} + P(X|A) \log P(X|A) \right) \\
&= \underset{\Theta_{k_1,k_2} \in \mathscr{M}_{k_1,k_2}}{\arg\max} \ (-D(P_A \parallel P_{\Theta_{k_1,k_2}}) - H(P_A))
\end{aligned}
$$

where $D$ is the Kullback-Leibler divergence between distributions, $H$ denotes entropy and, $P_A$ and $P_{\Theta_{k_1,k_2}}$ are the two families of Poisson distributions for the matrix elements of $X$. Since $H(P_A)$ is independent on $\Theta_{k_1,k_2}$, it does not affect the identification of the maximum likelihood model. Thus, the final program for finding the optimal model is

$$
\Theta^*_{k_1,k_2} = \underset{\Theta_{k_1,k_2} \in \mathscr{M}}{\arg\min} \ (D(P_A \parallel P_{\Theta_{k_1,k_2}})). \tag{4.9}
$$

The description length of a model $l(\Theta_{k_1,k_2} \in \mathcal{M}_{k_1,k_2})$ consists of the description length $l(V) + l(U)$ of the two partitions and the description length of the block average matrix $l(P(X))$. For the latter, we need to know only the integers presenting the block sums of $X$, since the denominator is known for a fixed partition $(U, V)$. The code lengths of such integers are, for large matrices, simply the logarithms of the integers. As a result, we end up with the following expression for the description length of the random multi-graph model $A$ using the modeling space $\mathcal{M}_{k_1,k_2}$:

$$l_{k_1,k_2}(A) = D(P_A \parallel P_{\Theta_{k_1,k_2}^*}) + l(V^*) + l(U^*)$$
$$+ \sum_{1 \le \alpha \le k_1; 1 \le \beta \le k_2} E(\log(e_{\alpha,\beta} + 1 \mid P_{\Theta_{k_1,k_2}^*}),$$

where

$$e_{\alpha,\beta} = \sum_{i \in V_\alpha^*, j \in U_\beta^*} x_{i,j}.$$

The star superscript refers to parameters corresponding to the solution of the program (4.9). The expectation of logarithm is not explicitly computable. However, we assume large matrices and blocks, and then Jensen's inequality provides a tight upper bound that can be used in practical computations. Thus, the final expression for the description length of $A$ is

$$l_{k_1,k_2}(A) = D(P_A \parallel P_{\Theta_{k_1,k_2}^*}) + l(V^*) + l(U^*) + \sum_{1 \le \alpha \le k_1; 1 \le \beta \le k_2} \log(a_{\alpha,\beta} + 1), \quad (4.10)$$

where

$$a_{\alpha,\beta} = \sum_{i \in V_\alpha^*, j \in U_\beta^*} a_{i,j}.$$

The full two-part MDL would now be realized by finding the global minimum of this expression over various $(k_1, k_2)$. Although a heuristic one, we believe that our method for matrices is both reasonable and easy to use and implement, see [84].

### 4.3.3 Refined MDL and asymptotic model complexity

Let us next consider Rissanen's *refined MDL* variant (see [47]). The idea is to generate just one distribution on $\Omega_n$, called the *normalized maximum likelihood distribution* $P_{nml}$. Then a graph $G \in \Omega_n$ has the description length $-\log P_{nml}(G)$ which is at most as large as the one given by the two-part code in (4.6). The function $P(\cdot \mid \hat{\Theta}_k(\cdot))$ maps graphs of size $n$ into $[0,1]$, and it is not a probability distribution, because $\sum_{G \in \Omega} P(G|\hat{\Theta}_k(G)) > 1$. However, a related true probability distribution can be defined as

$$P_{nml}(\cdot) = \frac{P(\cdot \mid \hat{\Theta}_k(\cdot))}{\sum_{G \in \Omega} P(G|\hat{\Theta}_k(G))}. \tag{4.11}$$

The problem with this is that a computation of the normalization factor in (4.11) is far too involved: finding a maximum likelihood parametrization for a single graph is a 'macroscopic' computational task by itself and it is not possible to solve such a problem explicitly for all graphs. Therefore, the two-part variant is a more attractive choice in a practical context. However, the refined MDL approach is useful as an idealized target object for justifying various approximations implementations of the basic idea. It appears that in an asymptotic sense the problem is solvable for large simple graphs. The logarithm of the normalization factor in (4.11) is called the *parametric complexity* of the model space $\mathscr{M}_{n/k}$:

$$COMP(\mathscr{M}_{n/k}) := \log\left(\sum_{G \in \Omega_n} P(G \mid \hat{\Theta}_k(G))\right). \tag{4.12}$$

In a finite modeling space case, like in ours, this can be considered as a definition of model complexity. We have now the following simple bounds:

**Theorem 6.**
$$\log\left(S2(n,k)\right) \leq COMP(\mathscr{M}_{n/k}) \leq m_k + 1,$$

*where we use the same notation as in Theorem 5.*

*Proof.* The lower bound follows from the fact that we can have at least this number of graphs that have likelihood 1 in $\mathscr{M}_{n/k}$. This corresponds to graphs for which the nodes can be partitioned into $k$ non-empty sets and inside each set we have a full graph and no links between the sets. Thus, for every partition there is at least one graph that has likelihood one and all such graphs are different from each other since there is a bijection between those graphs and partitions. For the upper bound, we notice that according to Theorem 5, there is a prefix coding with code lengths that correspond to the two-part code. As a result, Kraft's inequality yields that $\sum_{G \in \Omega_n} 2^{-l_k(G)} \leq 1$, or

$$1 \geq \sum_{G \in \Omega_n} 2^{-\lceil -\log P(G|\hat{\Theta}_k(G)) \rceil - m_k} \geq \sum_{G \in \Omega_n} 2^{\log P(G|\hat{\Theta}_k(G)) - 1 - m_k},$$

from which we get

$$\sum_{G \in \Omega_n} P(G \mid \hat{\Theta}_k(G)) \leq 2^{m_k+1}.$$

Taking logarithms, we arrive at the claimed upper bound.                    □

When considering large-scale structures corresponding to moderate $k$, the upper and lower bounds in Theorem 6 are asymptotically equivalent, and we have

**Theorem 7.** *Assume that $k > 1$ is fixed. Then*

$$COMP(\mathscr{M}_{n/k}) \sim n \log k, \quad n \to \infty.$$

*Proof.* Denoting the lower and upper bound of parametric complexity in Theorem 6 respectively by $b_l$ and $b_u$, we argue that $b_u \sim b_l \sim n \log k$ asymptotically when $n \to \infty$. This follows from the fact that the dominant asymptotic component of both $b_u$ and $b_l$ is $\log S2(n,k)$. Indeed, $S2(n,k) \sim \frac{k^n}{k!}$ for fixed $k$, the asymptotic of $\log S2(n,k)$ is linear in $n$, and all other terms of the asymptotics of both bounds are additive and at most logarithmic in $n$.                    □

**Remark 4.** *The speed of convergence of the upper and lower bounds in Theorem 6 is of type* $\log n / n$.

## 4.4   Regular decomposition

The previous sections developed both the two-part and refined variants of the MDL theory, as presented in [47], for the model space of stochastic block models. In the following, we describe a variant of regular decomposition (RD) algorithm, which was developed in works [87, 72, 79, 84, 86] for a generic matrix.

Consider a connected (finite, undirected) graph[1] $G$. If the original graph is not connected, we can first do a rough partitioning using the connected components. Here we assume that this simple task has already been carried out. Our goal is to partition a subset $V$ of $n$ nodes of the graph into $k$ disjoint nonempty sets called *communities*. Such a partition can be represented as an ordered list $(Z_1, \cdots, Z_n)$ where $Z_i \in [k]$ indicates the community of the $i$-th node in $V$. For convenience, we will also use an alternative representation of the partition as an $n$-by-$k$ matrix with entries

$$R_{iu} = \begin{cases} 1 & \text{if the } i\text{-th node of } V \text{ is in community } u, \\ 0 & \text{else.} \end{cases}$$

---

[1]Or strongly connected directed graph in a directed setting.

Such a matrix has binary entries, unit rows sums, and nonzero columns, and will be here called a *partition matrix*.

### 4.4.1    Statistical model for the distance matrix

The partitioning algorithm presented here is based on observed distances from a set of $m$ reference nodes to a (possibly overlapping) set of $n$ target nodes. Let $D_{ij}$ be the length of the shortest path from the $i$-th reference node to the $j$-th target node in the graph. The target is to find such a partition of nodes such that distances from any particular reference node $i$ to nodes in community $u$ are approximately similar, with minimal stochastic fluctuations. This modeling assumption can be quantified in terms of an $m$-by-$k$ matrix $(\Lambda_{iu})$ with nonnegative integer entries representing the average distance from the $i$-th reference node to nodes in community $u$. A simple model of a distance matrix in this setting is to assume that all distances are stochastically independent random integers such that the distance from the $i$-th reference node to a node in community $u$ follows a Poisson distribution with mean $\Lambda_{iu}$. This statistical model is parametrized by the $m$-by-$k$ average distance matrix $\Lambda$ and the $n$-by-$k$ partition matrix $R$, and corresponds to the discrete probability density function[2]

$$f_{\Lambda, R}(D) = \prod_{i=1}^{m} \prod_{j=1}^{n} e^{-\Lambda_{iZ_j}} \frac{\Lambda_{iZ_j}^{D_{ij}}}{D_{ij}!}, \quad D \in \mathbb{Z}_+^{m \times n},$$

having logarithm

$$\log f_{\Lambda, R}(D) = \sum_{i=1}^{m} \sum_{j=1}^{n} \sum_{v=1}^{k} R_{jv} \left( D_{ij} \log \Lambda_{iv} - \Lambda_{iv} \right) \\ - \sum_{i=1}^{m} \sum_{j=1}^{n} \log(D_{ij}!).$$

Such modeling assumption does not assume any particular distribution of distance matrix, question is about approximating the given distance matrix with a random matrix with parameters that give the best fitting. Such particular models are used because it results in a simple program, as we see in Algorithm 1. We have also tested it in our previous works with various data, showing good practical performance, [79, 84].

Having observed a distance matrix $D$, standard maximum likelihood estimation looks for $\Lambda$ and $R$ such that the above formula is maximized. For any fixed $R$, maximizing with respect to the continuous parameters $\Lambda_{iv}$ is easy. Differentiation shows that the map $\Lambda \mapsto \log f_{\Lambda,R}(D)$ is concave and attains its unique maximum at $\hat{\Lambda} = \hat{\Lambda}(R)$ where

$$\hat{\Lambda}_{iv}(R) = \frac{\sum_{j=1}^{n} D_{ij} R_{jv}}{\sum_{j=1}^{n} R_{jv}} \tag{4.13}$$

is the observed average distance from the $i$-th reference to nodes in community $v$. As a consequence, a maximum likelihood estimate of $(\Lambda, R)$ is obtained by minimizing the function

$$L(R) = \sum_{i=1}^{m} \sum_{j=1}^{n} \sum_{v=1}^{k} R_{jv} \left( \hat{\Lambda}_{iv}(R) - D_{ij} \log \hat{\Lambda}_{iv}(R) \right) \tag{4.14}$$

subject to $R \in \{0,1\}^{n \times k}$ having unit row sums and nonzero column sums, where $\hat{\Lambda}_{iv}(R)$ is given by (4.13).

---

[2]We could omit terms with $i = j$ from the product because of course $D_{ii} = 0$, but this does not make a big difference for large graphs.

## 4.4.2  Recursive algorithm

Minimizing (4.14) is a nonlinear discrete optimization problem with an exponentially large input space of order $\Theta(k^n)$. Hence an exhaustive search is not computationally feasible. The objective function can alternatively be written as $L(R) = \sum_{j=1}^{n} \ell_{jZ_j}(R)$, where

$$\ell_{jv}(R) = \sum_{i=1}^{m} \left( \hat{\Lambda}_{iv}(R) - D_{ij} \log \hat{\Lambda}_{iv}(R) \right). \tag{4.15}$$

This suggests a way to find local maximum by selecting a starting value $R^0$ for $R$ at random, and greedily updating the rows of $R$ one by one as long as the value of the objective function decreases. A local update rule for $R$ is achieved by a mapping $\Phi : \{0,1\}^{n \times k} \to \{0,1\}^{n \times k}$ defined by $\Phi(R)_{jv} = \delta_{Z_j^* v}$ where

$$Z_j^* = \arg\min_{v \in [k]} \ell_{jv}(R). \tag{4.16}$$

Algorithm 6 describes a way to implement this method. This is in spirit of the EM algorithm where the averaging step corresponds to an E-step and the optimization step to an M-step. The algorithm iterates these steps by starting from a random initial partition matrix $R^0$, and recursively computing $R^t = \Phi(R^{t-1})$ for $t \geq 1$. The runtime of the local update is $O(km + kn)$, so that as long as the number of communities $k$ and the parameters $s_{max}, t_{max}$ are bounded, the algorithm finishes in linear time with respect to $m$ and $n$ and is hence well scalable for very large graphs. The output of Algorithm 6 is a local optimum. To approximate a global optimum, parameter $s_{max}$ should be chosen as large as possible, within computational resources.

Finally, we describe how the rest of nodes are classified into $k$ groups or communities, after the optimal partition $R^*$ for a given target group and reference group is found. Let $i$ denote a node out of original target group. First we must obtain distances of this node to all reference nodes

$$(D_{i,j})_{1 \le j \le m}$$

Then the node i is classified into group number $\alpha$ according to

$$\alpha = \arg \min_{1 \le \beta \le k} \sum_{j=1}^{m} \left( \hat{\Lambda}_{j\beta}(R) - D_{ji} \log \hat{\Lambda}_{j\beta}(R^*) \right).$$

The time complexity of this task is dominated by the computations of distances of all nodes to the reference nodes, because for bounded $m$ the above optimization is done in a constant time. According to Dijkstra-algorithm computation of distances from all $N$ nodes to the target nodes takes $mO(|E| + N \log N)$. In a sparse graph, that we assume, $|E| \sim N$. Thus, if $m$ is bounded, the overall time complexity is $O(N \log N)$, which is only slightly over the best possible $O(N)$, which is needed just to enlist a partition. This is because the classification phase takes only $O(N)$ time for all nodes.

### 4.4.3 Estimating the number of groups

The regular decomposition algorithm presented in the previous section requires the number of groups $k$ as an input parameter. However, in most real-life situations this parameter is not a priori known and needs to be estimated from the observed data. The problem of estimating the number of groups $k$ can be approached by recasting the maximum likelihood problem in terms of the minimum description length (MDL) principle [90, 47] where the goal is to select a model which allows the minimum coding length for both the data and the model, among a given set of models.

When the set of models equasl the Poisson model described in Sec. 4.4.1, then the $R$-dependent part of the coding length equals the function $L(R)$ given by (4.14), and a MDL-optimal partition $R^*$, given $k$, corresponds to the minimal coding length

$$R^* = \arg \min_R L(R).$$

It is not hard to see that $L(R^*)$ is monotonously decreasing as a function of $k$, and in MDL a balancing term, the model complexity, is added to select the model that best explains the

---

**Algorithm 6** Regular decomposition algorithm

---

**Require:** Distance matrix $D \in \mathbb{Z}_+^{m \times n}$, integers $k, s_{\max}, t_{\max}$
**Ensure:** Partition matrix $R^* \in \{0,1\}^{n \times k}$

 1: **function** REGULARDECOMPOSITION($D, k, s_{\max}, t_{\max}$)
 2:      $L_{\min} \leftarrow \infty$
 3:      **for** $s$ in $1, \cdots, s_{\max}$ **do**
 4:          $R \leftarrow$ random $n$-by-$k$ partition matrix
 5:          **for** $t$ in $1, \cdots, t_{\max}$ **do**
 6:              $R \leftarrow$ LOCALUPDATE($R, D$)
 7:          $L \leftarrow L(R)$ given by equation (4.14)
 8:          **if** $L < L_{\min}$ **then**
 9:              $R^* \leftarrow R$
10:              $L_{\min} \leftarrow L$
11:      **return** $R^*$

12: **function** LOCALUPDATE($R, D$)
      *Averaging step*
13:      **for** $v$ in $1, \cdots, k$ **do**
14:          $n_v \leftarrow \sum_{j=1}^n R_{jv}$
15:          **for** $i$ in $1, \cdots, m$ **do**
16:              $\hat{\Lambda}_{iv} \leftarrow \sum_{j=1}^n D_{ij} R_{jv} / n_v$
17:      **for** $j$ in $1, \cdots, n$ **do**
18:          $\ell_{jv} \leftarrow \sum_{i=1}^m (\hat{\Lambda}_{iv} - D_{ij} \log \hat{\Lambda}_{iv})$
      *Optimization step*
19:      **for** $j$ in $1, \cdots, n$ **do**
20:          $Z_j^* \leftarrow \underset{v \in [k]}{\arg\min} \, \ell_{jv}$
21:          **for** $v$ in $1, \cdots, k$ **do**
22:              $R_{jv}^* \leftarrow \mathbb{1}(Z_j^* = v)$
23:      **return** $R^*$

---

observed data. However, in all of our experiments, the negative log-likelihood as a function of $k$ becomes essentially a constant above some value $k^*$. Such a knee-point $k^*$ is used as an estimate for the number of groups in this work. Thus, we are using a very simplified version of MDL, since it was found sufficient in our cases of examples. In a more accurate analysis one should use model complexity in higher detail. Some early work towards this direction includes [86].

## 4.5 Theoretical considerations

### 4.5.1 Planted partition model

A stochastic block model (SBM) with $n$ nodes and $k$ communities is a statistical model parametrized by a nonnegative symmetric $k$-by-$k$ matrix $(W_{uv})$ and a $n$-vector $(Z_i)$ with entries in $[k]$. The SBM generates a random graph where each node pair $\{i, j\}$ is linked with probability $W(Z_i, Z_j)$, independently of other node pairs. For simplicity, we restrict the analysis to the special case with $k = 2$ communities where the link matrix is of the form

$$W = \begin{bmatrix} a/n & b/n \\ b/n & a/n \end{bmatrix}$$

for some constants $a, b > 0$. This model, also known as the *planted partition model*, produces sparse random graphs with link density $\Theta(n^{-1})$, and is a de facto benchmark for testing the performance of community detection algorithms. As usual, we assume that the underlying partition is such that both communities are approximately of equal size, so that the partition matrix $R_{iu} = \delta_{Z_i,u}$ satisfies $\sum_{i=1}^{n} R_{iu} = (1 + o(1))\frac{n}{2}$. If $a > b$, there are two communities that have larger internal link density than link density between them. A well-known result states that for $n \gg 1$, partially recovering the partition matrix from an observed adjacency matrix is possible if

$$(a - b)^2 > 2(a + b), \tag{4.17}$$

and impossible if the above inequality is reversed. This result, called Kesten–Stigum threshold, was obtained in semi-rigorous way [30] and then proved rigorously in [70].

### 4.5.2 Expected and realized distances

Our aim is to have analytical formulas for distances $D_{ij}$ in a large graph generated from SBM. This question was addressed in [12] using spectral approach, where limiting average distances were found. We need the next to the leading term of the average distance. Although these calculations are not rigorous, it is well-known that in case of classical random graph similar approach produces a distance estimate that is asymptotically exact. That is why we believe that such an analysis makes sense in case of SBM as well.

To analyze distances, let us first investigate the growth of the neighborhoods from a given node as a function of the graph distance. Let us denote the communities by $V_u = \{i : Z_i = u\}$ for $u = 1, 2$. Fix a node $i \in V_1$ and denote by $n_u(t)$ the expected number of nodes in community $u$ at distance $t$ from $i$. Note that each node has approximately $a/2$ neighbors in the same community and approximately $b/2$ neighbors in the other community. Moreover, due to sparsity, the graph is locally treelike, and therefore we get the approximations

$$
\begin{aligned}
n_1(t) &\approx \frac{1}{2}an_1(t-1) + \frac{1}{2}bn_2(t-1) \\
n_2(t) &\approx \frac{1}{2}bn_1(t-1) + \frac{1}{2}an_2(t-1).
\end{aligned}
$$

Writing $N(t) = (n_1(t), n_2(t))^T$, this can be expressed in matrix form as $N(t) \approx AN(t-1)$, where

$$
A = \frac{1}{2}\begin{pmatrix} a & b \\ b & a \end{pmatrix}.
$$

As a result, $N(t) \approx A^t N(0)$ with $N(0) = (1,0)^T$. The matrix $A$ has a pair of orthogonal eigenvectors with eigenvalues:

$$
e_1 = \frac{1}{\sqrt{2}}\begin{pmatrix} 1 \\ 1 \end{pmatrix}, \quad \lambda_1 = \frac{a+b}{2}
$$

and

$$
e_2 = \frac{1}{\sqrt{2}}\begin{pmatrix} -1 \\ 1 \end{pmatrix}, \quad \lambda_2 = \frac{a-b}{2}.
$$

According to the spectral theorem, we can diagonalize the matrix $A$ and conclude that its powers are given by

$$
A^t = \lambda_1^t e_1 e_1^T + \lambda_2^t e_2 e_2^T.
$$

As a result, the expected numbers of nodes of types $u = 1, 2$ at distance $t$ from a node of type 1 are approximated by

$$
N(t) \approx A^t N(0) = \frac{1}{2}\begin{pmatrix} \lambda_1^t + \lambda_2^t \\ \lambda_1^t - \lambda_2^t \end{pmatrix}.
$$

Moreover, if $m_u(t) = \sum_{s=1}^{t} n_u(s)$, then

$$m_1(t) \approx \frac{1}{2}\left(-2 + \frac{\lambda_1}{\lambda_1 - 1}\lambda_1^t + \frac{\lambda_2}{\lambda_2 - 1}\lambda_2^t\right),$$

$$m_2(t) \approx \frac{1}{2}\left(-2 + \frac{\lambda_1}{\lambda_1 - 1}\lambda_1^t - \frac{\lambda_2}{\lambda_2 - 1}\lambda_2^t\right).$$

Next we want to find and estimate for the average distance $d_1$ (resp. $d_2$) from a node in $V_1$ to another node in $V_1$ (resp. $V_2$). We use the heuristic that the distances from a node to all nodes in the same group are well concentrated and close to each other. Under this assumption, we expect that $d_1$ and $d_2$ approximately solve the equations $m_1(d_1) = n/2$ and $m_2(d_2) = n/2$.

We get the equations for the distances:

$$\frac{\lambda_1^{d_1+1}}{\lambda_1 - 1} + \frac{\lambda_2^{d_1+1}}{\lambda_2 - 1} - 2 = n \tag{4.18}$$

$$\frac{\lambda_1^{d_2+1}}{\lambda_1 - 1} - \frac{\lambda_2^{d_2+1}}{\lambda_2 - 1} - 2 = n.$$

We are interested in leading order of difference of $d_2 - d_1$ for $n \to \infty$. Because $\lambda_1 > \lambda_2$ due to $a > b$, and $d_1, d_2 \to \infty$, we can use following iterative solution scheme. For $d_1$, we have:

$$\lambda_1^{d_1} = \frac{\lambda_1 - 1}{\lambda_1}n + 2\frac{\lambda_1 - 1}{\lambda_1} - \frac{\lambda_2(\lambda_1 - 1)}{\lambda_1(\lambda_2 - 1)}\lambda_2^{d_1}.$$

as a result, the equation we want to iterate is:

$$d_1 \log \lambda_1 = \log\left(\frac{\lambda_1 - 1}{\lambda_1}n\right) + \log\left(1 + \frac{2}{n} - \frac{\lambda_2}{\lambda_2 - 1}\frac{\lambda_2^{d_1}}{n}\right).$$

By expanding the second logarithm in series of powers of $1/n$, we get the leading terms of the solution:

$$d_1 \approx \frac{\log\left(\frac{\lambda_1 - 1}{\lambda_1}n\right)}{\log \lambda_1} - cn^{\alpha - 1},$$

where $c > 0$ is a constant and

$$\alpha = \frac{\log \lambda_2}{\log \lambda_1}, \quad c = \frac{1}{\log \lambda_1}\frac{\lambda_2}{\lambda_2 - 1}\lambda_2^\beta, \quad \beta = \frac{\log\frac{\lambda_1 - 1}{\lambda_1}}{\log \lambda_1}.$$

A similar procedure yields:

$$d_2 \approx \frac{\log\left(\frac{\lambda_1 - 1}{\lambda_1} n\right)}{\log \lambda_1} + c n^{\alpha - 1}.$$

Because $\alpha < 1$, both $d_1/\log n$ and $d_2/\log n$ have the same limit $1/\log \lambda_1$.

We conjecture that above the Kesten–Stigum threshold (4.17) the cost function, used in RD to partition graph distance matrix of the giant component of the graph generated from two part SBM, has a deep minimum corresponding to correct partition. More precisely, the cost of misplacing one node from the correct partition grows to infinity as $n \to \infty$.

First we conjecture that the found distance estimates of $d_1$ and $d_2$ are asymptotically equal to expected distances in a random graph corresponding to the Planted Partition model. For a node $i \in V_1$ and nodes $j \in V_1 \setminus \{i\}$ and $j' \in V_2$,

$$\mathbb{E}D_{i,j} \approx d - \delta, \quad \mathbb{E}D_{i,j'} \approx d + \delta,$$

where $d = \frac{\log\left(\frac{\lambda_1 - 1}{\lambda_1} n\right)}{\log \lambda_1}$ and $\delta = c n^{\alpha - 1}$, corresponding to the approximations in the previous section.

We also conjecture that for any $i \in V_1$, we have with high probability,

$$\sum_{j \in V_1} D_{i,j} = \frac{n}{2}(d - \delta) + O(\sqrt{n}),$$
$$\sum_{j \in V_2} D_{i,j} = \frac{n}{2}(d + \delta) + O(\sqrt{n}),$$

which is quite plausible if the first conjecture is true. The error term $O(\sqrt{n})$ can be neglected if $\alpha > \frac{1}{2}$, which is equivalent to being above the Kesten–Stigum threshold (4.17), which we assume from now on. If all nodes of the graph are partitioned correctly, then the cost (4.15) of target node $j$ in community $V_1$ is approximately

$$\ell_j \approx \frac{n}{2}(d + \delta - (d + \delta)\log(d + \delta) + d - \delta - (d - \delta)\log(d - \delta)).$$

If we switch the community $j$ to be $V_2$ then this cost changes into

$$\ell'_j \approx \frac{n}{2}(d + \delta - (d - \delta)\log(d + \delta) + d - \delta - (d + \delta)\log(d - \delta)).$$

As a result,

$$
\begin{aligned}
\ell'_j - \ell_j &\approx n\delta \log\left(\frac{d+\delta}{d-\delta}\right) \\
&\approx 2n\frac{\delta^2}{d} \\
&= 2\log(\lambda_1)c^2 nn^{2\alpha-2}/\log n \\
&= \frac{2\log(\lambda_1)c^2}{\log n}n^{2\alpha-1}.
\end{aligned}
$$

As a result if $\alpha > \frac{1}{2}$ or equivalently (4.17), the difference has infinite limit. This heuristic derivation suggests that the regular decomposition algorithm is capable of reaching the fundamental limit of resolution of the planted bipartition model.

## 4.6 Experiments with simulated data

### 4.6.1 Planted partition model

We investigate empirically the performance of the regular decomposition algorithm to synthetic data sets generated by the planted partition model described in Sec. 4.5.1. This is an instance of a very sparse graph with bounded degrees and with only two groups. In this case we argue that uniform random sampling of reference nodes will do. Here it is possible to compute full distance matrix up to sizes of 10000 nodes and sampling is not necessary.

For our test, we generated a graph with parameters $n = 2000$, $a = 20$, and $b = 2$. Another similar experiment was done with 10000 nodes. Next we computed the shortest paths between all pairs of nodes and formed a distance matrix $D$. RD was able to detect the structure with around 1 percent error rate. The average of one regular group shows that the distance has quite high level of noise, see Fig. 4.1. The reason why the communities become indistinguishable is probably in the increasing level of the variance. Below the threshold it is always too large, no matter how large $n$ is and above the threshold the communities can be detected provided $n$ is large enough. This is the conclusion of experiments not shown in this work.

Next we did experiments with 10000 nodes. In this particular case it looks that our method works better than a standard community detection algorithm of Girvan-Newman type. See Fig. 4.2 for graphical presentation.

As a sanity check we also used a usual community detection algorithm found in Wolfram Mathematica library. It was not capable of finding the true communities, see Fig. 4.4. The RD algorithm using the *D*-matrix, was able to find the communities correctly, with only a handful of misclassified nodes.

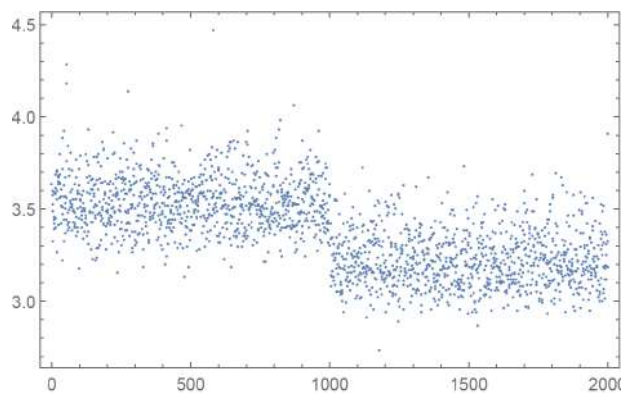

Fig. 4.1 Average distance ($y$-axis) $\hat{\Lambda}_{i2}$ to nodes in group 2 from all nodes $i$ (on $x$-axis) in the planted partition model. The $x$-axis nodes with indexes from 1 to 2000. The first 1000 nodes belong to the first group, and rest to the other one. The right half of the points, correspond to the distances within the same community, the left half of points corresponds to the distances between nodes in different communities. As can be seen, intra-group distances are systematically lower, although with some substantial variance. This is a case of tolerable variance which permits structure to be found.

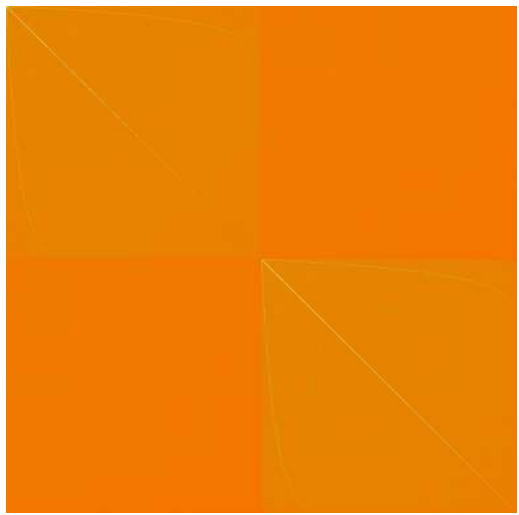

Fig. 4.2 Distance matrix of a graph with 10000 nodes sampled from the planted partition model, when nodes are labeled according to two true groups. The intra-block distances (average 4.75) are smaller than the inter-block distances (average 5.0).

### 4.6.2    Sampled distance matrices

To investigate experimentally how many reference nodes are needed to obtain an accurate partitioning of a set of *n* target nodes, we sampled a set of *m* reference nodes uniformly at random, and ran the regular decomposition algorithm on the corresponding *m*-by-*n* distance matrix.

It appears that even a modest sample of about $m = 400$ reference nodes is enough to have almost error free partitioning, see Fig. 4.3. It appears that with $m \geq 400$ reference nodes, a set of $n = 100$ target nodes can be accurately partitioned into $k = 2$ communities using RD, with error rate less than 1%. For larger sets of target nodes, the results appear similar. This suggests that such a method could work for very large graphs using this kind of sampling.

The regular decomposition algorithm also produces an estimated *m*-by-*k* average difference matrix ($\hat{\Lambda}_{iu}$). This model can be used to classify all nodes in the graph in linear time. To do this, we must compute distances to the $m = 400$ reference nodes, compute the negative log-likelihood for two groups based on ($\hat{\Lambda}_{iu}$), and place the node into the class with a smaller negative log-likelihood. All computations take just a constant time and that is why the linear scaling.

As a conclusion, we conjecture that for very large and sparse networks the distance matrix RD could be an option to study community structures.

The RD method seems to have better resolving power than community detection algorithms based on adjacency matrix and could work with sparse samples of data and thus scaling to extremely large networks. For a rather flat topology the uniform sampling method for distance matrix might be sufficient.

### 4.6.3    Preferential attachment models

The degree distributions of many social, physical, and biological networks have heavy tails resembling a power law. For testing network algorithms and protocols on variable instances of realistic graphs, synthetic random graph models that generate degrees according to power laws have been developed [104].

The main purpose of this exercise is to test sampling approach versus the full analysis. We used an instance of a preferential attachment model (Barabasi-Albert random graph) with 5000 nodes. The construction starts from a triangle. Then nodes are added one-by-one. Each incoming node makes 3 random links to existing nodes, and the probability of link is proportional to the degree of a node (preferential attachment). The result is somewhat comparable to the Gnutella network. However, in Gnutella networks instead of hubs, we had some more complicated dense parts.

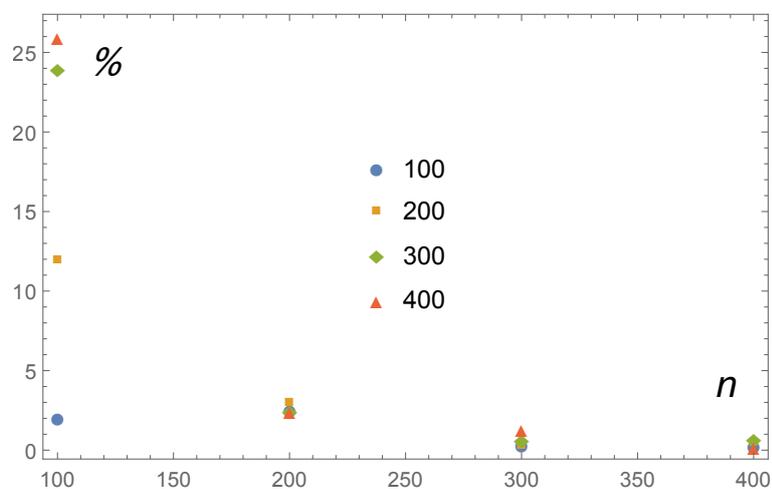

Fig. 4.3 Misclassification rates (*y*-axis) for synthetic data generated using a planted bipartition model shown in Fig. 4.2. The number of (randomly selected) target nodes has values $100, 200, 300, 400$, indicated by colored markers. For each case a random sample of *m* reference nodes was selected. The target node set was partitioned into $k = 2$ blocks and compared with the ground truth classification. When $m = 400$, the error rate is less than 1% in all cases.

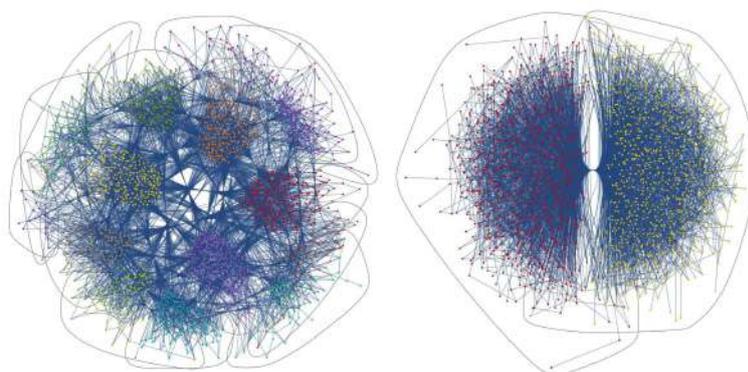

Fig. 4.4 Left: the community structure found with the Mathematica's FindGraphCommunities, (that uses, to our knowledge, Girvan-Newman algorithm) applied to our case of planted bipartition graph. It completely fails in detecting the right communities; instead of two correct 15 communities are found. Right: the almost correct communities found by RD. However, the Mathematica command was not forced to find just two communities.

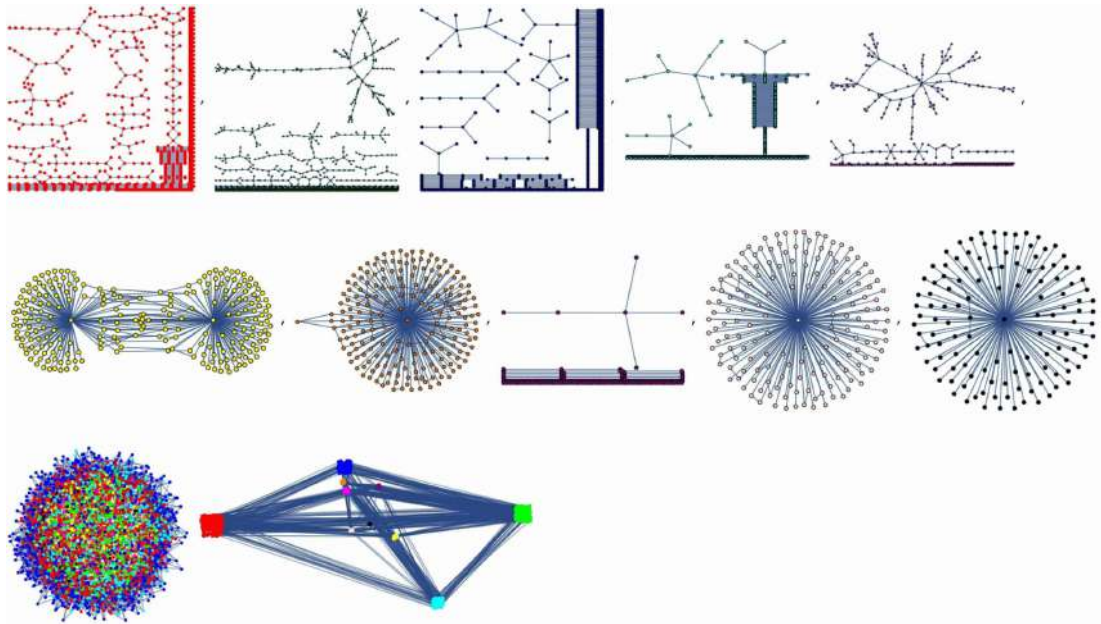

Fig. 4.5 At the bottom are the original network and its RD in 10 groups. Above are the internal structure of 10 sub-networks found with RD. Notably there are 4 groups that are "hubs", having one or two high degree nodes in a star-like topology. In a power-law graph such hubs are known to be essential.

To achieve scalability, instead of using full distance information between all $n = 5000$ nodes, we wish to restrict to distances to a small number of reference nodes. A main problem with the sampling of reference nodes is that high-degree core nodes are unlikely to show up in uniformly random samples. This is why decided to investigate the following nonuniform sampling scheme. The set of reference nodes was generated as the set of $m \approx 1000$ nodes which appeared in shortest paths between a randomly chosen set of 100 pairs of nodes. Distances to such reference nodes of high betweenness centrality are a strong indicator about distances between any two nodes because most short paths traverse through the central nodes. Next we ran the regular decomposition algorithm with $m$ reference nodes to partition the set of $n$ target nodes into $k = 10$ blocks. We get a quite similar result as the one for the entire distance matrix, see Fig 4.5 and Fig 4.6.

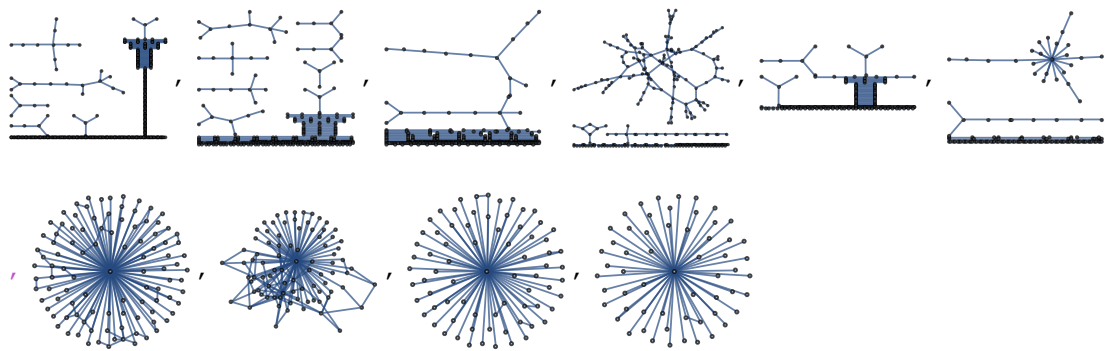

Fig. 4.6 Induced subgraphs of a preferential attachment graph corresponding to 10 groups discovered using the regular decomposition algorithm using a small set of $m$ reference nodes of high betweenness centrality. Although the found groups are not identical to the ones found with the full distance matrix ($m = n$), they are correlated to them, and the hub-like subnetworks are also found.

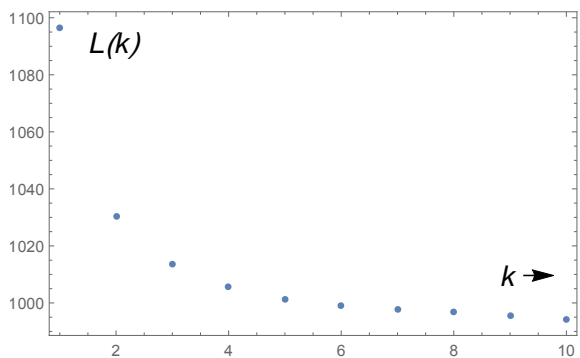

Fig. 4.7 Negative log-likelihood $L(k)$ as a function of the number of communities $k$. This plot is used to to find sufficiently optimal value of $k$. The right value of $k$ is in the range of 6 to 10, because for larger values $L$ is approximately a constant.

## 4.7   Experiments with real data

### 4.7.1   Gnutella network

We studied a Gnutella peer-to-peer network [89] with 10876 nodes representing hosts and 39994 directed links representing connections between the hosts. The graph is sparse because the link density is just about $3.4 \times 10^{-4}$. We extracted the largest strongly connected component which contains $n = 4317$ nodes and ran the regular decomposition algorithm for the corresponding full distance matrix ($m = n$) for different values for the number of communities in the range $k = 1, 2, \cdots, 10$. From the corresponding plot (Fig. 4.7) of the negative log-likelihood function we decided that $k = 10$ is valid choice.

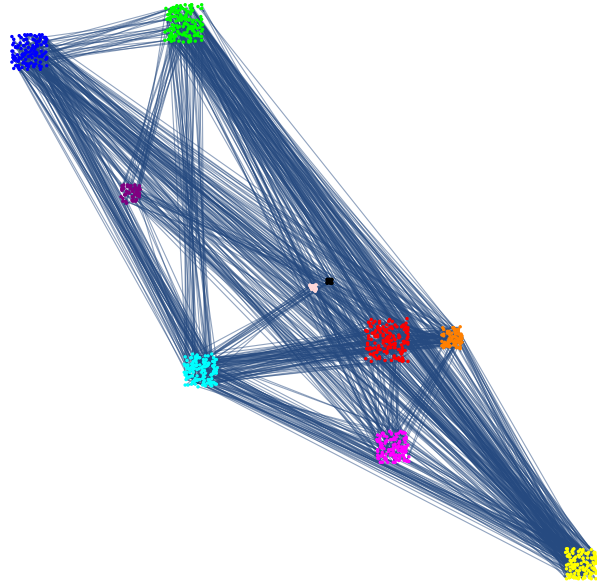

Fig. 4.8 The strongly connected component of a directed Gnutella network partitioned into 10 communities.

Fig. 4.8 illustrates the inter-community structure of the partitioned graph into $k = 10$ communities, and Fig. 4.9 describes the subgraphs induced by the communities. The induced subgraphs are internally quite different from each other. The high degree core-like parts form their own communities and they play a central role in forming paths through the network. Together these two figures provide a low-dimensional summary of the network as a weighted directed graph on 10 nodes with self-loops and weights corresponding to link densities in both directions.

### 4.7.2   Internet autonomous systems

The next example is a topology graph of Internet's Autonomous Systems [1] obtained from traceroute measurements, with around 1.7 million nodes and 11 million undirected links. This graph was analyzed using a HPC cluster.

We used a simplified scheme to analyze this graph. This was dictated by limited time and also we wanted to test some heuristic ideas to speed-up regular decomposition even further. First we computed shortest paths between a hundred randomly selected pairs of nodes. Then 30 most frequently appearing nodes in those shortest paths were selected as reference nodes. These nodes also appeared at the top of the link list provided by the source [1]. That is why we assume that such an important ordering of nodes is used in this source data set. Next we took 2000 top nodes from the source list and 3000 uniformly random nodes from the set of all nodes. A distance matrix from the $m = 30$ reference nodes to the selected $n = 5000$ target nodes was computed. Then the regular decomposition algorithm was run on this distance matrix for different values of $k$. From the negative log-likelihood function plot an optimal number of communities was estimated to be $k = 15$. As a result, we get a partition of the selected 5000 nodes into 15 communities.

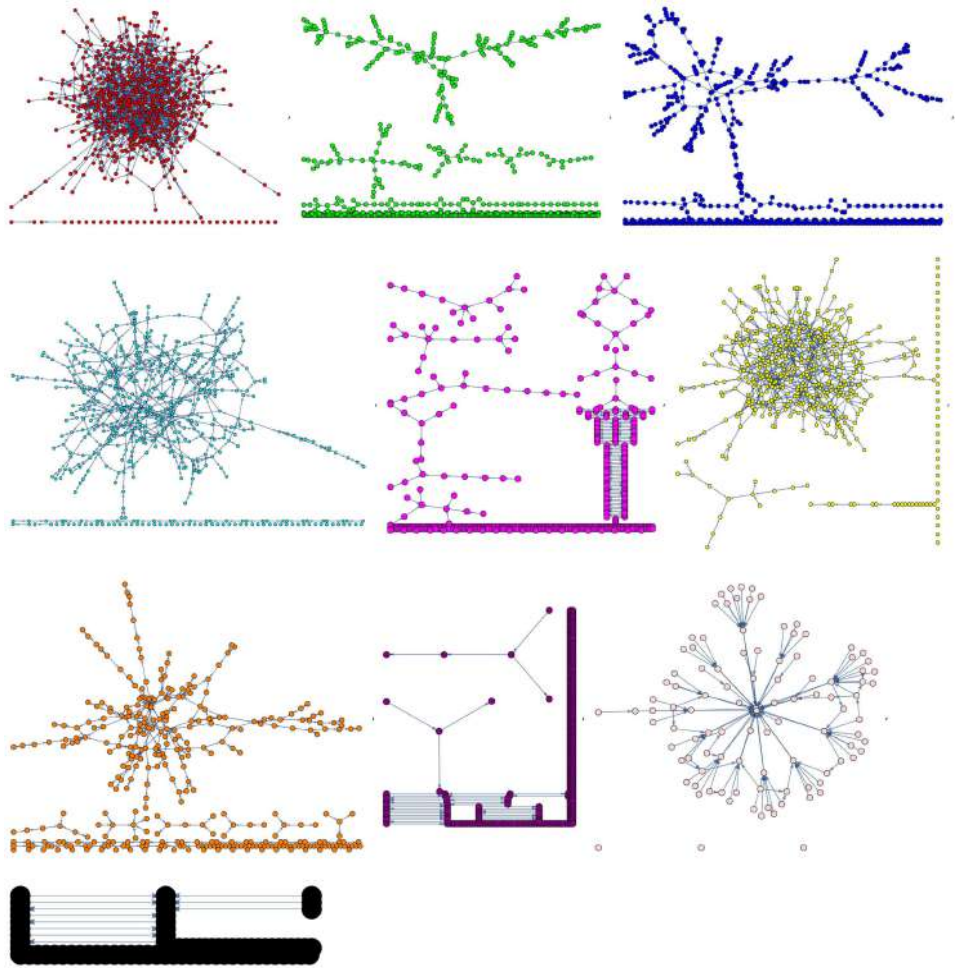

Fig. 4.9 Subgraphs induced by the 10 regular groups of the Gnutella network. The subgraphs are structurally significantly different from each other. For instance, the directed cycle counts of the subgraphs (ordered row by row from left to right) are 139045, 0, 0, 2, 0, 15, 3, 0, 0, 0. The first community might be identified as a core of the network. Corresponding sizes of regular groups can be seen in Fig. 4.8

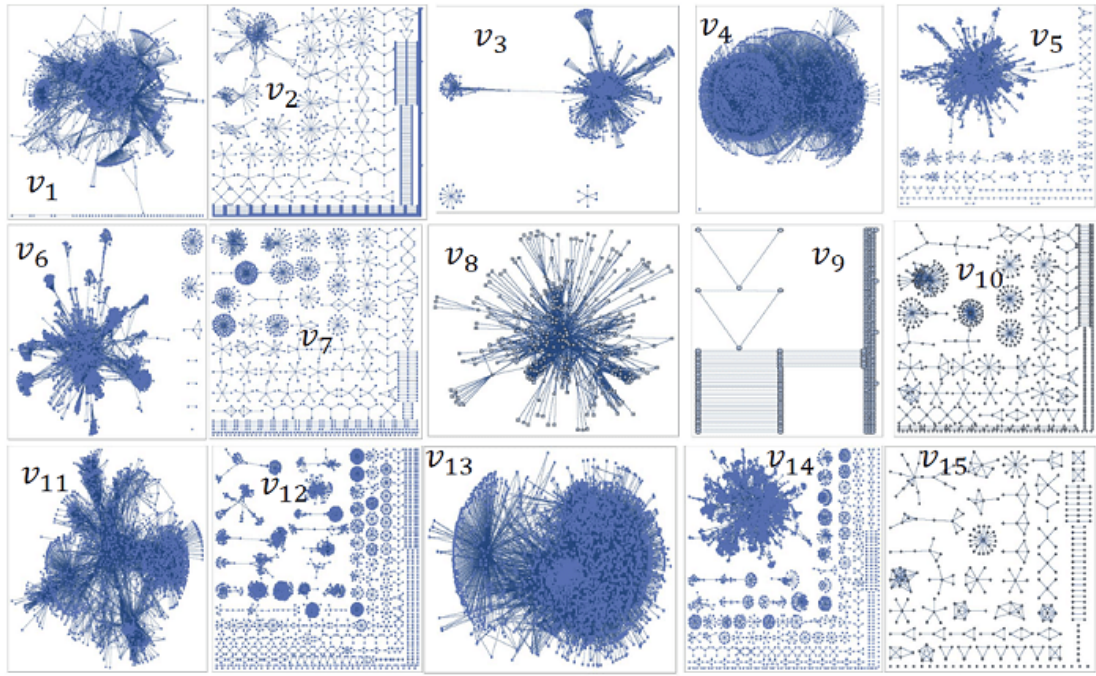

Fig. 4.10 Subgraphs of AS graph induced by 15 regular groups $v_1, \cdots, v_{15}$. The biggest 4 subgraphs are represented by subgraphs induced by around 10 percent of the nodes of the group (these groups are $v_1, v_4, v_{11}, v_{13}$) and the rest of 11 groups are fully depicted.

To enlarge the communities we used the following heuristic. For each node belonging to one of the communities, we include all neighbors of the node to the same group. This can be justified, since such neighbors should have very similar distance patterns as the root nodes. In this way a large proportion of nodes were included in the communities, more than 30 percent of all nodes. The result is partially shown in Fig. 4.10, some of the groups were very large, having around $3 \times 10^5$ nodes, and only part of them are plotted. The found subgraphs are structurally heterogeneous and thus informative. For comparison, most subgraphs induced by a random samples of 1000 nodes contained no links in our experiments.

## 4.8   Concluding remarks

This chapter introduced a new approach for partitioning graphs using observed distances instead of usual path and cycle counts. By design, the algorithm easily scales to very large data sets, linear in the number of target nodes to be partitioned. First experiments presented here with real and synthetic data sets suggest that this method might be quite accurate, and possibly capable of reaching the Kesten-Stigum threshold. However, to be convinced about this, more detailed theoretical studies and more extensive numerical experiments are needed. We also need to estimate quantitatively accuracy of the low-dimensional approximation in synthetic cases like the random power-law graphs.

Spectral methods utilizing the distance matrix as a basis of network analysis are of broader interest, see [15]. We are also interested in finding relations of our concept with graph limits in the case of sparse networks [19], and extending the analytical result to sparse random graph models with nontrivial clustering. We aim to study stochastic block models with more than two groups and the actual distance distributions in such random graphs [54].

We will also find real-life applications for our method in machine learning such as highly topical multilabel classification, [83, 31, 11]. For instance, in case of natural language documents like news release, we can use deep-learning to embed words or paragraphs into points in a vector space. Our graph method could be used to analyze networks of large volumes of such documents. Each document has usually more than one meaningful labeling. We will study possibilities of aiding such a multilabel classification using RD of the training data.

# Chapter 5

# Conclusions

> This is not the end. It is not even the
> beginning of the end. But it is, perhaps,
> the end of the beginning.
>
> Winston Churchill

In this thesis we introduced a principled framework for summarizing large graphs, which has been founded on *Szemerédi's regularity lemma*[99]. The key idea of this work has been to harness the power of the regularity lemma to tackle two of the main challenges of graph summarization: determine in a principled way the cut off between interesting and uninteresting information, and separate the interesting structural information from noise, which is often contained in real-word networks. The strength of the regularity lemma is corroborated by the key lemma [56, 55], which states that, under certain conditions, the partition resulting from the regularity lemma gives rise to a summary, which inherits many of the essential structural properties of the original graph. In particular, the key lemma ensures that every small subgraph of the summary is also a subgraph of the original graph. Hence, these results provide us with a principled way to obtain a concise representation of a large graph by revealing its interesting structural patterns, while filtering out the noise which corrupts these structures.

The original proof of the regularity lemma [99] is not constructive and the algorithmic solutions developed so far have been focused exclusively on *exact* algorithms which have a hidden tower-type dependence on an accuracy parameter. Therefore, with this thesis we describe a new heuristic algorithm based on the exact Alon et al.'s algorithm [4], who proposed the first constructive version of the regularity lemma. The proposed heuristic is an improvement, in terms of the summary quality and noise robustness, of the one introduced by Sperotto and Pelillo [97]. An extensive series of experiments demonstrated the effectiveness and the scalability of our approach. Along this path, we show how the notion of regular partition can provide fresh insights into old pattern recognition and machine learning problems by using our summarization method to address graph-based clustering and image segmentation tasks. In addition, we have successfully validated our framework both on synthetic and real-world graphs showing that it surpasses the state-of-the-art in term of noise robustness.

Being able to build a concise representation of a large graph, we also addressed the graph similarity search problem exploiting our summaries. Since noise is common in any real-world dataset, the biggest challenge in graph search is developing efficient algorithms suited for dealing with large graphs containing noise in terms of missing and adding spurious edges. In our approach, all the graphs contained in a database are compressed off-line, while the query graph is compressed on-line. Thus, graph search can be performed on the summaries, and this allows us to speed up the search process and to reduce storage space. The experimental results showed that our framework is tailored for efficiently dealing with databases containing a high number of large graphs, and, moreover, it is principled robust against noise. This achievement seems of particular interest since, to the best of our knowledge, we are the first to devise a graph search algorithm which satisfies all the above requirements together.

In the last part of the thesis, we studied the linkage among the regularity lemma, the stochastic block model and the minimum description length with the aim of devising an algorithm for analyzing sparse large networks. To this end, we decomposed a sparse graph fitting a stochastic block model by means of the likelihood maximization method. Stochastic Block Model is an important paradigm in network research, see e.g.[3], which is usually resolved around the concept of 'communities'. Our method is able to deal with other type of networks that do not fit well to such a community structure. In particular, we found that our algorithm seems to circumvent a famous problem in community detection: in a sparse network there is a well defined region on graph parameters where it is impossible to find communities, even in the limit of infinite graph size, although definite communities exist by construction [30]. Using a distance matrix-approach, we can find communities even in the case of modest graph sizes. In our approach distances within communities tend to be shorter than distances between communities. This may open a more effective new way for finding strong and even weak communities in large networks.

## Strengths and weaknesses

Although the topic addressed by this thesis, namely a graph summarization framework, is very practical, the foundation of our work is instead deep theoretical. In effect, we harness the power of the regularity lemma, which is one of the great triumphs of the "Hungarian approach" to mathematics: "pose very difficult problems, and let deep results, connections between different areas of math, and applications, come out as byproducts of the search for a solution". From our point of view this is a strength of this work, since it shows that our framework is built on solid foundations. Further points of strength are the ability of our algorithm to determine in a principled way the cut off between interesting and uninteresting information, and the ability of separate structural information from randomness, which are two major challenges of graph summarization. Any weak point of our work should instead be sought in the applications of our framework. Indeed, a weakness of our method is that it is designed to deal only with dense graphs, since the regularity lemma required a dense graph as input. However, the experimental results obtained on sparse real-world networks show that our method performed as well as the stat-of-the-art. Furthermore, it is worth noting that we introduce a principled regular decomposition algorithm which is suited to reveal the main structural pattern even in sparse graphs. Another weak point is related to the time complexity of our algorithms which required a pre-sampling strategy in order to deal with networks of millions of nodes. This demands for the development of a distributed version of the proposed heuristic algorithms.

## Future work

The work of this thesis makes room for further applications in the contexts of structural pattern recognition and of graph mining. In chapter 2, we proposed a two-stage strategy to address the clustering and image segmentation problem. It would be interesting to study how to use our framework for addressing the problem of graph isomorphism for large graphs. The problem of graph search under a similarity measure was addressed in chapter 3 by using the obtained summarieoints of strength are the ability of our algorithm to determine in a principled way the cut off between interesting and uninteresting information, and the ability of separate structural information from randomness, which are two major challenges of graph summarization. Any weak point of our work should instead be sought in the applications of our framework. Indeed, a weakness of our method is that it is designed to deal only with dense graphs, since the regularity lemma required a dense graph as input. However, the experimental results obtained on sparse real-world networks show that our method performed as well as the stat-of-the-art. Furthermore, it is worth noting that we introduce a principled regular decomposition algorithm which is suited to reveal the main structural pattern even in sparse graphs. Another weak point is related to the time complexity of our algorithms which required a pre-sampling strategy in order to deal with networks of millions of nodes. This demands for the development of a distributed version of the proposed heuristic algorithms.

## Future work

The work of this thesis makes room for further applications in the contexts of structural pattern recognition and of graph mining. In chapter 2, we proposed a two-stage strategy to address the clustering and image segmentation problem. It would be interesting to study how to use our framework for addressing the problem of graph isomorphism for large graphs. The problem of graph search under a similarity measure was addressed in chapter 3 by using the obtained summaries. As future work, it would be a good idea to extend our summarization algorithm to deal with labeled large graphs. Furthermore, we think that its. As future work, it would be a good idea to extend our summarization algorithm to deal with labeled large graphs. Furthermore, we think that it is important to develop an efficient algorithm suited to deal with sparse graphs based on the version of the weak regularity lemma introduce by Fox et al. [41]. In chapter 4, we introduced a principled regular decomposition framework based on the interplay among the regularity lemma, the stochastic block model and the minimum description length. It would be interesting to theoretically study why the distance matrix based approach seems to be very efficient. Finally, it would be a good idea to study how to extend the proposed summarization and regular decomposition algorithms to deal with time-evolving graphs.

# Appendix A

# A Computer Vision System for the Automatic Inventory of a Cooler

In this chapter, we describe a system for beverage product recognition through the analysis of cooler shelf images. The extreme objects occlusion, the strong light influence and the poor quality of the images make this task a challenging one. To overcome these limitations, we rely on simple computer vision algorithms, like chamfer and color histogram matching and we introduce simple $3D$ modeling techniques.

## A.1 Introduction

This chapter is devoted to the description of a computer vision system for the automatic inventory of a commercial cooler. The goal is to count, for each brand, the number of beverage products (bottles and cans) contained in the cooler at any given moment in order to efficiently schedule a refill if necessary. This is done through the continuous analysis of the images of the cooler's shelves taken by (low-cost) wide-angle cameras.

Although at first glance the task looks trivial, as the objects to be recognized are clearly distinguishable, rigid and in a well-known static environment, it is in fact a challenging one due to a combination of several factors. In particular, a first difficulty arises from the severe occlusion conditions under which the system has to work. In fact, in a typical scenario involving densely packed shelves, visibility decreases row by row, the rear products being almost completely hidden from the front ones (see Figure A.1 for some typical examples). The items are also typically very close to each other and this makes segmentation and detection more difficult. Recognition is also complicated by the lighting conditions: light is not uniform in the images, not only due to the shadows generated by the shelves and by the products themselves, but also due to the influence of external light. As a result, our images have typically poorly defined edges and distorted color representation, thereby making segmentation and brand classification more difficult. Also, the system has to be flexible enough to recognize new products after software installation. These difficulties are exacerbated by the need to cut off production costs and by the consequent use of low-quality cameras and limited computational resources. Indeed the whole system has to run on an embedded low-performance computer and this poses serious limitations as to the kind of algorithms that can be used, as computationally intensive techniques are clearly not feasible.

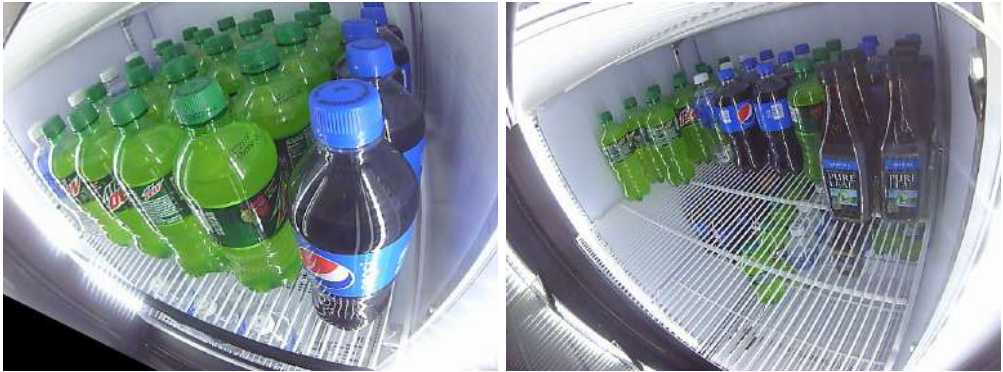

Fig. A.1 Typical images analyzed by our system.

The proposed system uses a combination of simple techniques to address these limitations. It is implemented into a pipeline of simple modules, as shown in Figure A.2. The pipeline begins with an edge detector which extracts the features that will be used by the distance transform module to construct a distance image. The next step in the pipeline is chamfer matching [18], which detects the shape of beverage products by shifting their templates at various locations of the distance image. A matching measure is used to detect a candidate beverage shape, which is then checked by a false positive elimination module. Finally, the brand of the beverage products is recognized using simple color histogram matching. The color histogram of the pixels which lie under a detected shape is compared with the color histograms build from the images of reference products. Despite the simplicity of the used techniques, preliminary results show the effectiveness of the proposed system in terms of both detection accuracy and computational time.

## A.2    The pipeline

The proposed pipeline is based on simple techniques applied in a cascaded way to enhance the recognition accuracy and to provide robustness. As previously mentioned, the pipeline begins with a learning-based edge detector [33] which extracts the most useful product edges that will be used to construct a distance image. This is used by the chamfer matching module [18] in order to detect candidate product shapes which will be checked by the false positive elimination module. The last module is the histogram matching that allows brand recognition. The algorithm is optimized by using 3*D* modeling techniques for template generation and by a space management system which allows faster image scan and avoids the need of a non-maximum suppression. Further accuracy is achieved by splitting a beverage into its main characterizing parts, processing them independently and considering the results as a whole. Occlusion is dealt by building an occlusion mask which keeps track of the image portions occupied by the detected beverages and masks the templates occluded parts. Figure A.2 shows the flow chart of the proposed pipeline.

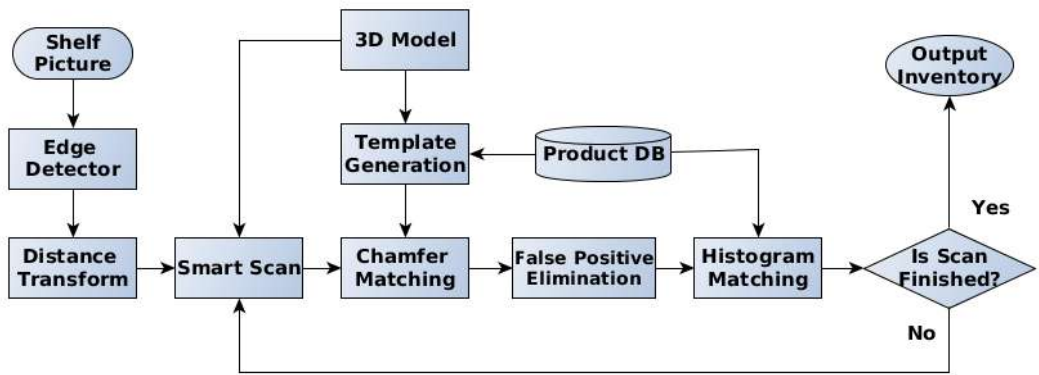

Fig. A.2 Flow-chart of the proposed system.

## A.2.1   Edge detection

Edge detection is the preprocessing stage of the pipeline. It relies on the OpenCV 3.2.0 [51] implementation of the fast edge detector proposed by Dollár and Zitnick [33], which is inspired by the work of Kontschieder et al [58]. It exploits the high interdependence of the edges in a local image patch. In particular, edges exhibit well-known patterns that can be used to train a structured learning model. Dollár and Zitnick's algorithm segments an image into local patches used to train a structured random forest model. This model provides a local edge mask which is applied to extract edges in an accurate and efficient way. Figure A.3 shows edge detection results obtained from Dollár and Zitnick's algorithm.

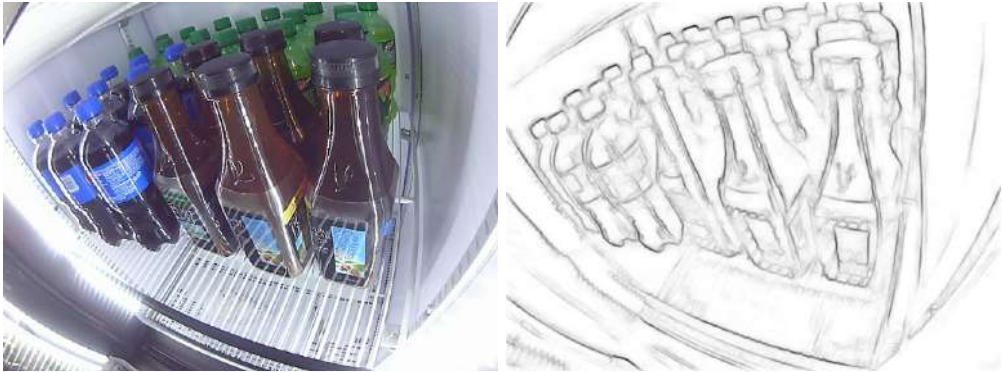

Fig. A.3 Example of the edge detection results: the original shelf image on the left and the edge image on the right.

## A.2.2   Shape detection

Template matching is the first stage of the proposed system in which beverage candidates are evaluated and discarded if they do not satisfy the shape requirements. It relies on a chamfer template matching [50] for the shape detection, on a 3*D* modeling for the template generation, on a smart sliding window for the space management and on a simple yet essential mechanism for the occlusion management.

Chamfer matching is a simple template matching algorithm which offers high performance and a robust detection as it is very flexible and more tolerant to low quality edges than other algorithms of the same kind. First, a morphological transformation, known as distance transform [36], is applied to the previously extracted edges. The resulting picture will be a gray-scale image in which each pixel will have the value of the distance from that pixel to the nearest edge. Finally, a query template is slided onto the distance image. At each position, a matching measure is computed by summing the pixel values of the distance transform image which lie under the edge pixels of the template. If the computed matching measure lies below a certain threshold, the target beverage shape is considered detected. The template threshold should be chosen to achieve a desired trade-off between false positives and false negatives.

Chamfer matching is very inefficient as all beverage templates of varying shape and size have to be tested at each locations of the distance image. Thus, a 3*D* model of the shelf is introduced to speed-up the matching process. It allows to check only one template per product at each location of the distance image avoiding to check, for each products, a bunch of templates of varying shape and size. To achieve this aim, we exploit the available information related to the objects, the cooler and the camera in order to render the shelf and to build the template for the shape matching. In particular, each object is accurately measured as follows: first the bottom diameter is measured, then, going up, for each change in the shape the value of the height and the corresponding diameter are collected. In this way we sum up the product contour as a collection of diameter discontinuities and their relative heights. The beverage partition into contour and horizontal parts can reproduce well most of the bottles and cans, even those which are not circular based, with a little error. Furthermore, camera's intrinsic parameters are collected, while real position of the camera and rotation angles are measured. For this purpose we introduce an artificial reference points in the picture: a special sheet of paper with a printed grid is laid on the shelf, while the same grid is rendered in a 3$D$ representation of that shelf, using the cooler information. At the beginning the virtual grid is in a random position but, using special buttons on the keyboard, a user is able to modify the camera position and the rotation angles in order to match as close as possible the virtual grid with the real grid. When the grids match, we obtain the camera position and the orientation with a good accuracy. This whole process should be done only once, when the camera is installed. Figure A.4 shows the calibration process. Finally, after the calibration step, the template of each product is rendered at any desired point of the shelf (Figure A.5).

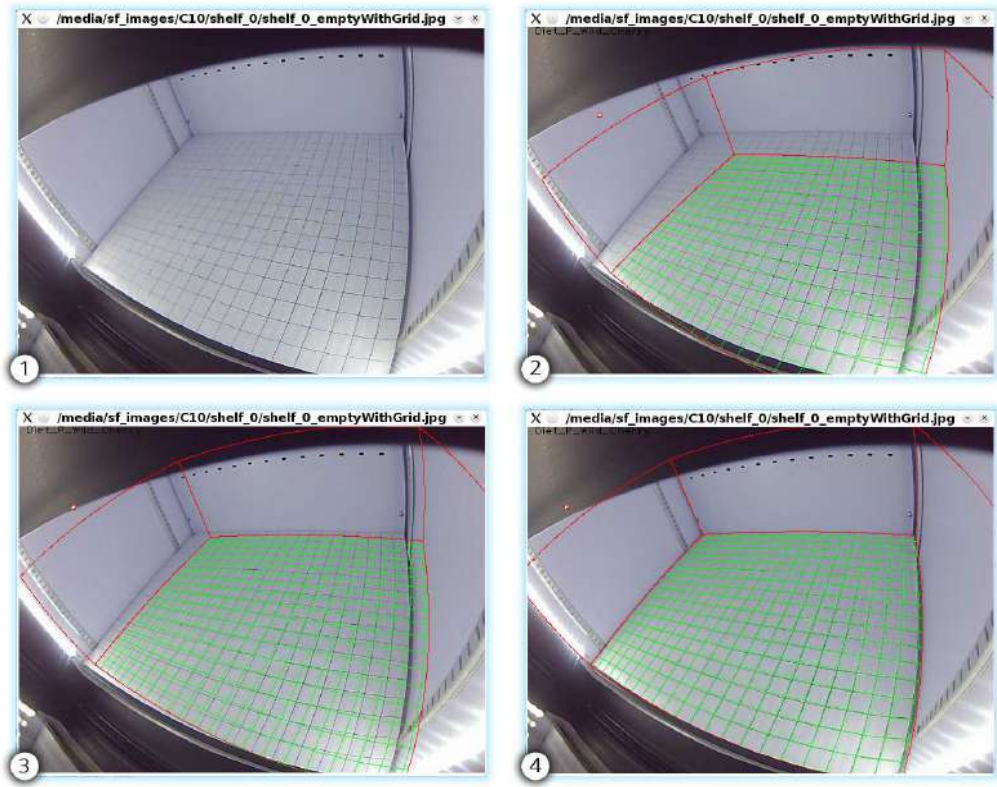

Fig. A.4 Calibration procedure: the goal is to match the grid on the shelf. (1): Real grid in the shelf. (2): Starting virtual grid with predefined camera position and orientation. (3): Close match of the grids. (4): Good grid match; now the camera parameters are known.

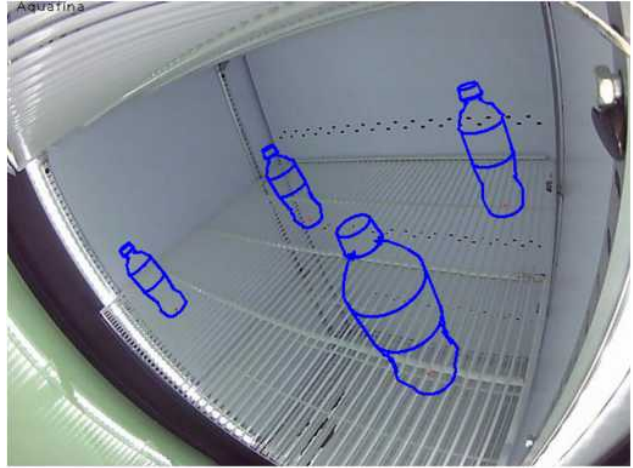

Fig. A.5 Examples of templates generated by the 3D modeling.

To further speed-up the matching process, a smart sliding window for the space management, named smart scan module, is introduced. It relies on a $3D$ shelf model which allows to switch from virtual coordinates (pixels of the image) to physical ones (millimeters of the real shelf) (Figure A.6). The scan is then performed referring to the physical shelf position $(x, z)$ so that the spatial information can be exploited to avoid points in which the template cannot fit due to the lack of space. In particular, the scan starts from the lowest right angle ($x = maxLength$, $z = 0$) and goes up column-wise: at each detection step we keep the x fixed and we increase the $z$ by $step_z$, until the innermost part is reached; then we reset $z$ to 0, we shift left by $step_x$ ($x = x - step_x$) and we start increasing the $z$ again; this procedure goes over until the left highest corner is reached. Thus, the $3D$ model and the smart scan allow to check only one template per product at each permissible position $(x, z)$ speeding up the template matching phase.

To deal with the occlusion conditions, a binary image, called occlusion mask (see figure A.7), keeps track of the detections found at every step. The occlusion mask has the same size of the shelf image, and it can be thought as a sort of shelf shadow doublet: each time a detection is confirmed in an image point, the occlusion mask is updated accordingly by setting to zero all the pixels belonging to the filled template shape at that same point. In this way the occlusion mask will be a binary image in which black pixels denote the scan image space occupied by the products found until that moment, while white pixels denote the free space left. We then update the query template by masking it with the occlusion mask, so that only the visible template portion is used in the subsequent matching. If the remaining template portion is under a certain threshold, it is discarded as not reliable enough. This solution offers good performance while keeping the problem at a very simple level, but it is not always accurate enough as it is based on a strong assumption which sometimes does not hold: products are considered to be picked in order from the visible ones to the most occluded.

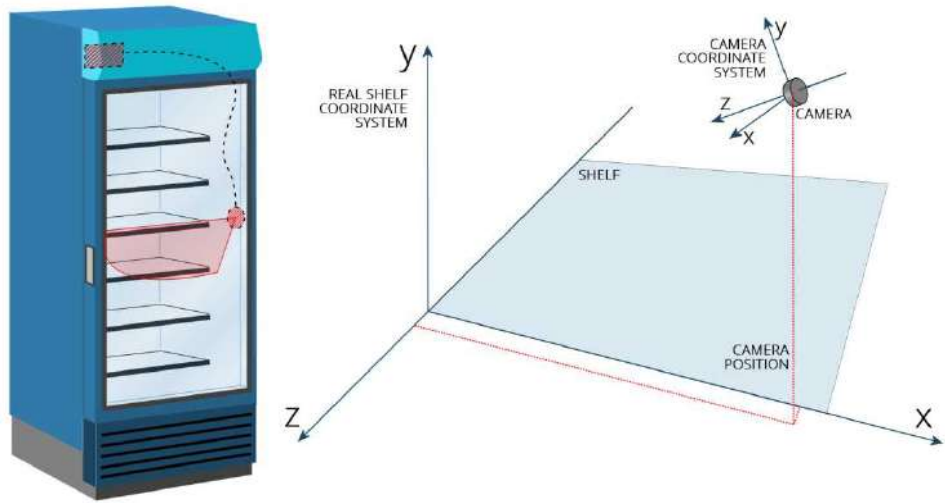

Fig. A.6 Real shelf and camera coordinate systems.

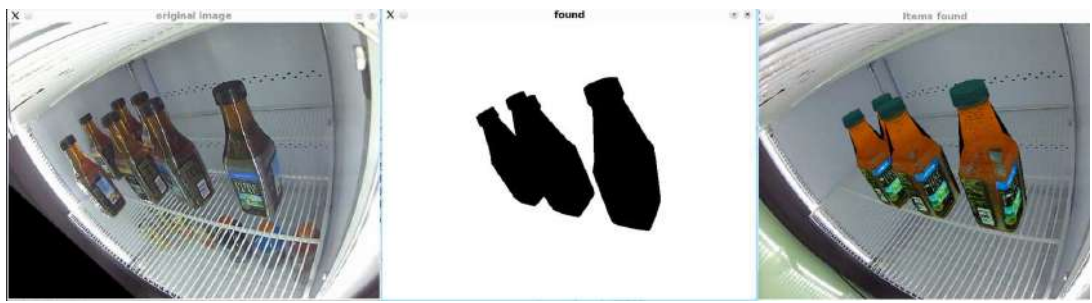

Fig. A.7 Example of the mask image during an ongoing detection. The source image is on the left, the occlusion mask is in the middle and the objects found until that moment are on the right.

Finally, to achieve better accuracy, a procedure known as false positive elimination is performed: each beverage part of a candidate detection is compared against the results achieved by the chamfer matching applied to a reference background image. If the results are too close to each other, the algorithm states that the match is a fake one (the match is a part of the background which is wrongly detected as a real object).

### A.2.3 Color classification

Histogram matching is the second and last stage of the proposed pipeline in which the brand of a previously detected shape is recognized. In particular, the histogram matching module exploits all the elements defining a visual beverage, i.e. shape and color, to enhance the correctness of the shape detection and to recognize the brand of previously detected shapes.

This module relies on the same distinction between the product parts done in the template matching one: a product is split into its main components (cap, bottle liquid and logo for the bottles, the top part of the can and the can surface for the cans) so that it is possible to focus on simple algorithms while keeping the spatial color information (as an example, the cap should be blue while the liquid is green, and not the opposite). It is worth nothing that in the same product part the color is often uniform, so there is no need to split the objects further.

The color analysis is based on simple color histograms [65, 24, 29] guided by the $3D$ model: only the image portion under the filled template is used to build the histogram. The color space is divided into $n$ sub-parts, called bins, covering specific color ranges. Three normalized color histograms, one for each channel, are then computed. Finally, the histogram of each product part is compared against histograms build from the products database in order to decide the fitness of the detection.

The product database contains reference photos of each product the algorithm should recognize. In particular, for each product, a series of photos are snapped in controlled conditions: the middle shelf of the reference fridge is divided into 9 zones and for each zone four pictures are snapped using 90 degrees rotation.

Histogram comparison is based on the following measure:

$$d(H(I), H(I')) = d_{mode}(H(I), H(I'))(1 - H(I) \cap H(I'))    \text{(A.1)}$$

where $H(I)$ and $H(I')$ is a pair of normalized histograms, each containing $n$ bins; $d_{mode}$ is the distance between the bins of each histogram having the highest frequency indexes and $H(I) \cap H(I')$ is the sum of the smallest corresponding bins between two histograms, i.e. the histogram intersection.

The measure (A.1) is a weighted distance which is robust against color distortion because of the modes, while keeping a deeper histogram comparison because of the intersection.

## A.3   Experimental results

We have performed a series of experiments to verify the performances and the accuracy level that can be obtained by our system. All the module of the pipelines have been implemented using GNU C++ and have been run on dual core CPU with 1.6GHz/core and 1 GB of RAM.

The results here presented are divided into two sections:

- the first section shows examples of products placed at random in the shelf;

- the second section shows examples of real cooler cases, where a shelf is filled by columns and each column will contain only bottles/cans of the same brand.

The experiments have been conducted in a $654 \times 594$ mm cooler shelf with 10 beverage brands. For each test it is shown: the original shelf image (on the left); the beverage edge image where detected caps are highlighted in red (in the middle) and, finally, the $3D$ rendering of the products detected by the pipeline (on the right).

### A.3.1   Random shelf configurations

Figure A.8 shows some examples of products randomly placed in the shelf and a few products placed at the rear. The recognition is high, even if some Lipton cans are seen as Kickstarter, since they are very similar; we can also note that the difference between the cans themselves is very little, as just a little part of the logo is different. It is worth to note that Gatorade are detected despite having a different shape from the one in our database: this highlights the algorithm is flexible enough to recognize even unknown products sharing similar properties to the known ones. As for the cans, the Lipton bottle brands (brown bottles) are so similar that it is almost impossible to distinguish between them. Finally, Pepsi and MtnDew (green bottles) have a distinctive color, hence we can achieve a good accuracy on them.

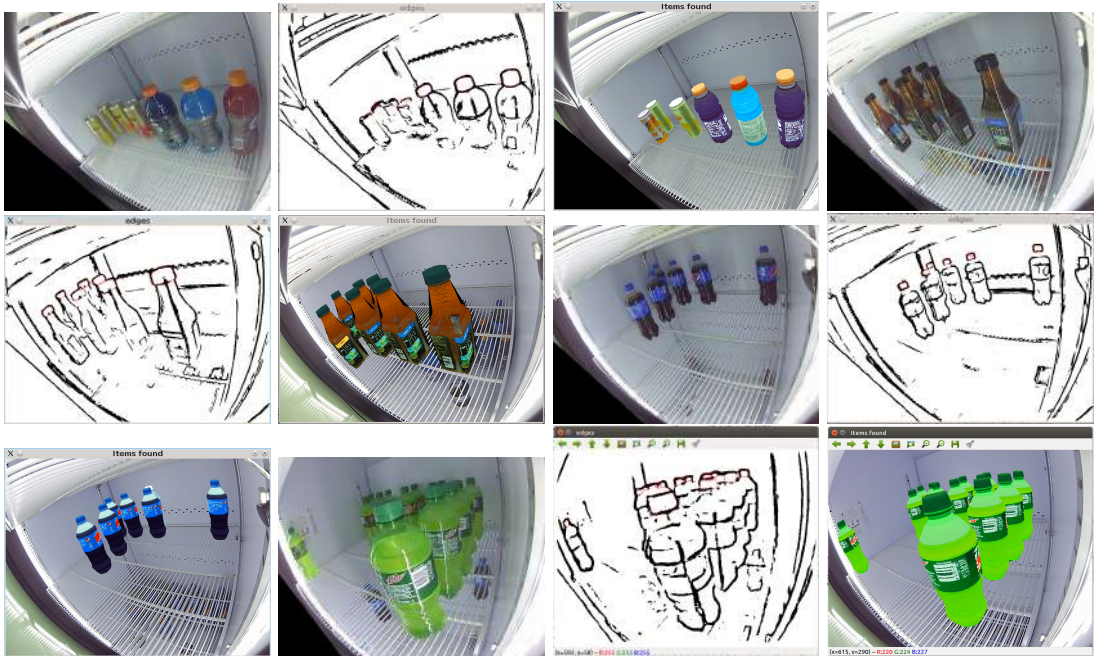

Fig. A.8 Examples of products randomly placed in the shelf and a few products placed at the rear.

## A.3.2 Ordered shelf configurations

Figure A.9 shows some examples of real cooler cases, where a shelf is filled by columns and each column will contain only bottles/cans of the same brand. In the first row there are two tea bottles placed in the rear of an almost empty fridge which are correctly recognized, while the second row there are two Gatorade and three Lipton cans which are correctly recognized too. The cooler is recognized to be almost empty in both cases. In the third row there are some missed Pepsi. This is due to weak edges which are not recognized by the template matching. In the last row, there is a shelf full of bottles and, in this case, some products are missed.

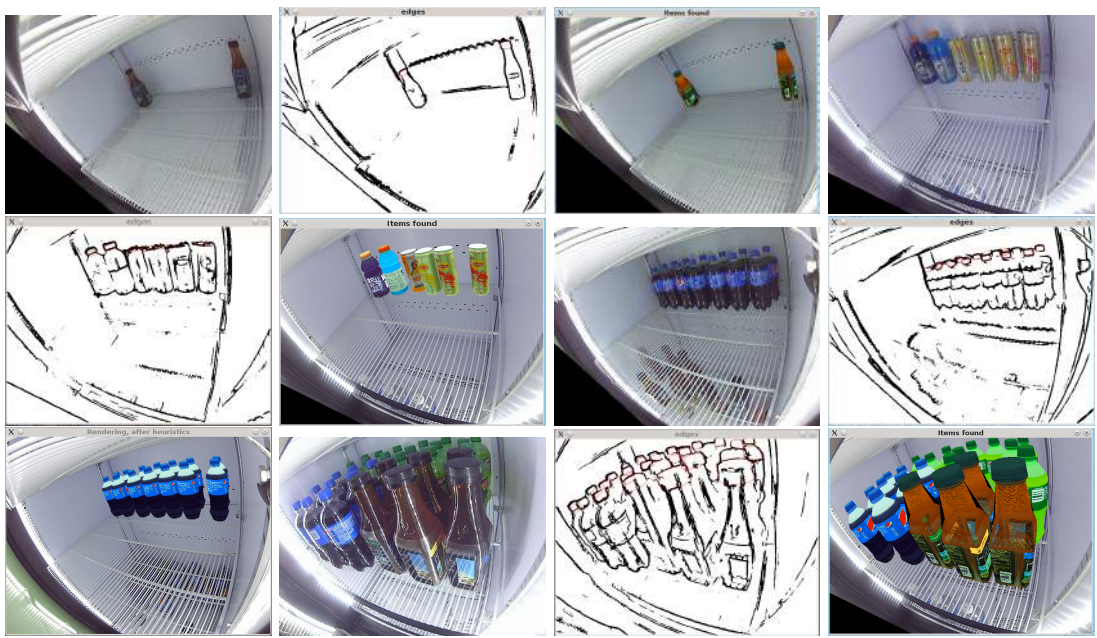

Fig. A.9 Examples of real cooler cases.

From the analysis of 100 experiments we can state that:

- the overall average accuracy level we have obtained is over 80%. In particular, an empty shelf can be identified with 100% precision, while the accuracy decreases to 70% if the shelf is almost full, because of the product occlusion that forces the algorithm to rely only on the top part of the product instead of considering it in its entirety.

- Since the system should send a cooler inventory every 10 minutes, the performances are quite satisfactory, as the whole scan of a 654 × 594 mm cooler shelf takes approximatively 100 seconds.

- Some products are more easily detectable than others since the colors of beverages like Pepsi, MtnDew, Gatorade create a well defined contrast with the background and are very different from the colors of other products. By contrast, Aquafina is very difficult to be identified because of its transparent bottle and its white cap which blends into the background.

## A.4   Concluding remarks

We have described a simple yet effective system for monitoring the content of a commercial cooler through the visual analysis of the shelves' images taken with low-cost wide-angle cameras. The difficulty of this task lies mainly in the challenging set-up in which it has to be carried out, such as severe or almost complete occlusion, uneven lighting conditions, poor image quality, and low-cost hardware. The proposed solution combines simple techniques which effectively work under these challenging conditions.

Despite the simplicity of the used techniques, we achieved a satisfactory accuracy level, being able to detect from 70% to 95% of the whole shelf in 100 images. Since the system should send a cooler inventory every 10 minutes, the computational performances are acceptable as a full shelf scan takes approximately 100 seconds using limited computational resources. Finally, the system is very flexible, as it needs just a simple and quick learning phase to add new products.

In future, we are planning to better handle irregular light intensity and color distortion in order to improve the recognition accuracy.